\documentclass[english,aps,pre,a4paper,floatfix, notitlepage, nofootinbib]{revtex4-1}
\usepackage{lmodern}

\usepackage[latin9]{inputenc}
\usepackage{color}
\usepackage{babel}
\usepackage{varioref}
\usepackage{refstyle}
\usepackage{amsmath}
\usepackage{amsthm}
\usepackage{amssymb}
\usepackage{graphicx}
\usepackage{wasysym}
\usepackage{esint}

\usepackage[unicode=true,pdfusetitle,
 bookmarks=true,bookmarksnumbered=true,bookmarksopen=true,bookmarksopenlevel=2,
 breaklinks=false,pdfborder={0 0 0},backref=false,colorlinks=true]
 {hyperref}

\makeatletter

\AtBeginDocument{\providecommand\secref[1]{\ref{sec:#1}}}
\AtBeginDocument{\providecommand\figref[1]{\ref{fig:#1}}}
\AtBeginDocument{\providecommand\eqref[1]{\ref{eq:#1}}}
\AtBeginDocument{\providecommand\tabref[1]{\ref{tab:#1}}}
\AtBeginDocument{}
\providecommand{\tabularnewline}{\\}
\RS@ifundefined{subref}
  {\def\RSsubtxt{section~}\newref{sub}{name = \RSsubtxt}}
  {}
\RS@ifundefined{thmref}
  {\def\RSthmtxt{theorem~}\newref{thm}{name = \RSthmtxt}}
  {}
\RS@ifundefined{lemref}
  {\def\RSlemtxt{lemma~}\newref{lem}{name = \RSlemtxt}}
  {}

  \theoremstyle{plain}
  \newtheorem{prop}{\protect\propositionname}

\@ifundefined{date}{}{\date{}}

\newref{app}{%
	name      = \RSappendixname,
	names     = \RSappendicesname,
	Name      = \RSAppendixname,
	Names     = \RSAppendicesname,
	rngtxt    = \RSrngtxt,
	lsttwotxt = \RSlsttwotxt,
	lsttxt    = \RSlsttxt}

\renewcommand*\refstepcounter[1]{\stepcounter{#1}%
	\protected@edef\@currentlabel{%
		\csname p@#1\expandafter\endcsname
		\csname the#1\endcsname
	}%
}
\renewcommand{\p@subsection}[1]{\thesection.\thesubsection}

\newref{sub}{name = {Subsection~}, refcmd = {\ref{#1}}}

\renewcommand\[{\begin{equation}}
\renewcommand\]{\end{equation}}

\def \Z {\mathcal{Z}}

\def \N {N}
\def \NN {D}

\iffalse
\def \twocolbr {\nonumber \\ & }
\else
\def \twocolbr {}
\fi

\def \bw {\begin{widetext}}
\def \ew {\end{widetext}}

\newif\ifthesis
\thesisfalse

\newif\ifdetailed
\detailedtrue

\usepackage[caption=false]{subfig}

\renewcommand{\p@subfigure}[1]{\thefigure(\thesubfigure)}

\newlength{\figurewidth}
\renewcommand\revised[1]{#1}

\detailedfalse

\newif\ifnumberalleqs
\numberalleqstrue

\usepackage[titletoc,title]{appendix}

\usepackage{etoolbox}

\ifnumberalleqs

\else

\usepackage{autonum}
\fi

\makeatletter
\newcommand{\restore@Environment}[1]{%
  \AtBeginDocument{%
    \csletcs{#1*}{#1}%
    \csletcs{end#1*}{end#1}%
  }%
}
\forcsvlist\restore@Environment{alignat,equation,gather,multline,flalign,align}

\detailedfalse

\makeatother

  \providecommand{\propositionname}{Proposition}

\begin{document}
	
\title{A Deal with the Devil: From Divergent Perturbation Theory to an Exponentially-Convergent Self-Consistent Expansion} 
	
	\author{Benjamin Remez}
	\author{Moshe Goldstein}
	\affiliation{Raymond and Beverly Sackler School of Physics and Astronomy, 
	Tel Aviv University, Tel Aviv 6997801, Israel}
	\date{\today}
	\begin{abstract}
For many nonlinear physical systems, approximate solutions are pursued by conventional perturbation theory in powers of the non-linear terms. Unfortunately, this often produces divergent asymptotic series, collectively dismissed by Abel as ``an invention of the devil.'' 
\revised{Although a lot of progress has been made on understanding the mathematics and physics behind this, new approaches are still called for.}  A related method, the self-consistent expansion (SCE), has been introduced by Schwartz and Edwards. Its basic idea is a rescaling of the zeroth-order system around which the solution is expanded, to achieve optimal results. 
While low-order SCEs have been remarkably successful in describing the dynamics of non-equilibrium	many-body systems (e.g., the Kardar-Parisi-Zhang equation), its convergence properties have not been elucidated before. 
To address this issue we apply this technique to the canonical partition function of the classical harmonic oscillator with a quartic $gx^{4}$ anharmonicity, for which perturbation theory's divergence is well-known. 
We  obtain the $N$th order SCE for the partition function, which is rigorously found to converge exponentially fast in $N$, and uniformly in $g\ge0$. 
We use our results to elucidate the relation between the SCE and the class of approaches based on the so-called ``order-dependent mapping.''
Moreover, we put the SCE to test against other methods that improve upon perturbation theory (Borel resummation, hyperasymptotics, Padé approximants, and the Lanczos $\tau$-method), and find that it compares favorably with all of them for small $g$ and dominates over them for large $g$. 
\revised{The SCE is shown to converge to the correct partition function for the double-well potential case,  even when expanding around the local maximum.} 
Our treatment is generalized to the case of many oscillators, as well as to any nonlinearity of the form $g|x|^{q}$ with $q\ge0$ and complex $g$. These results allow us to treat the Airy function, and to see the fingerprints of Stokes lines in the SCE.
	\end{abstract}

	\maketitle

\section{\label{sec:Introduction}Introduction}

A recurring theme in physics is the necessity of approximations. Exactly
solvable systems are rare in the landscape of modern science, and
the majority of physical problems cannot be assailed directly without
resorting to any estimates or simplifications. First-principle calculations
must often be replaced by phenomenological models, which are then
only solved approximately, to varying degrees of success. The ability
to make informed, strategic approximations that would allow us to
gain ground, but still capture the essential physics under consideration,
is as much an art as it is a science. To this end, over the last centuries,
practitioners of the natural sciences have collectively amassed an
ever-expanding arsenal of indispensable techniques and methods in
their professional toolbox.

Chief among these tools is the concept of perturbative expansion \cite{boyd1999devil}.
Its premise is simple and elegant: for a given difficult problem,
identify a related but easier one; solve the simpler system exactly;
and apply corrections to obtain an approximate solution to the original
problem. This time-tested principle is applied at all hierarchies
of engagement, from Taylor's theorem to quantum field theory. Broadly
speaking, linear systems, where the cumulative impact of additional
intricacies is additive, are considered ``simple,'' while nonlinear
systems are deemed ``complicated.'' An expansion of a nonlinear
system about its linear counterpart, in powers the non-linearity,
is recognized as a Perturbation Theory (PT). 

\revised{Unfortunately, PT frequently produces an asymptotic, divergent expansion.}
While this in practice provides sensible answers when applied to low
orders (quantum electrodynamics, for instance, is a remarkable example
\cite{feynman1949space}), in principle it denies us high-accuracy,
large-order results, especially for strongly coupled systems. \revised{Over time, the non-perturbative physical processes which stand behind
this behavior (tunneling, instantons, solitons, etc. \cite{ZinnJustinQFT2002})
have been identified, which have also allowed for the development
of sophisticated mathematical techniques for extracting correct physical
results from PT.} These efforts include resummation schemes, such as Borel \cite{costin2008asymptotics},
successive expansions of the PT remainder \cite{berry1990hyperasymptotics},
methods based on the theory of resurgence \cite{Cherman2015,aniceto2018primer},
and various numerical and asymptotic prescriptions \cite{BenderAndOrszag,boyd1999devil}.
\revised{However, these are usually technically difficult to implement, especially
to large order, and typically on a case-by-case basis.} This situation was surmised by Abel, who begrudgingly proclaimed
\cite[loosely translated]{Abel1826}\emph{ ``divergent series are
an invention of the devil, and it is shameful to base on them any
demonstration whatsoever.''} This implies an erroneous application
of a perturbative approach in systems where the fundamental phenomena
are non-perturbative. We stress that the issue is not the divergence
of individual perturbative terms, which is usually tamed by other
means (such as regularization and renormalization \cite{PeskinSchroeder1995}
for IR/UV catastrophes, and secular PT \cite{strogatz} for the time
dependence of nonlinear systems at long times), but rather the divergence
of the PT series itself, even if all its constituent terms are finite.

A related technique, the Self-Consistent Expansion (SCE) \cite{Schwartz2008,cohen2016self},
was introduced by Schwartz and Edwards \cite{Schwartz1992,schwartz1998peierls}
in their study of the Kardar--Parisi--Zhang (KPZ) problem of nonlinear
deposition \cite{KPZ1986}. \revised{The KPZ equation is not amenable to PT},
as Wiese showed \cite{Wiese1998} that above two dimensions, the KPZ
strongly-coupled phase is inherently inaccessible by a perturbative
expansion, regardless of its order. In contrast, the SCE, utilized
at low orders, was remarkably successful in reproducing exact critical
values in one dimension and agreeing with numerical values in higher
dimensions. This method was subsequently applied to more complicated
variants of the KPZ equation \cite{KatzavSchwarz1999,schwartz2002stretched,Katzav2003,KatzavSchwarz2004a,KatzavSchwarz2004b},
and to other problems such as fracture, turbulence, and the XY model
\cite{katzav2006roughness,katzav2007roughness,Edwards2002Lagrangian,Li1996vortex}.

The SCE\textquoteright s central idea is to smartly choose the zeroth-order
system around which one expands, so that the zeroth-order system is
as close as possible, in some sense, to the perturbed one. The SCE
prescription is \cite{Schwartz2008,cohen2016self}: (i) Decide on
the order $\N$ of the expansion. (ii) Split the potential into zeroth-
and first-order terms parametrically as a function of $\N$, namely
\[
V\left(x\right)=V_{0}\left(x;\N\right)+V_{1}\left(x;\N\right)\,.
\]
Fix this partitioning by demanding some ``self-consistent'' criterion,
which ensures that the expansion reproduces some physical property
of the system. Crucially, the self-consistency criterion must depend
on $\N$. This is akin to verifying that the new perturbation $V_{1}$
is ``small.'' Ideally, $V_{0}$ should be a problem which is solvable
exactly. (iii) Expand the solution around the problem $V_{0}$ in
powers of $V_{1}$, up to order $\N$.

The goal of this procedure is to optimize the choice of the zeroth-order
system such that the errors incurred by the expansion are minimal,
as the order $\N$ is increased. At its leading order, the SCE is
often equivalent to the variational method. However, at higher orders
the SCE constitutes a systematic improvement upon it, without introducing
any additional variational parameters. The actual expansion is technically
similar to PT, which is recovered if the separation of the potential
into two terms in step (ii) does not depend on the order of the expansion
$\N$. 

So far, the SCE has only been applied to low orders, and its phenomenal
empirical effectiveness \cite{Schwartz1992,KatzavSchwarz1999,Edwards2002Lagrangian,Katzav2003}
was left unexplained. The goal of this paper is to explore the large-order
properties of the method, by applying it to the toy model of the classical
anharmonic oscillator with a quartic $gx^{4}$ non-linearity, for
which PT diverges for any $g$ \cite{Negele1998}. \revised{As the difficulties presented by this divergence have long since been
resolved (i.e., by resummation, semi-classical analysis, instanton
effects, etc.), this model provides an ideal benchmark for the SCE.}

The SCE is related conceptually to several other schemes, such as
self-similar perturbation theory \cite{yukalov1976vi,Yukalov1999},
order-dependent mapping (ODM) \cite{ZinnJustinSeznec1979,ZinnJustinODM2010},
optimized perturbation theory (OPT) \cite{Stevenson1981}, and the
linear delta expansion (LDE) \cite{JonesLDE1990}, most of which were
introduced in contexts of high-energy physics, and which were subsequently
applied to problems such as scattering cross sections in quantum chromodynamics
\cite{mattinglyStevenson1994,stevenson2013optimization}, spontaneous
breaking of supersymmetry \cite{abdallaSUSYLDE2009}, the Gross-Neveu
and Nambu--Jona-Lasinio models \cite{kneur2007emergence,kneur2010NambuJonaLasinio},
critical $O\left(N\right)$ field theory \cite{BraatenFadescuOn2002},
melting and crystallization \cite{YukalovMelting1985}, Ising and
nonlinear $\sigma$ models \cite{Hirofumi2007LDEIsingNLSM}, and Bose-Einstein
condensation \cite{kneur2003OPTBoseEinsteinBEC,deSouzeCruz2001TransitionTemperatureBoseGas}.
Contrary to these methods, the SCE offers a flexibility in the condition
which determines the splitting of $V\left(x\right)$, and provides
criteria which are motivated physically rather than mathematically.
Thus, we study the SCE and find rigorously its regime of convergence
and convergence rate, and reveal its relationship with the other schemes
and their convergence properties (previously explored in Refs. \cite{ZinnJustinSeznec1979,BuckleyDuncanJonesZeroDimension,GuidaKonishiSuzuki1995,GuidaKonishiSuzuki1996}).
Moreover, our simplified approach will allow us to then extend our
treatment to the double well case, higher anharmonicities, many coupled
oscillators, and, most importantly, the complex coupling case vis-à-vis
the Stokes phenomenon. 

Our main result is a proof that by imposing self-consistency (i.e.,
determining the splitting of $V\left(x\right)$) through the moments
$\left\langle x^{2M}\right\rangle $ where $M=M\left(\N\right)$,
the SCE is a convergent approximation scheme, \revised{and thus, as opposed to PT, does not require non-perturbative corrections}.
We show this convergence is exponentially fast and uniform in $g>0$,
and provide a lower bounds on its rate at $10^{-c\N}$ with $c$ a
constant, if $M\left(\N\right)\sim N$, and also find the rates of
convergence for other scalings of $M\left(\N\right)$. \revised{This is extended to the double-well potential, showing the SCE can
also be formulated around the oscillator's origin and still converge
to the correct result, while a PT must instead be expanded around
the minima of the potential.} We also show that the SCE compares favorably against other asymptotic
and numerical approximation schemes, with striking supremacy in the
strongly-coupled regime. We then generalize our results to arbitrary
power-law $g\left|x\right|^{q}$ perturbations, as well as to many
coupled degrees of freedom. Applying the former results for $q=3$,
and extending our treatment to the complex parameter plane, we inspect
the Airy function $\mathrm{Ai}\left(z\right)$, a representative WKB
application, where we show that the SCE gives a correct description
of $\mathrm{Ai}\left(z\right)$ in the entire $z$ plane, and explore
the manifestation of the Stokes lines of $\mathrm{Ai}\left(z\right)$
in this new formalism.

The rest of this paper is organized as follows: In \secref{The-SCE-of-Z4}
we derive the explicit form of the SCE for the partition function
of the classical anharmonic oscillator. \secref{convergence_q_4}
provides the proof of the SCE's convergence, where we place bounds
on the remainder of the expansion, and identify the domain of convergence
of the self-consistency parameter $M\left(\N\right)$. \secref{numerical_results_q_4}
demonstrates the actual numerical performance of the SCE, showing
good agreement with our analytical results. In \secref{double_well}
we show SCE's success even in the case of a double-well potential.
We generalize our results to powers $q>0$ in \secref{General-q-convergence},
and apply it to SCE of $\mathrm{Ai}\left(z\right)$ in \secref{SCE-of-airy-function-Ai}.
We briefly sketch an argument for the SCE's convergence in the case
of many coupled oscillators in \secref{Multiple-DOF}. Lastly, we
offer our conclusions and outlook for future work in \secref{Conclusions-and-Outlook}.

\section{\label{sec:The-SCE-of-Z4}The SCE for the Anharmonic Oscillator}

\subsection{Divergence of PT}

\revised{A canonical example of the limitation of PT is found in the anharmonic
oscillator}. The simplest anharmonicity that keeps global stability is a quartic
perturbation, which we write
\begin{equation}
V(x)=\frac{1}{2}\gamma x^{2}+g_{0}x^{4}\,,\label{eq:anharmonic_potential_V}
\end{equation}
with $g_{0}>0$. This potential refines models of ideal binding interactions
and also lends itself to $\phi^{4}$ field theories. Coupled to a
thermal bath with inverse temperature $\beta$, the system is described
completely by a single parameter, the effective coupling $g=\frac{g_{0}}{\beta\gamma^{2}}$.
Its partition function is then given, up to constants, by\footnote{We assume $\gamma>0$. Otherwise, the sign of the quadratic term should
be flipped. The complementary case will be discussed in \secref{double_well}.} 
\begin{equation}
\Z\left(g\right)=\int_{-\infty}^{\infty}e^{-\left[\frac{1}{2}x^{2}+gx^{4}\right]}dx\,,\label{eq:Z_definition}
\end{equation}
It so happens that this partition function has a closed-form expression,
\begin{equation}
\Z\left(g\right)=\sqrt{\frac{1}{8g}}e^{\frac{1}{32g}}K_{\frac{1}{4}}\left(\frac{1}{32g}\right)\,,\label{eq:Z_analytic}
\end{equation}
where $K_{\nu}(x)$ is the modified Bessel function of the second
kind \cite{abramowitz1964handbook}. 

One may wish to make a perturbative expansion of $\Z$ in small $g$,
which would correspond to the asymptotic expansion of the Bessel function
for a large value of its argument. This, however, leads to an asymptotic
series which diverges for all $g$ \cite[Sec. 2]{Negele1998}, 
\begin{align}
\Z & \neq\sum_{n=0}^{\infty}\int_{-\infty}^{\infty}e^{-\frac{1}{2}x^{2}}\frac{\left(-gx^{4}\right)^{n}}{n!}dx\twocolbr=\sum_{n=0}^{\infty}\frac{\sqrt{2}}{n!}\left(-4g\right)^{n}\Gamma\left(2n+\frac{1}{2}\right)\,.\label{eq:Divergent_Z_expansion}
\end{align}
This occurs because the often-useful exchange of summation and integration,
$\int\left[\lim_{\N\rightarrow\infty}\sum_{n}^{\N}\right]\rightarrow\lim_{\N\rightarrow\infty}\int\sum_{n}^{\N}$
is invalid \cite{PerniceOleaga1998}. More generally, this divergence
was expected by virtue of an argument due to Dyson \cite{Dyson1952}:
The system is unstable for negative $g$ (as the potential would cease
to be binding), therefore its PT must diverge for any $g>0$. A similar
situation occurs when evaluating the ground state energy in the quantum
mechanical version of the problem \cite{BenderWu1969,BenderWu1973}.

\revised{Of course, the divergence of this model is well understood. In particular,
the divergent series (\ref{eq:Divergent_Z_expansion}) is Borel-resummable
to the correct result, and we enjoy the knowledge of the closed-form
solution (\ref{eq:Z_analytic}). However, for more interesting problems
this is oftentimes not the case, due to two counts: Chiefly, some
problems may not be resummable, (such as the double-well case which
we address in \secref{double_well}) unless PT is supplemented by
non-perturbative effects. Second, the application of the resummation
procedure presents its own challenges, such as knowledge of the late
PT terms, which may not be available. Therefore, we will use the anharmonic
oscillator as a test-bed and show that a convergent series may instead
be obtained by applying the SCE.}

\subsection{SCE Around a Modified Oscillator}

Ref. \cite{Schwartz2008} offers a treatment of the anharmonic oscillator
by expanding its Fokker-Planck equation of motion. Here we pursue
another approach: In the language of the SCE, instead of expanding
the system around the harmonic term, we expand around a modified harmonic
potential, whose strength is consistently varied to obtain an optimal
approximation. We thus split the potential into two terms\footnote{We denote the adjusted harmonic coefficient by $G$ instead of $\Gamma$
which was used in \cite{Schwartz2008}, because of the prevalence
of gamma functions in the following calculations. We also differ by
our convention for the coupling strength, denoting it by $g$ instead
of $\frac{g}{4}$.}
\[
V(x)=\frac{1}{2}Gx^{2}+\left[\frac{1}{2}\left(1-G\right)x^{2}+gx^{4}\right]=V_{0}\left(x\right)+V_{1}\left(x\right)\,,
\]
where $G$ is the coefficient of the new harmonic potential which
constitutes our zeroth-order system. However, for any choice of constant
$G\left(g\right)$, we run into the same difficulty as that of a naive
expansion, and obtain a divergent series. The crucial principle of
SCE is that $G$ should also depend on the order of the approximation,
so\footnote{We henceforth completely suppress the dependence of $G$ on $g$ .
We will frequently also drop the explicit dependence on $\N$, but
it is crucial to remember that it exists.} $G=G\left(\N,g\right)$ where $\N$ is the order of the expansion.
The SCE is then an expansion in powers of $V_{1}$ up to $\N$, along
with a proper choice of $G\left(\N\right)$.

A drawback of the approach employed in Ref. \cite{Schwartz2008},
which aims to calculate the moments $\left\langle x^{2k}\right\rangle $,
is that the successive terms in their expansions are obtained by a
recurrence relation. For a given order and moment, these entail the
calculation of all lower moments at the same order, as well as many
higher moments to lower order. We therefore concentrate on the partition
function itself, from which all moments may be derived. 

We wish to evaluate

\begin{equation}
\Z=\int_{-\infty}^{\infty}e^{-\frac{1}{2}Gx^{2}-\left[\frac{1}{2}\left(1-G\right)x^{2}+gx^{4}\right]}dx\,,\label{eq:SCE_original_integral}
\end{equation}
for which the choice of $G$ is still arbitrary. The perturbative
expansion would then be 
\begin{align}
\Z & =\int_{-\infty}^{\infty}\lim_{\N\rightarrow\infty}e^{-\frac{1}{2}G\left(\N\right)x^{2}}\sum_{n=0}^{\N}\frac{1}{n!}\left(-\left[\frac{1}{2}\left(1-G\left(\N\right)\right)x^{2}+gx^{4}\right]\right)^{n}dx\,,\label{eq:SCE_expanded_integral}
\end{align}
where $n$ enumerates the order of each term. Again, swapping integration
with the limit is not permissible. However, if we instead truncate
the expansion at a fixed order $\N$, we may compute a finite-order
approximation. Note that after truncation, we have lost the arbitrariness
of $G$, whose value now impacts the numeric efficacy of the approximation.
Binomial expansion then gives: \bw 
\begin{align}
\Z^{\left(\N\right)} & =\int_{-\infty}^{\infty}e^{-\frac{1}{2}G\left(\N\right)x^{2}}\sum_{n=0}^{\N}\frac{\left(-1\right)^{n}}{n!}\sum_{l=0}^{n}\binom{n}{l}\left[\frac{1}{2}(1-G\left(\N)\right)x^{2}\right]^{n-l}\left[gx^{4}\right]^{l}dx\nonumber \\
 & =\sum_{n=0}^{\N}\frac{\left(-1\right)^{n}}{n!}\sum_{l=0}^{n}\binom{n}{l}2^{l-n}\left(1-G\left(\N\right)\right)^{n-l}g^{l}\int_{-\infty}^{\infty}e^{-\frac{1}{2}G\left(\N\right)x^{2}}x^{2n+2l}dx\nonumber \\
 & =\sum_{n=0}^{\N}\frac{\left(-1\right)^{n}}{n!}\sum_{l=0}^{n}\binom{n}{l}2^{l-n}\left(1-G\left(\N\right)\right)^{n-l}g^{l}\left(\frac{G(\N)}{2}\right)^{-n-l-\frac{1}{2}}\Gamma\left(n+l+\frac{1}{2}\right)\,.
\end{align}
 \ew 

Thus, we find the $\N^{\text{th}}$ order expansion of $\Z$, \bw
\begin{align}
\Z^{\left(\N\right)} & =\sqrt{\frac{2}{G\left(\N\right)}}\sum_{n=0}^{\N}\left[1-\frac{1}{G\left(\N\right)}\right]^{n}\sum_{l=0}^{n}\binom{n}{l}\frac{\Gamma\left(n+l+\frac{1}{2}\right)}{n!}\left(\frac{\left(1-G\left(\N\right)\right)G\left(\N\right)}{4g}\right)^{-l}\,.\label{eq:SCE_moments_full}
\end{align}
 \ew

\subsection{The Self-Consistent Criteria for $G$}

Lastly, we require a choice of the function $G\left(\N\right)$. In
the ODM it is determined solely based on mathematical convergence
properties, while for the OPT and the LDE this is usually done by
one of two common criteria: the principle of minimal sensitivity (PMS),
or the principle of fastest apparent convergence (FAC) \cite{Stevenson1981}.
The PMS stipulates that since $G$ is a synthetically introduced parameter,
it should be fixed to a value at which the expansion is stationary,
\[
\left.\frac{d\Z^{\left(\N\right)}}{dG}\right|_{G_{PMS}\left(\N\right)}=0\,.
\]
In a related vain, FAC requires that $G$ be fixed so that two subsequent
approximations agree numerically, that is 
\[
\left.\Z^{\left(\N\right)}\right|_{G_{FAC}\left(\N\right)}=\left.\Z^{\left(\N-1\right)}\right|_{G_{FAC}\left(\N\right)}\,,
\]
which is equivalent to demanding that the final $n=\N$ term in \eqref{SCE_moments_full}
is zero. Since these schemes require inspection of the expansion at
high order, they are often difficult to implement explicitly.

In the SCE, $G$ is chosen to be ``self-consistent,'' in the sense
that as $\N$ increases, the expansion still reproduces faithfully
some physical feature of the system. To this end, in Ref. \cite{Schwartz2008}
the authors suggest picking $G$ for which the first order correction
to some even moment $\left\langle x^{2M}\right\rangle $ vanishes
(i.e., $\left\langle x^{2M}\right\rangle ^{\left(1\right)}=\left\langle x^{2M}\right\rangle ^{\left(0\right)}$).
As this criterion is first-order for any $\N$, it is easily evaluated,
and was found to be \cite[Eq. (21), up to change of notation of $g$]{Schwartz2008}
\begin{equation}
G\left(M\right)=\frac{1}{2}+\frac{1}{2}\sqrt{1+16\left(M\left(\N\right)+2\right)g}\,,\label{eq:First_order_G}
\end{equation}
where the dependence on $\N$ has been abstracted by the dependence
$M\left(\N\right)$. 

We may show that this result is reproduced in our formulation of the
SCE. To first order in the SCE perturbation, the partition function
is 
\begin{align}
\Z^{\left(1\right)} & =\sqrt{\frac{2}{G}}\Gamma\left(\frac{1}{2}\right)\left(1+\frac{1}{2}\left(\left(1-\frac{1}{G}\right)-\frac{4g}{G}\frac{3}{2}\right)\right)\,,\label{eq:First_order_Z}
\end{align}
while the corresponding moment $x^{2M}$ would be \bw 
\begin{align}
\left[\Z\cdot\left\langle x^{2M}\right\rangle \right]^{\left(1\right)} & =\left(\frac{2}{G}\right)^{\frac{1}{2}+M}\Gamma\left(M+\frac{1}{2}\right)\left(1+\left(M+\frac{1}{2}\right)\left(\left(1-\frac{1}{G}\right)-\frac{4g}{G}\left(\frac{3}{2}+M\right)\right)\right)\,,
\end{align}
 \ew so we can find the first order correction, \bw 
\begin{align}
\left\langle x^{2M}\right\rangle ^{\left(1\right)} & =\left\langle x^{2M}\right\rangle ^{\left(0\right)}\left[1+\left(M+\frac{1}{2}\right)\left(\left(1-\frac{1}{G}\right)-\frac{4g}{G}\left(\frac{3}{2}+M\right)\right)-\frac{1}{2}\left(\left(1-\frac{1}{G}\right)-\frac{4g}{G}\frac{3}{2}\right)\right]\,.
\end{align}
 \ew \ifdetailed
\else Demanding that
$\left\langle x^{2M}\right\rangle ^{\left(1\right)}=\left\langle x^{2M}\right\rangle ^{\left(0\right)}$,
we get \eqref{First_order_G}. \fi For convenience we will define
the symbol $K=M+2$, so $G\left(K\right)=\frac{1}{2}+\frac{1}{2}\sqrt{1+16gK}$.
The motivation for this notation will become apparent in \secref{General-q-convergence}.
A useful inequality for $G$ is then:
\[
\max\left(1,\,\frac{1}{2}+2\sqrt{gK}\right)<G\left(K\right)<1+2\sqrt{gK}\,.
\]

A particular convenience of this choice is that 
\[
\frac{(1-G)G}{4g}=\frac{1}{16g}\left(1-\sqrt{1+16Kg}\right)\left(1+\sqrt{1+16Kg}\right)=-K\,,
\]
so now the expansion takes the form \bw 
\begin{equation}
\Z^{\left(\N\right)}=\sqrt{\frac{2}{G\left(K\left(\N\right)\right)}}\sum_{n=0}^{\N}\left[1-\frac{1}{G\left(K\left(\N\right)\right)}\right]^{n}\sum_{l=0}^{n}\left(-1\right)^{l}\binom{n}{l}\frac{\Gamma\left(n+l+\frac{1}{2}\right)}{n!K^{l}\left(\N\right)}\,.\label{eq:SCE_partition}
\end{equation}
\ew The sum over $l$ may be expressed through the confluent hypergeometric
function $_{1}F_{1}$$\left(-n,\frac{1}{2}-2n,-K\right)$ \cite[Chap. 13]{NIST:DLMF},
which proves useful for numerical evaluation. This representation,
while not conducive to our proof of convergence, admits a direct error
estimate which is explored in \appref{Direct_1F1_estimation}.

It now remains to be seen whether this expansion is convergent, and
if so, whether it is equal to the required partition function, namely
if 
\begin{equation}
\lim_{\N\rightarrow\infty}\Z^{\left(\N\right)}\overset{?}{=}\Z\,.\label{eq:SCE_convergence_conjecture}
\end{equation}

Ref. \cite{Schwartz2008} employs the choice of $M\left(\N\right)=\N$.
While this produced good results empirically for small $\N$, the
convergence of this approximation as $\N$ tends to infinity was left
unexplored. We will proceed to show that the sequence $\Z^{\left(\N\right)}$,
with $M\propto\N$ and a sufficiently large proportionality constant,
converges to the exact result in the limit of $\N\rightarrow\infty$.
A ratio $M/\N$ of unity, as in Ref. \cite{Schwartz2008}, is adequate,
but not optimal.

\section{\label{sec:convergence_q_4}Convergence Properties of the SCE}

Let us state and prove our main result for the convergence properties
of the SCE; its relation to the convergence properties and their proofs
for the related schemes mentioned above will be discussed at the end
of this Section. We will show that the following proposition holds:
\begin{prop}
\label{Proposition_Combined}Let the self-consistently conserved moment\footnote{We will see in \secref{General-q-convergence} that while the moment
$M$ carries a physical interpretation, its mapping to $K$ is model-specific,
and depends on details such as system dimentionality and the exact
form of the perturbation. For more complicated problems, the exact
mapping might not be attainable explicitly. Therefore, our perspective
on the SCE's properties focuses on $M\left(\N\right)$, and not $K$. } $\left\langle x^{2M\left(\N\right)}\right\rangle $ scale as $M\left(\N\right)\sim\N^{p}$.
Then depending on the value of $p$,

1. For $0\le p<1$, the SCE is divergent, with $\lim_{\N\rightarrow\infty}\Z^{\left(\N\right)}=\left(-\right)^{\N}\infty$.

2a. For $1\le p<2$, the SCE converges to the correct result, $\lim_{\N\rightarrow\infty}\Z^{\left(\N\right)}=\Z$.
Convergence is uniform for any $g>0$, and its rate is bounded by
$\exp\left\{ -A\N^{2-p}-B\left(g\right)\N^{1-p/2}\right\} $, where
$A,\,B>0$ are independent of $\N$, and $A$ is independent of $g$. 

2b. The borderline case $p=1$ converges, uniformly and exponentially
fast, only if $\alpha\equiv M\left(\N\right)/N$ is sufficiently large.
Below we estimate the minimal satisfactory value, and find $A\left(\alpha\right)$
and $B\left(\alpha,g\right)$. 

3. For $p>2$, the expansion is convergent. Unfortunately, $\lim_{\N\rightarrow\infty}\Z^{\left(\N\right)}=0$.
\end{prop}
Case 1 is a trivial reproduction of the divergence of standard PT,
and will be shown in \appref{Direct_1F1_estimation}. We now concentrate
on cases 2 and 3, and most specifically on 2b. These results will
be generalized to any arbitrary anharmonicity $g\left|x\right|^{q}$
in \secref{General-q-convergence}, where case 2 corresponds to $1\le p<\frac{q}{q-2}=1+\frac{2}{q-2}$,
and the convergence rate will be replaced by $\exp\left\{ -A\N^{1-(p-1)(q-2)/2}-B\left(g\right)\N^{1-p\left(1-2/q\right)}\right\} $.

\subsection{Convergence to $\Z$ and Its Rate}

For our proof we will denote explicitly the limiting operations involved
in the definitions of the summations of infinite series, in order
to emphasize that we never stumble into the same pitfall as regular
perturbation theory. Our proof begins by examining: \bw 
\begin{align}
\lim_{\N\rightarrow\infty}\Z^{\left(\N\right)} & =\lim_{\N\rightarrow\infty}\int_{-\infty}^{\infty}e^{-\frac{1}{2}G\left(K\left(\N\right)\right)x^{2}}\sum_{n=0}^{\N}\frac{1}{n!}\left(-\left[\frac{1}{2}\left(1-G\left(K\left(\N\right)\right)\right)x^{2}+gx^{4}\right]\right)^{n}dx\nonumber \\
 & =\lim_{\N\rightarrow\infty}\sqrt{\frac{2}{G}}\int_{0}^{\infty}e^{-u}\sum_{n=0}^{\N}\frac{1}{n!}\left(-\left[\frac{1}{2}\left(1-G\right)\frac{2u}{G}+g\left(\frac{2u}{G}\right)^{2}\right]\right)^{n}\frac{du}{\sqrt{u}}\nonumber \\
 & =\lim_{\N\rightarrow\infty}\sqrt{\frac{2}{G}}\int_{0}^{\infty}e^{-u}\sum_{n=0}^{\N}\frac{1}{n!}\left(1-\frac{1}{G}\right)^{n}\left(u-\frac{4g}{G\left(G-1\right)}u^{2}\right)^{n}\frac{du}{\sqrt{u}}\nonumber \\
 & =\lim_{\N\rightarrow\infty}\sqrt{\frac{2K\left(\N\right)}{G\left(K\left(\N\right)\right)}}\int_{0}^{\infty}e^{-Kv}\sum_{n=0}^{\N}\frac{K^{n}\left(\N\right)}{n!}\left(1-\frac{1}{G\left(K\left(\N\right)\right)}\right)^{n}v^{n}\left(1-v\right)^{n}\frac{dv}{\sqrt{v}}\,.\label{eq:Z_limit_integral_form}
\end{align}
 \ew 

Indeed, this equation is similar to \eqref{SCE_expanded_integral},
apart from the flip of integral and limit. We stress that we do not
assume \emph{a priori }that this exchange preserves the value of the
expansion. The aim of this proof is to show that this is true only
under certain restrictions on $K\left(\N\right)=M\left(\N\right)+2$.
Inside the integral we have the exponential function's Taylor series.
It is evident that if this would be summed to infinity (i.e., the
limit was inside the integral), we would return to the original integral
(\ref{eq:SCE_original_integral}), whose result is independent of
$\N$. Thus, the error associated with the expansion is due to the
remainder of the Taylor series, which is truncated before integration. 

We note that the integrand has three distinct regions where its behavior
is qualitatively different:
\begin{gather}
\mathcal{D}_{1}=\left[0,1\right],\qquad\mathcal{D}_{2}=\left[1,\infty\right)=\mathcal{D}_{2A}\cup\mathcal{D}_{2B},\qquad\mathcal{D}_{2A}=\left[1,2\right],\qquad\mathcal{D}_{2B}=\left[2,\infty\right)\,.\label{eq:Definition_of_D_Domains}
\end{gather}
These domains correspond to the regions where each term in the potential
$V_{1}$ dominates: in $\mathcal{D}_{1}$ it is the harmonic term
while in $\mathcal{D}_{2}$ it is the quartic. We thus denote the
remainders due to domain $\mathcal{D}_{1}$ and $\mathcal{D}_{2}$
as $R_{1}^{\left(\N\right)}$ and $R_{2}^{\left(\N\right)}$, respectively,
so we have 
\[
\lim_{\N\rightarrow\infty}\Z^{\left(\N\right)}=\lim_{\N\rightarrow\infty}\left[\Z+R_{1}^{\left(\N\right)}+R_{2}^{\left(\N\right)}\right]\,.
\]

\subsection{The Domain $\mathcal{D}_{1}$}

In this domain, the remainder can be bounded explicitly. The error
is negative, and is the sum of the truncated terms in the exponent's
Taylor series: \bw 
\begin{align*}
-R_{1}^{\left(\N\right)} & =\sqrt{\frac{2K\left(\N\right)}{G\left(K\left(\N\right)\right)}}\int_{0}^{1}e^{-Kv}\left[\lim_{L\rightarrow\infty}\sum_{n=\N+1}^{L}\frac{K^{n}\left(\N\right)}{n!}\left(1-\frac{1}{G\left(K\left(\N\right)\right)}\right)^{n}v^{n}\left(1-v\right)^{n}\right]\frac{dv}{\sqrt{v}}\,.
\end{align*}
 \ew 

We note that the summand is positive for all $v$ in $\mathcal{D}_{1}$;
also, as the terms in the sum are independent of the upper limit $L$
(which, by construction, is not true when we sum up to $\N$, such
as in \eqref{Z_limit_integral_form}) we find that the partial sums
over $n$ constitute a monotonically increasing sequence in $L$.
Thus, by the Monotone Convergence Theorem, we may take liberty to
swap the order between integration and the limit of $L$, and integrate
term by term. We then find \bw
\begin{align}
-R_{1}^{\left(\N\right)} & =\sqrt{\frac{2K\left(\N\right)}{G\left(K\left(\N\right)\right)}}\sum_{n=\N+1}^{\infty}\left(1-\frac{1}{G\left(K\left(\N\right)\right)}\right)^{n}\frac{K^{n}\left(\N\right)}{n!}\int_{0}^{1}e^{-Kv}v^{n}\left(1-v\right)^{n}\frac{dv}{\sqrt{v}}\nonumber \\
 & <\sqrt{\frac{2K\left(\N\right)}{G\left(K\left(\N\right)\right)}}\sum_{n=\N+1}^{\infty}\left(1-\frac{1}{G\left(K\left(\N\right)\right)}\right)^{n}\frac{K^{n}\left(\N\right)}{n!}\int_{0}^{1}e^{-Kv}v^{n}e^{-nv}\frac{dv}{\sqrt{v}}\nonumber \\
 & <\sqrt{\frac{2}{G\left(K\left(\N\right)\right)}}\sum_{n=\N+1}^{\infty}\left(1-\frac{1}{G\left(K\left(\N\right)\right)}\right)^{n}\left[\frac{K\left(\N\right)}{K\left(\N\right)+n}\right]^{n+\frac{1}{2}}\frac{\Gamma\left(n+\frac{1}{2}\right)}{n!}\,.
\end{align}
 \ew 

Next, we use Gaustchi's inequality \cite[Eq. (5.6.4)]{NIST:DLMF}
$x^{1-s}<\frac{\Gamma\left(x+1\right)}{\Gamma\left(x+s\right)}$ for
$s<1$. With $x=n$ and $s=\frac{1}{2}$ it reads $\frac{\Gamma\left(n+\frac{1}{2}\right)}{\Gamma\left(n+1\right)}<n^{-\frac{1}{2}}$,
leaving us with \bw 
\begin{align}
-R_{1}^{\left(\N\right)} & <\sqrt{\frac{2}{G\left(K\left(\N\right)\right)}}\sum_{n=\N+1}^{\infty}\left(1-\frac{1}{G\left(K\left(\N\right)\right)}\right)^{n}\left[\frac{K\left(\N\right)}{K\left(\N\right)+n}\right]^{n+\frac{1}{2}}\frac{1}{\sqrt{n}}\nonumber \\
 & <\sqrt{\frac{2}{G\left(K\left(\N\right)\right)}}\frac{1}{\sqrt{\N+1}}\sum_{n=\N+1}^{\infty}\left(1-\frac{1}{G\left(K\left(\N\right)\right)}\right)^{n}\left[\frac{K\left(\N\right)}{K\left(\N\right)+\N'}\right]^{n+\frac{1}{2}}\nonumber \\
 & <\sqrt{\frac{2}{G\left(K\left(\N\right)\right)}}\sqrt{\frac{K^{2}}{\left(\N'\right)^{3}}}\left(1-\frac{1}{G\left(K\left(\N\right)\right)}\right)^{\N'}\left[\frac{K\left(\N\right)}{K\left(\N\right)+\N'}\right]^{\N'-\frac{1}{2}}\,,
\end{align}
\ew where we have introduced the shorthand notation $\N'=\N+1$.
This error indeed decays to zero regardless of the choice of $K\left(\N\right)$,
as the ratios which are exponentiated are both smaller than unity
for all $K$.

\subsection{The Domain $\mathcal{D}_{2}$}

In this domain, the integrand alternates in sign as $\left(-1\right)^{n}$,
which prevents us from using the Monotone Convergence Theorem. However,
the Taylor series in \eqref{Z_limit_integral_form} now represents
a negative exponent, so its remainder may be bounded with the Lagrange
remainder form $\left|e^{-x}-\sum_{n=0}^{\N}\frac{\left(-x\right)^{n}}{n!}\right|<\frac{x^{\N+1}}{\left(\N+1\right)!}$,
with $x=v\left(v-1\right)$. We thus have \bw
\begin{align}
\left(-1\right)^{\N}R_{2}^{\left(\N\right)} & <\sqrt{\frac{2K\left(\N\right)}{G\left(K\left(\N\right)\right)}}\int_{1}^{\infty}e^{-Kv}\frac{K^{\N'}\left(\N\right)}{\N'!}\left(1-\frac{1}{G\left(K\left(\N\right)\right)}\right)^{\N'}v^{\N'-\frac{1}{2}}\left(v-1\right)^{\N'}dv\nonumber \\
 & =\sqrt{\frac{2K\left(\N\right)}{G\left(K\left(\N\right)\right)}}e^{-K}\frac{K^{\N'}\left(\N\right)}{\N'!}\left(1-\frac{1}{G\left(K\left(\N\right)\right)}\right)^{\N'}\int_{0}^{\infty}e^{-Kv}\left(v+1\right)^{\N'-\frac{1}{2}}v^{\N'}dv\,.\label{eq:R2_for_q_4}
\end{align}
\ew 

Separating the integral into the domains corresponding to (the now
shifted) $\mathcal{D}_{2A}$ and $\mathcal{D}_{2B}$, in $\mathcal{D}_{2A}$
we have 
\begin{align}
\int_{0}^{1}v^{\N'} & \left(v+1\right)^{\N'-\frac{1}{2}}e^{-Kv}dv\twocolbr<\int_{0}^{\infty}v^{\N'}\left(2\right)^{\N'-\frac{1}{2}}e^{-Kv}dv=\frac{2^{\N'-\frac{1}{2}}}{K^{\N'+1}}\N'!\,.
\end{align}

For $\mathcal{D}_{2B}$, we write 
\begin{align}
\int_{1}^{\infty}v^{\N'} & \left(v+1\right)^{\N'-\frac{1}{2}}e^{-Kv}dv\twocolbr<\int_{1}^{\infty}v^{\N'}\left(v+1\right)^{\N'}e^{-Kv}dv\,.
\end{align}
We now wish to find the maximum of this integrand\footnote{Admittedly, we could have bounded the integral in a way similar to
$\mathcal{D}_{2A}$, namely $\int_{1}^{\infty}v^{\N'}\left(v+1\right)^{\N'}e^{-Kv}dv<\int_{1}^{\infty}v^{\N'}\left(2v\right)^{\N'}e^{-Kv}dv$,
and get a factor of $\left(2\N\right)!$ instead of $2^{\N}\N!$.
However, it turns out that this bound is a bit looser, and only shows
convergence for $M/N>1.04$ . We go the extra mile so we can show
that $M=\N$, as used in Ref. \cite{Schwartz2008}, leads to convergence
as well. }, which occurs at\footnote{For $K>2\N$, this maximum lies outside the domain of integration,
as $2\N/K<1$. However, this maximum still bounds the integrand from
above. In any case, bounding the convergence error is much easier
if $K>2\N$, so we proceed with the analysis while assuming $K<2\N$.} 
\begin{align*}
v & =\frac{1}{2}\left(\left(\frac{2\N'}{K}-1\right)+\sqrt{\left(\frac{2\N'}{K}\right)^{2}+1}\right)<\frac{2\N'}{K}\,.
\end{align*}
Next, we observe the fact that the function $\ln\left(v\right)+\ln\left(v+1\right)$
is strictly concave in the domain $\left[1,\infty\right)$. Thus,
the function is bounded from above by any line tangential to it at
any point $v_{0}$ of our choosing. We would like to optimally approximate
the peak of the integrand; however, that would produce a horizontal
tangent, and consequentially a divergent integral. Therefore, we pick
$v_{0}=2\N'/K$ which is in the vicinity, but to the right of the
peak, thus producing a negative slope for the tangential approximation,
effectively attenuating the integrand as $v$ tends to infinity. We
so proceed to bound 
\begin{gather*}
\ln\left(v\right)+\ln\left(v+1\right)\le\ln\left(v_{0}\right)+\ln\left(v_{0}+1\right)+\left(\frac{1}{v_{0}}+\frac{1}{v_{0}+1}\right)\left(v-v_{0}\right)\,.
\end{gather*}
In total, we now get \bw
\begin{align}
\int_{1}^{\infty}v^{\N'} & \left(v+1\right)^{\N'}e^{-Kv}dv\twocolbr\le\int_{1}^{\infty}v_{0}^{\N'}\left(v_{0}+1\right)^{\N'}e^{-Kv+\N'\frac{2v_{0}+1}{v_{0\left(v_{0}+1\right)}}\left(v-v_{0}\right)}dv\nonumber \\
 & =\frac{2}{K}\left(\frac{2\N'}{K}\right)^{\N'}\left(\frac{2\N'}{K}+1\right)^{\N'+1}e^{-2\N'+\frac{K}{2}-\frac{K^{2}}{2\N'+K}}\,.
\end{align}
\ew

Collecting the contributions of $\mathcal{D}_{2A}$ and $\mathcal{D}_{2B}$,
we finally find \bw
\begin{align*}
\left(-1\right)^{\N'}R_{2}^{\left(\N\right)}< & \sqrt{\frac{2}{G}}\left(1-\frac{1}{G}\right)^{\N'}\left\{ \frac{2^{\N'}e^{-K}}{\sqrt{2K}}+\frac{2}{K}\frac{\left(2\N'\right)}{\N'!}^{\N'}\left(\frac{2\N'}{K}+1\right)^{\N'+1}e^{-2\N'-\frac{K}{2}-\frac{K^{2}}{2\N'+K}}\right\} \,.
\end{align*}
\ew

\subsection{Domain of Convergence}

Summing the magnitudes of the remainders from both domains, we get
a total error bound of \bw
\begin{align}
R^{\left(\N\right)} & =\left|\Z^{\left(\N\right)}-\Z\right|=\left|R_{1}^{\left(\N\right)}+R_{2}^{\left(\N\right)}\right|<\left|R_{1}^{\left(\N\right)}\right|+\left|R_{2}^{\left(\N\right)}\right|\nonumber \\
 & <\sqrt{\frac{2}{G}}\left(1-\frac{1}{G}\right)^{\N'}\left\{ \sqrt{\frac{K^{2}}{\left(\N'\right)^{3}}}\left[\frac{K}{K+\N'}\right]^{\N'-\frac{1}{2}}+\frac{2^{\N'}e^{-K}}{\sqrt{2K}}+\frac{2}{K}\frac{\left(2\N'\right)}{\N'!}^{\N'}\left(\frac{2\N'}{K}+1\right)^{\N'+1}e^{-2\N'-\frac{K}{2}-\frac{K^{2}}{2\N'+K}}\right\} \,.\label{eq:Total_Error_Bound}
\end{align}
 \ew

We can see that all the terms above scale exponentially with $\N$.
Convergence would thus require a choice of $M\left(\N\right)=K\left(\N\right)-2$
for which the base of this exponent is smaller than unity. Apart from
the prefactor $\sqrt{\frac{2}{G}}(1-\frac{1}{G})^{\N'}$, all the
terms inside the braces are independent of $g$. Let us start from
the case $M(N)=\alpha\N$ for constant $\alpha$, and $\N,M\left(\N\right)\gg1$:

(i) The first term scales as $\left(\frac{M}{M+\N}\right)^{\N}$,
resulting from a geometric progression. This is evidently smaller
than unity for any choice of $M\left(\N\right)$. However, note that
if $M$ is super-linear in $\N$, then the ratio $\frac{M}{M+\N}$
would approach $1$ as $\N$ increases, so convergence would be hindered. 

(ii) The second term scales as $2^{\N}e^{-M}=\left(2e^{-\alpha}\right)^{\N}$.
This would require us to pick a minimal value of $\alpha>\ln2\approx0.693$.

(iii) The last term scales as 
\begin{align}
\frac{2^{\N}\N^{\N}}{\N!}\left(\frac{2\N}{M}+1\right)^{\N}e^{-2\N-\frac{M}{2}-\frac{M^{2}}{2\N+M}} & \sim\left(2\left(\frac{2}{\alpha}+1\right)e^{-1-\frac{\alpha}{2}-\frac{\alpha^{2}}{\alpha+2}}\right)^{\N}\,.
\end{align}
The exponentiated expression is monotonically decreasing with $\alpha$,
and reaches unity for the numerically obtained critical value of\footnote{We note that this does not imply that the expansion diverges for $\alpha<\alpha_{c}$;
it is simply the lowest value for which this proof is still applicable.
Numerically, one witnesses what seems to be convergent behavior for
$\alpha$ as low as $\sim0.8$.} 
\[
\alpha_{c}\approx0.976\,.
\]

Moreover, one now observes that the error bound above increases for
$\alpha$ which is too large, as the first term $\sim\left(\frac{\alpha}{\alpha+1}\right)^{\N}$
approaches $1$. It is then apparent that an optimum exists, where
the dominance in the bound shifts from the third term to the first.
This will occur when (neglecting the prefactors) 
\begin{equation}
2\left(\frac{2}{\alpha}+1\right)e^{-1-\frac{\alpha}{2}-\frac{\alpha^{2}}{\alpha+2}}\approx\frac{\alpha}{\alpha+1}\,.\label{eq:alpha_star_condition}
\end{equation}
This equation is satisfied for 
\begin{equation}
\alpha^{*}\approx1.317\,\text{,}\label{eq:alpha_star_value}
\end{equation}
 for which the exponential rate of convergence is at least $10^{-0.245\N}$. 

Additionally, we estimate the contribution of the factor $\left(1-\frac{1}{G}\right)^{\N}$,
\begin{equation}
\left(1-\frac{1}{G}\right)^{\N}<e^{-\frac{\N}{G}}<e^{-\frac{1}{4}\sqrt{\frac{1}{g}\frac{\N}{\alpha}}}\,,\label{eq:SCE_Stretched_Exponent}
\end{equation}
where we have assumed $\sqrt{gK}>\frac{1}{2}$. Thus, this factor
behaves as a stretched exponent. We note its sign corresponds to the
sign of the quadratic potential; for a double well ($\gamma<0$ in
\eqref{anharmonic_potential_V}), the argument of the stretched exponent
would be positive, but it would still be overwhelmed for large $\N$.
This case is discussed further in \secref{double_well}.

In total, we expect the large-$\N$ error to scale asymptotically
as 
\begin{equation}
R^{\left(\N\right)}=\mathcal{O}\left(10^{-A\left(\alpha\right)\N-B\left(\alpha,g\right)\sqrt{N}}\right)\,,\label{eq:Total_error_functional_form}
\end{equation}

where the bound on $A\left(\alpha\right)$ is 
\begin{equation}
A\left(\alpha\right)=\begin{cases}
-\log_{10}\left(2\left(\frac{2}{\alpha}+1\right)e^{-1-\frac{\alpha}{2}-\frac{\alpha^{2}}{\alpha+2}}\right)\,, & \alpha<\alpha^{*}\\
-\log_{10}\left(\frac{\alpha}{\alpha+1}\right)\,, & \alpha>\alpha^{*}
\end{cases}\,.\label{eq:A_coefficient_bound}
\end{equation}

Furthermore, since $1-1/G\le1$, we may loosen the bound and get \bw
\begin{align*}
\left|\Z^{\left(\N\right)}-\Z\right| & <\sqrt{\frac{2K^{2}}{\left(\N'\right)^{3}}}\left[\frac{K}{K+\N'}\right]^{\N'-\frac{1}{2}}+\frac{2^{\N'}e^{-K}}{K}+\frac{\sqrt{8}}{K}\frac{\left(2\N'\right)}{\N'!}^{\N'}\left(\frac{2\N'}{K}+1\right)^{\N'+1}e^{-2\N'-\frac{K}{2}-\frac{K^{2}}{2\N'+K}}\,,
\end{align*}
\ew which depends only on $\N$ and not on $g$. This implies that
for any desired level of accuracy, one may find to which order the
expansion needs to be evaluated, regardless of the coupling strength.
This concludes the proof that the expansion is uniformly convergent,
case 2b of Proposition \ref{Proposition_Combined}.

Finally, if $M\sim\N^{p}$ with $p>1$, then $\alpha\left(\N\right)$
is increasing and will eventually surpass $\alpha^{*}$. The error
will then be dominated by the first term, and have the functional
form
\begin{align*}
\sqrt{\frac{2}{G}}\sqrt{\frac{M^{2}}{\N^{3}}}\left(1-\frac{1}{G}\right)^{\N}\left[\frac{M}{M+\N}\right]^{\N} & \sim\N^{-\frac{p}{4}}\N^{p-\frac{3}{2}}\left(1-\frac{1}{\N^{p/2}}\right)^{\N}\left(1-\frac{\N}{\N^{p}}\right)^{\N}\sim\N^{\frac{3}{4}\left(p-2\right)}e^{-\N^{1-p/2}}e^{-\N^{2-p}}\,.
\end{align*}
This shows the convergence of the method in case 2a of Proposition
\ref{Proposition_Combined}. 

Note that for $p>2$, this error is algebraically divergent due to
the first factor, and recalling that $R_{1}^{\left(\N\right)}$ is
negative, we have that $\Z^{\left(\N\right)}\le\Z$. However, this
bound does not imply that $\Z^{\left(\N\right)}$ could not converge
after all. In order to prove case 3, we proceed by a different approach:
Consider the inner summation over $l$ in \eqref{SCE_partition}.
The maximal term in the sum would be at the index $l$ which is the
integer closest to the solution of 
\[
\frac{\left(n-l\right)\left(2n+2l+1\right)}{2K\left(l+1\right)}=1\,,
\]
which is quadratic in $l$. Only one of its roots is positive, so
there exists at most a single peak for $0\le l\le n$. We find the
value of $K$ for which this peak occurs at $l=0$ to be
\[
K=M+2>n^{2}+\frac{1}{2}n\,,
\]
 which would be satisfied for all values of $n=0,1,\dots,\N$ if we
take 
\[
M\left(\N\right)>\N^{2}+\frac{1}{2}\N\,.
\]

If $M\left(\N\right)\sim\N^{p}$ with $p>2$, then for sufficiently
large $\N$, the condition above will be satisfied. We now have, for
a given $n$, a monotonically decreasing alternating sum in $l$.
It is bounded by the first term in the sum, $l=0$ (which is positive),
so we have
\begin{align}
\Z^{\left(\N\right)} & <\sqrt{\frac{2}{G\left(K\left(\N\right)\right)}}\sum_{n=0}^{\N}\left[1-\frac{1}{G\left(K\left(\N\right)\right)}\right]^{n}\frac{\Gamma\left(n+\frac{1}{2}\right)}{n!}\nonumber \\
 & <\sqrt{\frac{2}{G\left(K\left(\N\right)\right)}}\sum_{n=0}^{\N}\frac{\Gamma\left(n+\frac{1}{2}\right)}{n!}\nonumber \\
 & \overset{!}{=}\sqrt{\frac{2}{G\left(K\left(\N\right)\right)}}\frac{2\Gamma\left(\N+\frac{3}{2}\right)}{\N!}\nonumber \\
 & <\sqrt{\frac{2}{G\left(K\left(\N\right)\right)}}\frac{2\left(\N+1\right)!/\sqrt{\N+1}}{\N!}\nonumber \\
 & <\sqrt{\frac{4\left(\N+1\right)}{\left(gM\right)^{1/2}}}\nonumber \\
 & \sim\N^{-\frac{p-2}{4}}\rightarrow0\,,
\end{align}
since we assumed $p>2$. Thus, we have proved all cases of Proposition
\ref{Proposition_Combined}.

\subsection{Comparison with Results for Related Methods}

It is instructive to compare the results we have shown here with those
obtained for the ODM/OPT/LDE schemes. Following arguments by Zinn-Justin
and Seznec \cite{ZinnJustinSeznec1979}, Buckley, Duncan, and Jones
showed \cite{BuckleyDuncanJonesZeroDimension} that the sequence $\Z^{\left(\N\right)}$
converges to $\Z\left(g\right)$ if the modified harmonic coefficient
$G$ scales as $\sim\sqrt{g\N}$, and that $G\approx2.30\sqrt{g\N}$
is the asymptotic solution to the PMS condition, so that the PMS ensures
convergence. This scaling would correspond to $M\left(\N\right)\sim\N$
in Proposition \ref{Proposition_Combined}, and to an optimal $\alpha^{*}=1.325$.
Convergence was subsequently extended to a wider scaling range, equivalent
to $M\sim N^{p}$ with $1\le p<2$, by Guida, Konishi, and Suzuki
\cite[Appendix B]{GuidaKonishiSuzuki1996}, who also showed that at
large order the FAC criterion produces a similar condition for $G$
as the PMS. However, these sources neither provided estimated rates
of convergence away from the PMS solution, nor a precise functional
form of the expansion error for $p\neq1$. Furthermore, these proofs
relied on knowledge of the approximate location of the PMS solution,
and subsequently also on the analytic structure of the expanded function
$\Z\left(g\right)$; this information might not be available in systems
other than the most simple.

Though we have obtained slightly looser bounds for the error at $G$
which satisfies the PMS (i.e., our optimal $\alpha$), our proof offers
several improvements over that of Buckley \emph{et al}.: (i) We estimate
the minimal proportionality constant $M/\N$ necessary for convergence.
(ii) Our error bound is applicable to any monotonic arbitrary function
$M\left(\N\right)$, and thus any $G\left(\N\right)$. (iii) Our proof
covers even values of $\N$, and dispenses with the need that the
approximants satisfy $\Z^{\left(\N\right)}<\Z$. (A simplified argument,
which recovers the convergence rate of Ref. \cite{BuckleyDuncanJonesZeroDimension}
but for all $\N$, is given in \appref{Direct_1F1_estimation}.) (iv)
The derivation does not require an \emph{a priori} knowledge of the
required scaling of $G$. (v) We solve for the general case of a non-zero
harmonic term in the original potential. 

Thanks to the simplified analysis, in the following Sections our results
will extend naturally to the many-oscillators and double-well cases,
and generalize to arbitrarily high anharmonicities and the complex
$g$ plane.

To conclude this discussion, the treatment above establishes the equivalence
of the SCE with the other methods in the large order limit. Namely,
the SCE optimum for $G$ coincides with that of the PMS and FAC criteria.
However, the SCE procedure is easily posed and solved explicitly,
and its implementation generates its own effective coupling $1-1/G$.
Moreover, we will see that the SCE has additional intrinsic appeal,
as the linear relation $M=\alpha\N$ will repeatedly yield the optimal
convergence rate, even for different anharmonicities.

\section{\label{sec:numerical_results_q_4}Numerical Results}

In order to demonstrate the properties of SCE, the expansion in \eqref{SCE_partition}
was evaluated in \emph{Mathematica} \cite{Mathematica}, and was compared
against a direct evaluation of \eqref{Z_analytic}\emph{. Mathematica}
was chosen by virtue of its ability to evaluate both to arbitrary
numerical precision \cite{Mathematica}. However, this precluded the
usage of floating-point values of $g$ and $\alpha$; instead we only
evaluate rational values. In particular, instead of evaluating at
$\alpha^{*}\approx1.317$ {[}cf. \eqref{alpha_star_value}{]}, we
use $\tilde{\alpha}^{*}=\frac{4}{3}$. It is also worth noting that
the summand in the summation over the index $l$ in \eqref{SCE_partition}
is wildly alternating, in the sense that delicate cancellations occur
between terms of opposite signs, and the total sum for a given $n$
may be many orders of magnitude smaller than any individual term.
Thus, for a given total precision of the expansion, we empirically
find that intermediate calculations need to be carried out with roughly
$\sim3$ times as many significant digits. 

\begin{figure*}[tp]
\begin{centering}
\subfloat{\includegraphics[width=\figurewidth]{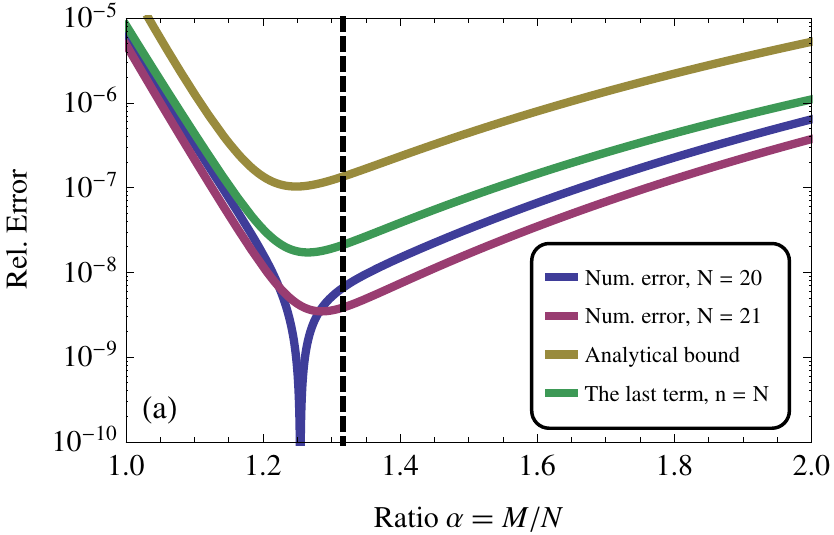}\label{fig:Error_vs_alpha}}\hspace*{\fill}\subfloat{\includegraphics[width=\figurewidth]{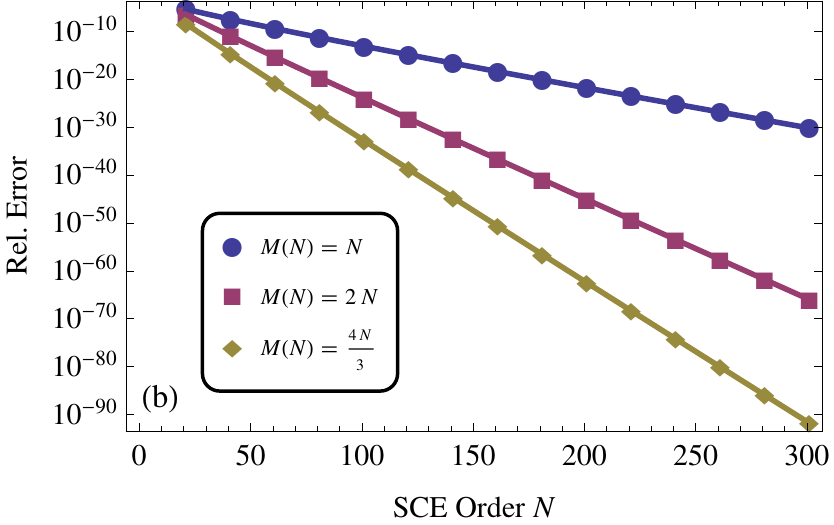}\label{fig:Error_vs_N_classical}}
\par\end{centering}

\caption[Figure \thefigure]{Performance of the SCE versus the parameter $\alpha$ and order $\N$.
(a) Relative error of the SCE as a function of $\alpha=M\left(\N\right)/\N$.
$g$ is fixed at $1$. Shown are measured errors for $\N=20$ and
$21$. Also plotted is the analytical bound from \eqref{Total_Error_Bound},
and the last, $n=\N$ term in \eqref{SCE_partition} (after summation
over $l$, for $\N=20$). The analytical bound is quite loose, while
the $n=\N$ term is a much better estimator of the error of the approximation,
especially for $\alpha<\alpha^{*}$ --- this property is explored
in \appref{Direct_1F1_estimation}. All the graphs exhibit an optimum
choice in the vicinity of $\alpha=1.3$, close to our estimated $\alpha^{*}\approx1.317$
{[}\eqref{alpha_star_value}{]}, marked by the vertical dashed line;
note that this value was calculated in the limit of $\N\gg1$. The
different behavior of $\N=20$ and $\N=21$ at optimum is explained
in the text. (b) Convergence of the SCE at high order. Plotted are
the relative errors of the SCE expansion of the partition function
$\Z^{(\N)}$ versus the order $\N$ (markers). These are plotted for
$M=\alpha\N$ with $\alpha=1$, $2$, and $\frac{4}{3}$, in order
of merit (blue, purple, and yellow, respectively). The coupling $g$
is fixed at $1$. The vertical axis reflects the number of significant
digits successfully reproduced by the expansion. The data was fitted
with a trend of the form $10^{-A\N-B\sqrt{\N}+C}$, corresponding
to our analytical bound (continuous lines). The fits achieve a $\chi^{2}$
value of $\sim10^{-2}$ ($\chi^{2}$ is defined in the text). Though
even orders admit better accuracy, we evaluate at odd orders to mitigate
the sensitivity of the fitted parameters to the exact choice of $\alpha$
near its optimum (cf. the left panel). The fitted values for the coefficient
$A$ are $0.072$, $0.200$, and $0.283$, for $\alpha=1$ , $2$,
and $\frac{4}{3}$, respectively, representing the minimal amount
of decimal places obtained per increment of $\N$. The lower bounds
for $A\left(\alpha\right)$ as predicted by \eqref{A_coefficient_bound}
are $0.018$, $0.176$ and $0.243$, in agreement with the numerical
results. The minimal error shown is roughly $1.5\times10^{-92}$.
Note that in order to obtain this many accurate digits, the intermediate
calculations had to be performed with over $300$ digit precision. }
\end{figure*}

\figref{Error_vs_alpha} depicts the convergence properties of the
SCE as a function of $\alpha$. It shows the error of the expansion
for two orders, $\N=20$ and $21$, as well as the bound on the error
given by \eqref{Total_Error_Bound}, for a moderate coupling strength
$g=1$. Lastly, the final $n=\N$ term of the expansion is plotted
for $\N=20$. The analytical bound captures the qualitative behavior
of the error, but the last term provides an estimate for the error
which is much tighter. This is explored further in \appref{Direct_1F1_estimation}.

All four error measures exhibit an optimum in the vicinity of $\alpha\approx1.3$,
close to the analytic value we deduced for $\alpha^{*}$. We argued
that this point occurs when the dominant domain in the error shifts
from $\mathcal{D}_{1}$ to $\mathcal{D}_{2}$. Recall that $R_{1}^{\left(\N\right)}$
is always negative, while $R_{2}^{\left(\N\right)}$ alternates in
sign as $\left(-1\right)^{\N}$. If these are continuous functions
of $\alpha$, then at $\alpha^{*}$ they should have equal magnitudes.
This would imply that for $\N$ which is even, at the optimum point
they cancel each other to achieve zero error. Indeed, in \figref{Error_vs_alpha},
the expansion with $\N=21$ is somewhat better than $\N=20$ far from
$\alpha^{*}$ (due to the exponential convergence in $\N$), but close
to $\alpha^{*}$, $\N=20$ affords a better approximation. In fact,
we note that the location of $\alpha^{*}$ represents the solution
of the PMS condition, since (our bound for) the remainder is stationary
there for odd $\N$; however, whereas the SCE works equally well for
even or odd $\N$, the PMS breaks down for even $\N$ \cite{BuckleyDuncanJonesZeroDimension},
despite the fact that the remainder could be canceled completely at
even orders.

Next, in \figref{Error_vs_N_classical} we push the SCE to large order,
evaluating it up to $\N=301$. This is performed for $g=1$ and with
$\alpha=1$ (as in Ref. \cite{Schwartz2008}), $2$ and $\frac{4}{3}$,
as an approximation to our estimate for $\alpha^{*}$ and the location
of the optima visible in \figref{Error_vs_alpha}. The resulting relative
errors are fitted with a curve of the form $10^{C-A\N-B\sqrt{\N}}$,
according to \eqref{Total_error_functional_form}. These fits all
achieve a value of $\chi^{2}\sim10^{-2}$, defined as $\chi^{2}=\left[\frac{1}{L}\sum_{\N}\left(C-A\N-B\sqrt{\N}-\log_{10}\left|R^{\left(N\right)}\big/\Z\right|\right)^{2}\right]^{\frac{1}{2}}$,
where $R^{\left(N\right)}/\Z$ is the relative error, the sum is over
all the data points, and $L$ is their number. Discarding the stretched
exponent by setting $B=0$, $\chi^{2}$ jumps to about $\sim0.2$.
The fitted values for the parameter $A$ are $0.072$, $0.200$ and
$0.283$, for $\alpha=1$ , $2$, and $\frac{4}{3}$, respectively.
These are in agreement with the bounding values $0.018$, $0.176$,
and $0.243$, given by \eqref{A_coefficient_bound}.

We continue with a comparison of the SCE with other asymptotic and
numerical approximation schemes. These include the methods of super
and hyperasymptotics \cite{boyd1999devil,berry1990hyperasymptotics,berry1991hyperasymptotics},
Borel remainder summation \cite{Negele1998,costin2008asymptotics},
Padé approximants \cite{BenderAndOrszag}, and the Chebyshev $\tau$
method due to Lanczos \cite{boyd2001chebyshev}. For an outline of
these methods, refer to \appref{Competing_Methods}.

\begin{figure}[tp]
\centering{}\includegraphics[width=\figurewidth]{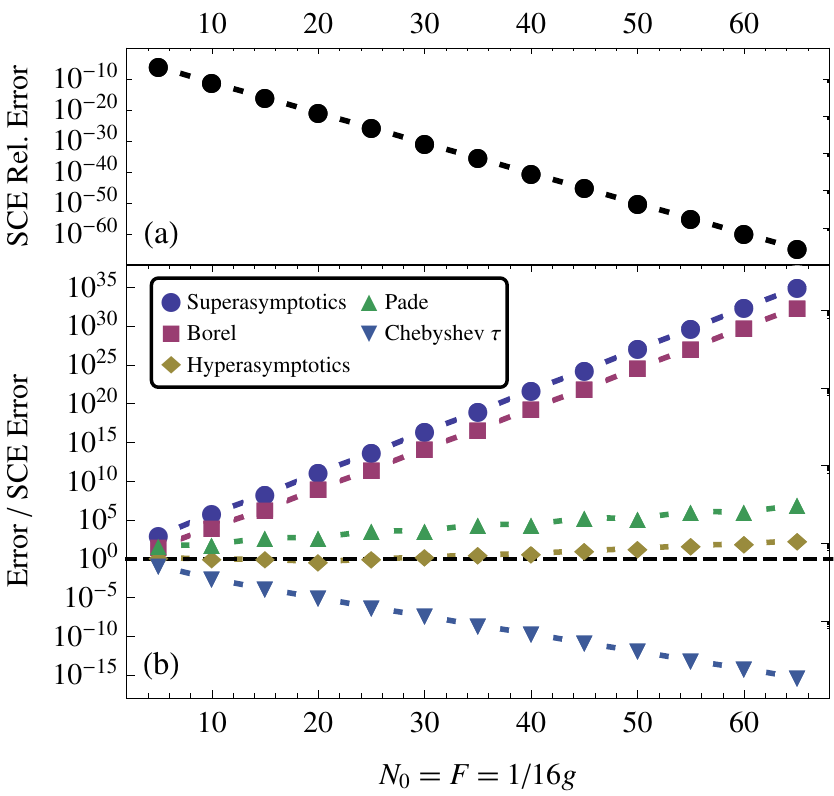}\caption[Figure \thefigure]{\label{fig:SCE_vs_hyperasymptotics}A comparison of the SCE with
other asymptotic and numeric schemes, in order of merit (see \appref{Competing_Methods}):
Superasymptotics (i.e. truncation of PT at the least term) \cite{boyd1999devil};
Borel resummation of the superasymptotic remainder \cite[Sec. 7.2]{Negele1998};
Padé approximants of PT \cite{BenderAndOrszag}; $3^{\text{rd}}$
level hyperasymptotics \cite{berry1990hyperasymptotics,berry1991hyperasymptotics};
SCE (with $\alpha=4/3$); and Lanczos's Chebyshev $\tau$ approximations
of $\Z\left(g\right)$ \cite{boyd2001chebyshev}. Since the Padé,
SCE and Lanczos methods converge and thus can be evaluated to arbitrary
order, we choose to compare them with the asymptotic methods at the
same order $\N=\N_{0}=\frac{1}{16g}$ as that used in the superasymptotic
expansion. (a) The relative error of the SCE. The horizontal axis
is the singulant $F=\frac{1}{16g}$ (cf. Subsection \ref{sub:Hyperasymptotics}),
and at each point the SCE is evaluated to order $\N_{0}=F$, which
is the same truncation as the superasymptotic scheme. The line segments
are a guide to the eye. (b) The errors produced by the other methods,
relative to that produced by the SCE (i.e., the data was normalized
by that shown in the top panel). The line segments are a guide to
the eye. When evaluated at the same order, the SCE outperforms superasymptotics,
Borel tail resummation, and the Padé approximants. Note the extremely
similar errors produced by the SCE and hyperasymptotics, which is
formally of order $2\N_{0}$; this is discussed in the text. Lastly,
the Chebyshev $\tau$ method appears more accurate than the SCE, but
only because of the specific truncation at $1/16g$. In the large-$\N$
limit, the SCE is more rapidly convergent, as demonstrated in \figref{SCE_vs_Pade_Lanczos_vs_N}.
The dominance of the SCE in the strongly-coupled regime is not manifest
in this comparison, since for $g\gg1$ the superasymptotic truncation
occurs at $\N_{0}=0$; however, for fixed and finite $\N$, the SCE
is vastly superior in this scenario, as depicted in \figref{SCE_vs_Pade_Lanczos_vs_g}.}
\end{figure}

A comparative plot of all techniques is shown in \figref{SCE_vs_hyperasymptotics}
for different values of weak $g$. By definition, superasymptotics
are evaluated up to the least term of PT, which occurs at $\N_{0}=\frac{1}{16g}$
(round down). Hyperasymptotics are evaluated up to level 3 (or less,
if the procedure halts before). To make a ``fair'' comparison, the
SCE, Padé, and $\tau$ methods are evaluated at $\N=\N_{0}$, though
in principle they converge as $\N\rightarrow\infty$.

A striking result of \figref{SCE_vs_hyperasymptotics} is the similarity
of the errors produced by SCE (at order $\N_{0}=\frac{1}{16g}$) and
hyperasymptotics. This can be explained by the error estimates of
both: Revisiting \eqref{SCE_Stretched_Exponent}, we can bound 
\begin{align*}
\left(1-\frac{1}{G\left(\N_{0}\right)}\right)^{\N_{0}} & =\left(1-\frac{1}{\frac{1}{2}+\frac{1}{2}\sqrt{1+16g\left(\frac{\alpha}{16g}+2\right)}}\right)^{\N_{0}}\twocolbr\approx\left(1-\frac{1}{\frac{1}{2}+\frac{1}{2}\sqrt{1+\alpha}}\right)^{\N_{0}}\approx e^{-1.57\N_{0}}\,,
\end{align*}
where we have substituted $\alpha=\frac{4}{3}$. Notably, at this
truncation, we do not yet witness the stretched-exponential behavior
which occurs at large $\N$, but rather a regular exponent. Adding
the factor $\left(\frac{\alpha}{\alpha+1}\right)^{\N_{0}}=e^{-0.56\N_{0}}$
{[}\eqref{Total_error_functional_form} and \eqref{A_coefficient_bound}{]},
we find that the SCE at order $\frac{1}{16g}$ has an error bound
that scales as $\sim e^{-2.12\N_{0}}$. Substituting the numerically
fitted uniform factor $10^{-0.283\N}=e^{-0.65\N}$ {[}cf. \figref{Error_vs_N_classical}{]},
an even smaller error of $\sim e^{-2.22\N_{0}}$ is obtained. In comparison,
examining the exponential factors in \eqref{Anharmonic_Z4_Hypersymptotic_Error},
we see a scaling of $\exp\left\{ -\left(1+\sum_{s=1}^{S}\frac{\ln2}{2^{s-1}}\right)\N_{0}\right\} $
with $S=-\log_{2}\left(16g\right)$. For $2$ and $3$ stages, the
scaling so obtained is $e^{-2.04\N_{0}}$ and $e^{-2.21\N_{0}}$,
respectively, very close to the scaling for infinitesimal $g$ (when
$S$ is very large), $e^{-\left(1+2\ln2\right)\N_{0}}\approx e^{-2.386\N_{0}}$
\cite{berry1990hyperasymptotics,berry1991hyperasymptotics}. This
implies that the SCE to order $\N_{0}$ produces an error comparable
with the hyperasymptotic trans-series at levels $2$ and $3$. However,
the SCE requires only $\N_{0}$ terms to achieve this, compared to
$\sim\frac{3}{2}\N_{0}$ or $\sim\frac{7}{4}\N_{0}$ terms required
by hyperasymptotics at these stages. If compared with hyperasymptotics
when carried through to its conclusion, halting at roughly $2\N_{0}$
terms with an error of order $e^{-2.386\N_{0}}$, then at order $2\N_{0}$
the SCE would result in an error of order $\left(1-\frac{1}{\frac{1}{2}+\frac{1}{2}\sqrt{1+2\times\alpha}}\right)^{2\N_{0}}\left(\frac{\alpha}{\alpha+1}\right)^{2\N_{0}}\approx e^{-3.44\N_{0}}$,
which is about $44\%$ faster in terms of its dependence on $\N_{0}$. 

\begin{figure*}[tp]
\begin{centering}
\subfloat{\includegraphics[width=\figurewidth]{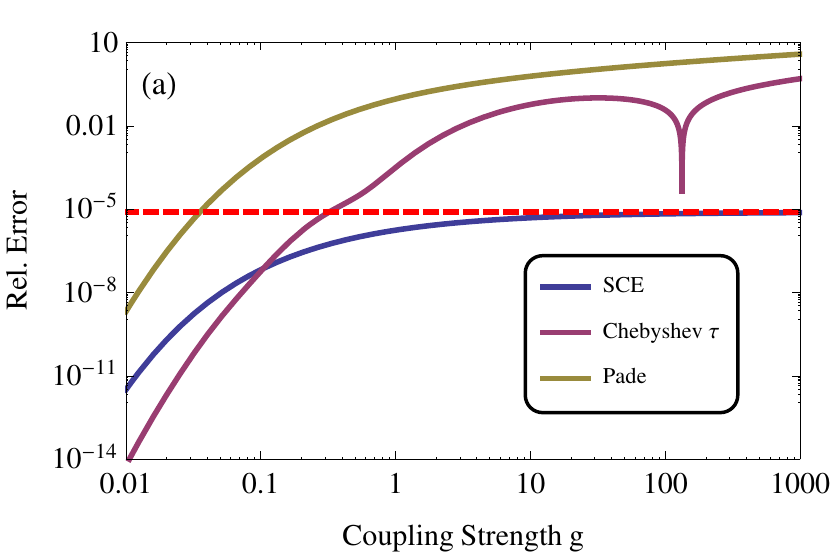}\label{fig:SCE_vs_Pade_Lanczos_vs_g}}\hspace*{\fill}\subfloat{\includegraphics[width=\figurewidth]{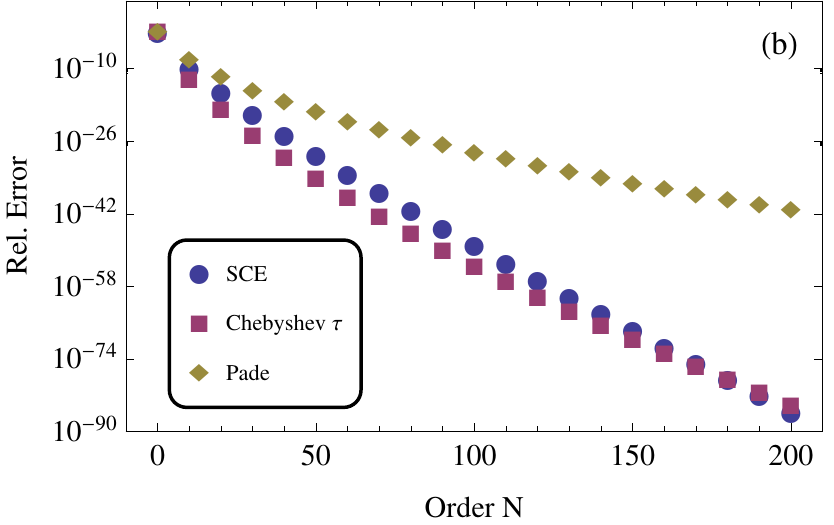}\label{fig:SCE_vs_Pade_Lanczos_vs_N}}
\par\end{centering}

\caption[Figure \thefigure]{Comparison of the SCE against numerical methods. (a) The relative
error produced by SCE, Padé approximants \cite{BenderAndOrszag} and
the Chebyshev $\tau$ method \cite{boyd2001chebyshev} versus $g$
(see \appref{Competing_Methods}). All methods are evaluated at $\N=13$;
the SCE is evaluated with $\alpha=\frac{4}{3}$. The SCE outperforms
the Padé approximants across a wide range of values for $g$, and
outperforms the $\tau$ method for $g\gtrsim0.1$. This holds for
other values of $\N$ as well. As $g$ grows larger, the relative
errors of the Padé and $\tau$ approximations diverge, though the
approximations themselves are finite. The dashed red line shows the
ultimate error of the SCE for $g\rightarrow\infty$, about $7.83\times10^{-6}$.
This value was calculated by comparing the coefficients of $g^{-\frac{1}{4}}$
in the large-$g$ expansions of \eqref{Z_analytic} {[}which is $\frac{1}{2}\Gamma\left(\frac{1}{4}\right)${]}
and \eqref{SCE_partition} {[}obtained by taking $1-1/G\to1$ and
$\sqrt{\frac{2}{G}}\to\left(gK\right)^{-\frac{1}{4}}${]}. The sharp
dip in the error of the $\tau$ method near $g\approx150$ is due
to the the error switching signs across that value, so at the crossover
the relative error vanishes. This does not occur with the SCE since
we evaluate at an odd order $\N$ {[}cf. \figref{Error_vs_alpha}{]}.
(b) The relative error produced by the same methods versus the approximation
order $\N$, up to order $200$. All methods are evaluated at $g=0.01$;
the SCE is evaluated with $\alpha=\frac{4}{3}$. The SCE is consistently
more accurate than the Padé approximants, and it supersedes the $\tau$
method at roughly $\N=180$, which depends on the value of $g$.}

\end{figure*}

We also wish to compare the SCE with convergent methods of approximation.
For this comparison, we can again decouple $g$ and the truncation
order $\N$. From \figref{SCE_vs_hyperasymptotics} one may deduce
that the SCE is numerically comparable to the Padé approximants, and
inferior to the Chebyshev $\tau$ method, but that is misleading.
First, the immediate advantage of the SCE over the other methods is
its uniform convergence in $g$ for all positive values, and in particular
for $g\gtrsim1$, a domain which was not represented in \figref{SCE_vs_hyperasymptotics},
since there the superasymptotic series is trivial (truncated at the
zeroth term) for $g>\frac{1}{16}$. Both the Padé approximants and
the $\tau$ approximations are rational functions of $g$, where the
numerator and denominator are of identical order in $g$; as such,
for a fixed order $\N$, they will tend to a constant as $g\rightarrow\infty$.
$\Z\left(g\right)$ decays to zero for $g\rightarrow\infty$ (the
anharmonicity compresses the oscillator's spatial distribution), so
the relative error of these schemes will diverge for larger $g$.
The SCE's supremacy over these techniques for larger couplings is
demonstrated in \figref{SCE_vs_Pade_Lanczos_vs_g}. 

Second, we contend that even for $g\ll1$, while the SCE is initially
less accurate, it overtakes other methods at high enough orders. We
show this numerically in \figref{SCE_vs_Pade_Lanczos_vs_N} for $g=\frac{1}{100}$.
The SCE converges markedly faster than the Padé approximants for any
$\N$, while for smaller $\N$ it is outperformed by the $\tau$ method.
However, at sufficiently large order --- $\N\approx180$, in this
case --- the SCE overtakes it. This crossover point depends on $g$;
for example, for $g=\frac{2}{100}$, it occurs as early as $\N=90$.
It is also worth mentioning that at a given order, evaluating the
SCE is much quicker than any of the other methods.

\section{\label{sec:double_well}Non-Perturbative SCE: The Double-Well Potential}

Now that the convergent nature of the SCE has been observed, we can
examine a more intricate case, that of the double-well potential,
corresponding to a negative quadratic part of the potential ($\gamma<0$
in \eqref{anharmonic_potential_V}). This has the effect of flipping
the sign of the quadratic term in \eqref{Z_definition}. \revised{It is an interesting test case, since the quadratic Hamiltonian alone
is not stable, and thus the PT expansion must be performed about the
minima of the potential, instead of around $g=0$ and the $x$ origin.
However, we will show that the SCE captures the correct partition
function even when expanding around a harmonic zeroth order approximation. }

The partition function is now given by 
\[
\Z\left(g\right)=\frac{\pi}{\sqrt{16g}}e^{\frac{1}{32g}}\left[I_{\frac{1}{4}}\left(\frac{1}{32g}\right)+I_{-\frac{1}{4}}\left(\frac{1}{32g}\right)\right]\,,
\]
with $I_{\nu}\left(x\right)$ the modified Bessel functions of the
first kind \cite{abramowitz1964handbook}. Carrying through the same
analysis as in \secref{The-SCE-of-Z4}, we find that now the self-consistent
harmonic coefficient $G$ becomes
\begin{equation}
G_{DW}\left(M\right)=-\frac{1}{2}+\frac{1}{2}\sqrt{1+16g\left(M+2\right)}>0\,,\label{eq:Double_Well_G}
\end{equation}
where we have chosen the positive root for $G$, so that the integrals
of $\exp\left(-\frac{1}{2}Gx^{2}\right)$ are convergent. We then
find that the SCE for the double-well partition function is the same
power series as (\ref{eq:SCE_partition}), apart from the replacement
$\left(1-1/G\right)\to\left(1+1/G\right)>1$.

\begin{figure*}[tp]
\centering{}\subfloat{\includegraphics[width=\figurewidth]{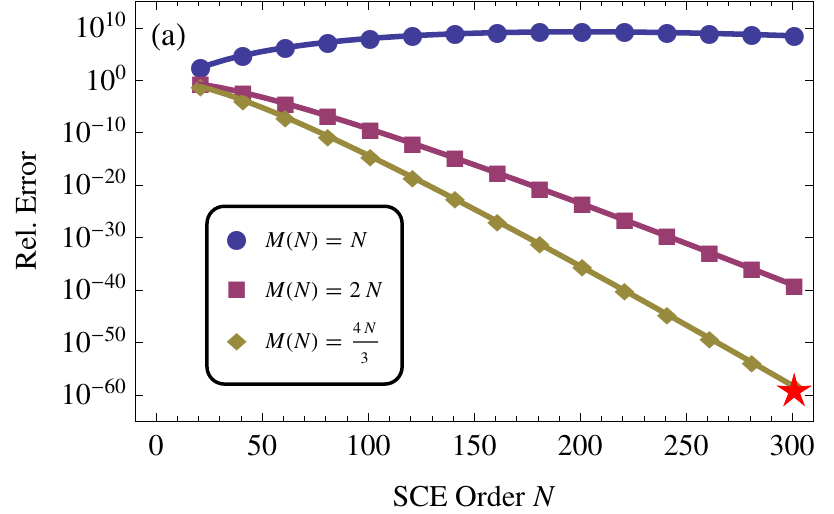}\label{fig:Double_Well_Convergence}}\hspace*{\fill}\subfloat{\includegraphics[width=\figurewidth]{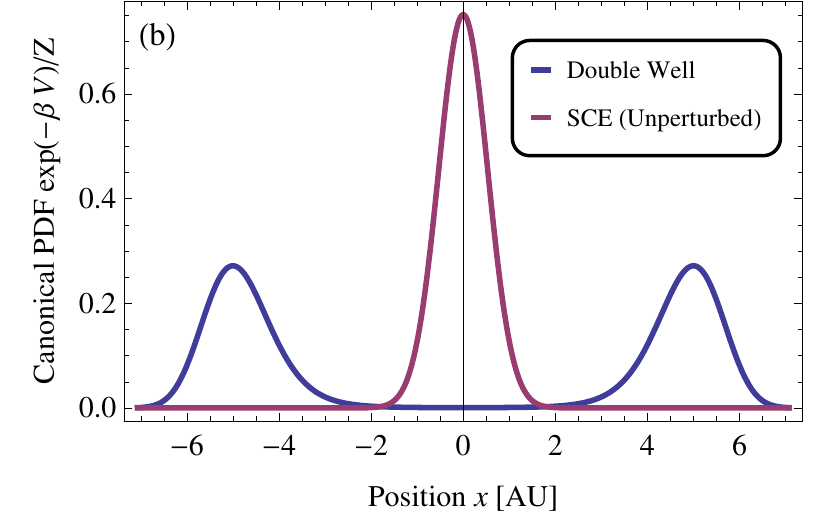}\label{fig:Double_Well_Potential}}\caption[Figure \thefigure]{\label{fig:Double_Well}The convergence of the SCE for the case of
a double-well potential. We examine the barely binding case of $g=0.01$,
so that the differences between this case and the previous Sections
are accentuated. (a) The double-well counterpart of \figref{Error_vs_N_classical}:
Shown is the numerical accuracy of the SCE, evaluated for $\alpha=1$,
$2$, and $\frac{4}{3}$ (markers). Note that all three trends converge,
though for the $\alpha=1$ case, the exponential convergence has barely
begun to rein in the divergent stretched exponent at $\N\approx200$,
showing that a proper choice of $\alpha$ is critical to the effectiveness
of the approximation. All points are evaluated at odd order {[}cf.
\figref{Error_vs_alpha}{]}. Against the data we fit a trend line
of the form $10^{C-A\N+B\sqrt{\N}}$ (solid lines), and the fitted
values of $A$ are $0.075,$ $0.197$, and $0.282$ for $\alpha=1$,
$2$, and $\frac{4}{3}$, respectively, essentially the same as those
fitted in \figref{Error_vs_N_classical}. (b) The Boltzmannian (normalized)
distributions $e^{-\beta V\left(x\right)}/\mathcal{Z}$ of the double-well
potential $V\left(x\right)=-\frac{1}{2}x^{2}+gx^{4}$, and the zeroth-order
approximation $V\left(x\right)=+\frac{1}{2}Gx^{2}$ around which the
SCE is expanded. This plot corresponds to the point marked by a red
star in the left panel, where $\N=301$ and $\alpha=\frac{4}{3}$.
Note how the SCE's reference system bears little resemblance to the
approximated one.}
\end{figure*}

In \secref{convergence_q_4} we established that the exponential convergence
of the SCE was due to the large-$\N$ scaling of its coefficients.
For the double-well potential, these coefficients are unchanged and
exponential convergence is preserved. However, the factor $\left(1-1/G\right)^{\N}$,
which previously contributed a decaying stretched exponent, now becomes
a divergent $\left(1+1/G\right)^{\N}$ factor. Furthermore, note that
the small-$Mg$ limit of \eqref{Double_Well_G} gives $G_{DW}=0$
and not $1$, so that at low orders, $G^{-1}\approx\left(4gM\right)^{-1}$
is very large. However, at large order, this factor behaves again
as a stretched exponent, which will eventually be overwhelmed by the
exponential convergence for sufficiently large $\N$, once {[}cf.
\eqref{SCE_Stretched_Exponent} and \eqref{A_coefficient_bound}{]}
\begin{equation}
\N\ge\N_{c}\left(\alpha,g\right)=-\sqrt{\frac{1}{16g\alpha}}\Bigg/\ln\left(\frac{\alpha}{\alpha+1}\right)\,.\label{eq:Double_well_critical_N}
\end{equation}
For our optimal $\alpha^{*}\approx\frac{4}{3}$, this is satisfied
already at order $\N=1$ for $g\gtrsim0.2$. This convergence is demonstrated
in \figref{Double_Well_Convergence}. The positivity of the stretched
exponent implies that for the double-well system, convergences is
not uniform for all $g>0$, but only for $g\ge\epsilon>0$, i.e. on
any interval of finite $g$. This is of course unavoidable, since
at $g=0$ the potential is not bound from below and the partition
function cannot be defined.

We attribute the initial divergence of the SCE at low orders to an
incompatibility between two conflicting goals. The first is that the
SCE zeroth order potential $V_{0}\left(x\right)=\frac{1}{2}Gx^{2}$
preserve the moments $\left\langle x^{2M}\right\rangle $; the second
goal is that it should revert to the original system when $g\rightarrow0$,
i.e. reduce to $V\left(x\right)=-\frac{1}{2}x^{2}$. Of course, the
latter potential has ill-defined moments, as the original system relies
on the quartic term for stability. Faced with this conflict, our formulation
of the SCE favors the first goal. Let us also note that the solution
in \eqref{Double_Well_G} is but one of two for $G$, borne of a quadratic
equation, with the other being $G=-\frac{1}{2}-\frac{1}{2}\sqrt{1+16g\left(M+2\right)}$,
which does revert to the correct potential when $g=0$; however, this
$G$ is strictly negative, contrary to the requirement that integrals
of $\exp\left(-\frac{1}{2}Gx^{2}\right)$ converge.

Lastly, we examine the behavior of the SCE if the quadratic coefficient
$\gamma$ is varied continuously. Let us begin with a positive $\gamma$,
and steadily decrease it. The effective coupling $g=\frac{g_{0}}{\beta\gamma^{2}}$
starts to increase, while the solution for $G$ is on the branch given
by \eqref{First_order_G}. As $\gamma\rightarrow0^{+}$, $g\rightarrow\infty$
and so does $G$. As $\gamma$ crosses zero, $G\rightarrow\infty$
for $\gamma\rightarrow0^{-}$, but as $\left|\gamma\right|$ increases,
$G$ descends from infinity along the double-well branch in \eqref{Double_Well_G}.

\revised{Strikingly, we have shown that the SCE provides a means to write down
a perturbative expansion to a non-perturbative system (in the sense
that the quartic contribution is essential to the stability of the
harmonic term).} Furthermore, this is despite the fact that the reference harmonic
model around which the SCE is performed is very different from the
original system, as illustrated in \figref{Double_Well_Potential}.
However, the SCE manages to coerce the calculation into the correct
result. For the ODM/OPT/LDE this was previously stated in Ref. \cite{BuckleyDuncanJonesZeroDimension}
and shown in Ref. \cite{GuidaKonishiSuzuki1996}, though they derive
looser functional forms for the convergence rate (i.e., stretched
instead of normal exponential convergence).

\revised{Naturally, a proper perturbative treatment of the problem would be
to expand around the two minima of the double-well, with the associated
degenerate ground states and tunneling between them. Of course, a
standard PT around the minima would still be formally divergent and
would require incorporating instantons (tunneling events) \cite{ZinnJustinQFT2002}.
However, our goal here was to demonstrate the power of the SCE, and
in particular its flexibility in formulating an exact approximation
scheme; note that the SCE is convergent to the correct result, even
when expanding around the local maximum ($x=0$) and without the inclusion
of non-perturbative contributions, such as instantons. What is more,
the discrete mapping between the anharmonic and double-well potentials
is instructive to analyze and gain intuition about, before we generalize
it to a continuous transition in the complex plane in \secref{SCE-of-airy-function-Ai}.}

\section{\label{sec:General-q-convergence}General Power-Law Perturbations }

A SCE may be performed in the case of a general perturbation $g\left|x\right|^{q}$.
We assume that $q>2$, and that $g$ may be complex but has a positive
real part. Expanding again around a modified $\frac{1}{2}Gx^{2}$
harmonic oscillator, the partition function is now \bw 
\[
\Z^{\left(\N\right)}=\sqrt{\frac{2}{G}}\sum_{n=0}^{\N}\left(1-\frac{1}{G}\right)^{n}\sum_{l}^{n}\frac{1}{n!}\binom{n}{l}\Gamma\left(\frac{1}{2}+n+\left(\frac{q}{2}-1\right)l\right)\left[\frac{\left(1-G\right)G^{\frac{q-2}{2}}}{2^{\frac{q}{2}}g}\right]^{-l}\,.
\]
 \ew One can then show that the self-consistency condition for the
moment $\left\langle x^{2M}\right\rangle $ is
\begin{equation}
\left(\frac{G}{2}\right)^{\frac{q}{2}}-\frac{1}{2}\left(\frac{G}{2}\right)^{\frac{q}{2}-1}-\frac{g}{M}C_{q}\left(M\right)=0\,,\qquad C_{q}\left(M\right)=\frac{\Gamma\left(M+\frac{q}{2}+\frac{1}{2}\right)}{\Gamma\left(M+\frac{1}{2}\right)}-\frac{\Gamma\left(\frac{q}{2}+\frac{1}{2}\right)}{\Gamma\left(\frac{1}{2}\right)}\,,\label{eq:G_equation_general_q}
\end{equation}
which has at least one solution for $G>1$. Let us first examine the
limiting behaviors of the solution, which we will use below. Asymptotically,
for large $M$ and fixed $g$ and $q$, $C_{q}\left(M\right)\sim M^{\frac{q}{2}}$
and thus 
\begin{equation}
G=2g^{\frac{2}{q}}M^{1-\frac{2}{q}}+\frac{2}{q}+\mathcal{O}\left(1/g^{\frac{2}{q}}M^{1-\frac{2}{q}}\right)\,.\label{eq:G_large_order_general_q}
\end{equation}
For small $\frac{g}{M}C_{q}$, $G$ is close to unity, or approximately
\begin{equation}
G\approx1-\frac{1}{q-2}+\sqrt{\left(\frac{1}{q-2}\right)^{2}+2^{1+\frac{q}{2}}\frac{gC_{q}\left(M\right)}{M\left(q-2\right)}}\,,\label{eq:G_general_q_small_g}
\end{equation}
which is exact for $q=4$. If $G-1$ is small but $M\gg q$ (i.e.,
$gM^{\frac{q}{2}-1}$ is small but $M$ is large), then a simpler
limit applies, 
\begin{equation}
G\approx1+2^{\frac{q}{2}}gM^{\frac{q}{2}-1}\,.\label{eq:G_general_q_superasymptotic}
\end{equation}
Going back to the expansion for $\Z$, it now becomes
\[
\Z^{\left(\N\right)}=\sqrt{\frac{2}{G}}\sum_{n=0}^{\N}\left(1-\frac{1}{G}\right)^{n}\sum_{l=0}^{n}\binom{n}{l}\frac{\left(-1\right)^{l}}{n!}\left(\frac{C_{q}\left(M\right)}{M}\right)^{-l}\Gamma\left(\frac{1}{2}+n+\left(\frac{q}{2}-1\right)l\right)\,.
\]

To first prove the possibility of non-divergence, we can generalize
case 3 of Proposition \ref{Proposition_Combined}. The dominant term
in the $l$ summation will occur at the index $l=0$ if 
\begin{align*}
1 & \ge\frac{M}{C_{q}}\frac{n-l}{l+1}\frac{\Gamma\left(n+\frac{q-2}{2}\left(l+1\right)+\frac{1}{2}\right)}{\Gamma\left(n+\frac{q-2}{2}l+\frac{1}{2}\right)}\Biggr|_{l=0}\twocolbr=\frac{M}{C_{q}}n\frac{\Gamma\left(n+\frac{q-2}{2}+\frac{1}{2}\right)}{\Gamma\left(n+\frac{1}{2}\right)}\,.
\end{align*}
The rhs scales as $n^{q/2}M^{1-q/2}$, hence the critical $M$ scales
as $M\sim\N^{\frac{q}{q-2}}$. Now, the partition function is once
more bounded by the sum of the $l=0$ terms, which again produces
the bound 
\[
\Z^{\left(\N\right)}\apprle\sqrt{\frac{N}{G}}\sim\N^{\frac{1}{2}\left(1-p\left(1-\frac{2}{q}\right)\right)}\rightarrow0\,,
\]
if $M\sim\N^{p}$ and $p>\frac{q}{q-2}$, in agreement with the critical
value we had just obtained. 

Let us now proceed to generalize cases 2a and 2b. The error can once
again be calculated by \bw 
\begin{align}
\lim_{\N\rightarrow\infty}\Z^{\left(\N\right)} & =\lim_{\N\rightarrow\infty}\int_{-\infty}^{\infty}e^{-\frac{1}{2}Gx^{2}}\sum_{n=0}^{\N}\frac{1}{n!}\left(-\left[\frac{1}{2}\left(1-G\right)x^{2}+gx^{q}\right]\right)^{n}dx\nonumber \\
 & =\lim_{\N\rightarrow\infty}\sqrt{\frac{2}{G}}\int_{0}^{\infty}e^{-u}\sum_{n=0}^{\N}\frac{1}{n!}\left(1-\frac{1}{G}\right)^{n}\left(u-\frac{Gg}{\left(G-1\right)}\left(\frac{2}{G}\right)^{\frac{q}{2}}u^{\frac{q}{2}}\right)^{n}\frac{du}{\sqrt{u}}\nonumber \\
 & =\lim_{\N\rightarrow\infty}\sqrt{\frac{2K}{G}}\sum_{n=0}^{\N}\left(1-\frac{1}{G}\right)^{n}\frac{K^{n}}{n!}\int_{0}^{\infty}e^{-Kv}v^{n}\left(1-v^{r}\right)^{n}\frac{dv}{\sqrt{v}}\,,
\end{align}
 \ew with $K=\left(\frac{C_{q}\left(M\right)}{M}\right)^{1/r}$ and
$r=\frac{q-2}{2}$. For $q=4$, we have $r=1$ and $C_{4}\left(M\right)=M^{2}+2M$,
and thus $K=M+2$, which is consistent with its definition in the
previous Sections. Apart from the power of $v$ in the third factor
of the integrand, this equation is the same as that for $q=4$ {[}see
\eqref{Z_limit_integral_form}{]}. In particular, the domains defined
in \eqref{Definition_of_D_Domains} still apply for the analysis of
the error of this expansion. Our aim is again to show that this expansion
can be constructed to converge exponentially with a suitable choice
of $M\left(\N\right)=\alpha\N$. Note that for any fixed value of
$q$, in the limit of very large $M\gg q$, $K=M+\frac{q^{2}}{4\left(q-2\right)}+o\left(1\right)$,
so up to constants they are asymptotically equivalent at large order.
Realizing this, we choose not to reformulate the convergence criteria
in terms of $K$ (i.e., we do \emph{not} redefine $\alpha=K/\N$),
since the moment $M$ carries a physical interpretation, while its
mapping to $K$ is not universal, as it clearly depends on the detailed
structure of the theory.

\subsection{The Domain $\mathcal{D}_{1}$ }

In this domain, we again bound the sum of truncated terms with the
aid of the Monotone Convergence Theorem, 
\begin{align}
-R_{1,q}^{\left(\N\right)} & =\sqrt{\frac{2K}{G}}\sum_{n=\N'}^{\infty}\frac{K^{n}}{n!}\left(1-\frac{1}{G}\right)^{n}\int_{0}^{1}e^{-Kv}v^{n}\left(1-v^{r}\right)^{n}\frac{dv}{\sqrt{v}}\twocolbr<\sqrt{\frac{2K}{G}}\sum_{n=\N'}^{\infty}\frac{K^{n}}{n!}\left(1-\frac{1}{G}\right)^{n}\int_{0}^{1}e^{-Kv-nv^{r}}v^{n-\frac{1}{2}}dv\,.
\end{align}
Our goal now is to bound the power $v^{r}$ in the exponent from below
by a linear approximation. If $q<4$ then $r<1$, and $v^{r}$ is
a concave function which on the interval $\left[0,1\right]$ is simply
bounded by $v$. This will reproduce the eventual $\left(\frac{K}{K+N'}\right)^{\N}$
result obtained for $q=4$. Let us then focus on $q>4$, so henceforth
$r>1$: As $v^{r}$ is now a convex function, it is bounded by any
line drawn tangential to it. Denoting by $v_{0}>0$ the arbitrary
point where we draw this tangent, we can bound
\begin{align*}
v^{r} & >rv_{0}^{r-1}\left(v-v_{0}\right)+v_{0}^{r}=rv_{0}^{r-1}v-\left(r-1\right)v_{0}^{r}\,.
\end{align*}
Plugging this in, the error is bounded by \bw 
\begin{align}
-R_{1,q}^{\left(\N\right)} & <\sqrt{\frac{2K}{G}}\sum_{n=\N'}^{\infty}\frac{K^{n}}{n!}\left(1-\frac{1}{G}\right)^{n}e^{n\left(r-1\right)v_{0}^{r}}\int_{0}^{\infty}e^{-\left(K+nrv_{0}^{r-1}\right)v}v^{n-\frac{1}{2}}dv\nonumber \\
 & =\sqrt{\frac{2K}{G}}\sum_{n=\N'}^{\infty}\frac{K^{n}}{n!}\left(1-\frac{1}{G}\right)^{n}e^{n\left(r-1\right)v_{0}^{r}}\frac{\Gamma\left(n+\frac{1}{2}\right)}{\left(K+nrv_{0}^{r-1}\right)^{n+\frac{1}{2}}}\nonumber \\
 & <\sqrt{\frac{2K}{G\N'\left(K+\N'rv_{0}^{r-1}\right)}}\left(1-\frac{1}{G}\right)^{\N'}\sum_{n=\N'}^{\infty}\left(\frac{Ke^{\left(r-1\right)v_{0}^{r}}}{K+\N'rv_{0}^{r-1}}\right)^{n}\,,
\end{align}
 \ew This is a geometric series which converges only if its quotient
is smaller than one, namely if 
\begin{equation}
K\left(e^{\left(r-1\right)v_{0}^{r}}-1\right)-\N'rv_{0}^{r-1}<0\,.\label{eq:tangent_convergence_criteria}
\end{equation}
This condition is satisfied trivially for $r=1$ (i.e. $q=4$), where
we saw that $\mathcal{D}_{1}$ is necessarily convergent. For $r>1$,
we note that for $v_{0}=0$ the left-hand side of \eqref{tangent_convergence_criteria}
equals $0$. However, since $e^{\left(r-1\right)v_{0}^{r}}$ can be
expanded in powers of $v_{0}^{r}$, the first non-vanishing derivative
of the lhs is the $\left(r-1\right)$-th, giving $\N'\left(-r!\right)<0$.
Thus, this equation is satisfied at least in the neighborhood of $0^{+}$.
Indeed, for $v_{0}\ll1$, we can expand 
\[
K\left(r-1\right)v_{0}^{r}-\N'rv_{0}^{r-1}<0\,,
\]
giving us a self-consistent solution as long as $v_{0}<\frac{r}{\left(r-1\right)}\frac{\N'}{K}$. 

Next, we wish to optimize the bound. Clearly, the quotient diverges
at large $v_{0}$ due to the exponent $e^{\left(r-1\right)v_{0}^{r}}$,
so the interval over which it is smaller than unity is finite, and
thus contains a minimum. Differentiating with respect to $v_{0}$
and setting to zero, we find 
\begin{equation}
\frac{K}{\N'}+rv_{0}^{r-1}=v_{0}^{-1}\,.\label{eq:optimal_tangent}
\end{equation}
For $v_{0}\rightarrow0$ the rhs diverges, while for large $v_{0}$
the lhs is dominant. Since at $v_{0}=\frac{\N'}{K}$ the lhs reads
$\frac{K}{\N'}+r\left(\frac{K}{\N'}\right)^{1-r}$ and the rhs is
$\frac{K}{\N'}$, this implies that the root of \eqref{optimal_tangent}
satisfies $v_{0}<\frac{\N'}{K}$. For $r=2$, it is solved explicitly
by
\[
v_{0}=\frac{1}{4}\left(\sqrt{\left(\frac{K}{\N'}\right)^{2}+8}-\frac{K}{\N'}\right)\sim\frac{\N'}{K}-2\left(\frac{\N'}{K}\right)^{3}\,.
\]

From now we set $v_{0}$ to be the root of \eqref{optimal_tangent}.
Inserting it into the error bound, we have
\begin{align*}
-R_{1,q}^{\left(\N\right)} & <\sqrt{\frac{2K}{G\N'\left(K+\N'rv_{0}^{r-1}\right)}}\left(1-\frac{1}{G}\right)^{\N'}\sum_{n=\N'}^{\infty}Q_{1}^{n}\twocolbr=\sqrt{\frac{2Kv_{0}}{G\left(\N'\right)^{2}}}\left(1-\frac{1}{G}\right)^{\N'}\frac{Q_{1}^{\N'}}{1-Q_{1}}\,,
\end{align*}
\begin{equation}
Q_{1}=\frac{Kv_{0}}{\N'}e^{\frac{r-1}{r}\left(1-\frac{K}{\N'}v_{0}\right)}<1\,.\label{eq:Quotient_D1_general_q}
\end{equation}

\subsection{The Domain $\mathcal{D}_{2}$}

In this domain, the error is bounded by the Lagrange remainder, 
\begin{align*}
\left|R_{2,q}^{\left(\N\right)}\right| & <\sqrt{\frac{2K}{G}}\frac{K^{\N'}}{\N'!}\left(1-\frac{1}{G}\right)^{\N'}\int_{1}^{\infty}e^{-Kv}v^{\N'}\left(v^{r}-1\right)^{\N'}\frac{dv}{\sqrt{v}}\,.
\end{align*}

For $q>4$ we can perform a bound which was too loose for $q=4$ but
is tight for larger powers. We may write
\begin{align}
\left|R_{2,q}^{\left(\N\right)}\right| & =\sqrt{\frac{2K}{G}}\frac{K^{\N'}}{\N'!}\left(1-\frac{1}{G}\right)^{\N'}\int_{1}^{\infty}e^{-Kv}v^{\N'}\left(v^{r}-1\right)^{\N'}\frac{dv}{\sqrt{v}}\nonumber \\
 & <\sqrt{\frac{2K}{G}}\frac{K^{\N'}}{\N'!}\left(1-\frac{1}{G}\right)^{\N'}\int_{1}^{\infty}e^{-Kv}v^{\left(r+1\right)\N'}\frac{dv}{\sqrt{v}}\nonumber \\
 & <\sqrt{\frac{2K}{G}}\frac{K^{\N'}}{\N'!}\left(1-\frac{1}{G}\right)^{\N'}\frac{\Gamma\left(\left(r+1\right)\N'+\frac{1}{2}\right)}{K^{\left(r+1\right)\N'+1/2}}\nonumber \\
 & <\sqrt{\frac{2}{G}}\left(1-\frac{1}{G}\right)^{\N'}\frac{\left(\left(r+1\right)\N'\right)!}{K^{r\N'}\N'!}\,.
\end{align}

For $2<q<4$, we can write

\begin{align}
\left|R_{2,q}^{\left(\N\right)}\right| & \Bigg/\sqrt{\frac{2K}{G}}\frac{K^{\N'}}{\N'!}\left(1-\frac{1}{G}\right)^{\N'}=\int_{1}^{\infty}e^{-Kv}v^{\N'}\left(v^{r}-1\right)^{\N'}\frac{dv}{\sqrt{v}}\nonumber \\
 & =e^{-K}\int_{0}^{\infty}e^{-Kv}\left(v+1\right)^{\N'-\frac{1}{2}}\left(\left(v+1\right)^{r}-1\right)^{\N'}dv\nonumber \\
 & <e^{-K}r^{\N'}\int_{0}^{\infty}e^{-Kv}\left(v+1\right)^{\N'-\frac{1}{2}}v^{\N'}dv
\end{align}
where he have used Bernoulli's inequality $\left(1+x\right)^{r}\le1+rx$
for positive $x$ and $0\le r\le1$. Note that we now have to evaluate
the same expression for $R_{2}^{\left(\N\right)}$ as we did for $q=4$
in \eqref{R2_for_q_4}, only multiplied by $r^{\N'}$. Since $q<4$,
we have $r=\frac{q-2}{2}<1$, and thus $r^{\N'}<1$. This means that
the requirements imposed on $K\left(\N\right)$, and consequently
on $M\left(\N\right)$, are more relaxed than for $q=4$, and thus
the expansion can be made convergent for any $2<q<4$.

\subsection{Rates and Domains of Convergence}

Collecting the contributions from all domains and cases, we find the
total bound on the error to be \bw 

\begin{align}
R_{q}^{\left(\N\right)} & <\sqrt{\frac{2}{G}}\left(1-\frac{1}{G}\right)^{\N'}\begin{cases}
\sqrt{\frac{Kv_{0}}{\left(\N'\right)^{2}}}\frac{Q_{1}^{\N'}}{1-Q_{1}}+\frac{\left(\left(r+1\right)\N'\right)!}{K^{r\N'}\N'!}\,, & r>1\\
\sqrt{\frac{K^{2}}{\left(\N'\right)^{3}}}\left(\frac{K}{K+\N'}\right)^{\N'-\frac{1}{2}}+\frac{\left(2r\right)^{\N'}}{\sqrt{2K}}e^{-K}+\frac{2}{K}\frac{\left(2r\N'\right)}{\N'!}^{\N'}\left(\frac{2\N'}{K}+1\right)^{\N'+1}e^{-2\N'-\frac{K}{2}-\frac{K^{2}}{2\N'+K}}. & r\le1
\end{cases}\label{eq:total_error_general_q}
\end{align}
 \ew 

This bound is similar in structure to that in \secref{convergence_q_4},
and can be shown to converge if $K/\N$ is sufficiently large. Let
us again start from the case $M=\alpha\N$ for constant $\alpha$
(case 2b). Recalling that $K=\left(\frac{C_{q}\left(M\right)}{M}\right)^{1/r}=M+\frac{q^{2}}{8r}+o\left(1\right)>M$,
we see that picking such $M$ ensures $K/\N>\alpha$. Thus, by replacing
$K/\N'\rightarrow\alpha$, we can find the minimal value of $\alpha$
required for convergence in the selection of the self-consistent moment
$\left\langle x^{2M}\right\rangle $.

Specifically, all error components are exponential in $\N$, that
is, scale as some $Q^{\N}$. For $r>1$, from $\mathcal{D}_{1}$ we
have $Q_{1}<1$ given by \eqref{Quotient_D1_general_q}, while from
$\mathcal{D}_{2}$ we have

\begin{equation}
Q_{2}=\alpha^{-r}\left(\frac{r+1}{e}\right)^{r+1}e\Rightarrow\alpha_{c}=\frac{1}{e}\left(r+1\right)^{1+\frac{1}{r}}\,.\label{eq:critical_alpha_D2_large_r_Q2}
\end{equation}

For $r\le1$, $Q_{1}=$$\frac{\alpha}{\alpha+1}$ and $\mathcal{D}_{2}$
has two components, which are
\begin{equation}
Q_{2A}=e^{-\alpha}2r\Rightarrow\alpha_{c,A}=\ln\left(2r\right)\,,\label{eq:Q2A_small_r}
\end{equation}
\begin{equation}
Q_{2B}=2r\left(\frac{2}{\alpha}+1\right)e^{-1-\frac{\alpha}{2}-\frac{\alpha^{2}}{\alpha+2}}\,,\label{eq:Q2B_small_r}
\end{equation}
 and $\alpha_{c,B}$ can be found numerically for each $q$. Note
that $Q_{2A,B}$ are both scaled by $r<1$ relative to their values
for $r=1$ while $Q_{1}$ is unchanged, so all interesting features
will occur for values of $\alpha$ smaller then those in \secref{convergence_q_4},
namely below $\alpha^{*}=1.317$ {[}see \eqref{alpha_star_value}{]}.
$Q_{2A}<Q_{2B}$ for all $\alpha<1.42$ (found numerically), thus
for all $r<1$, $Q_{2B}$ determines both the critical $\alpha_{c}$
or the optimal $\alpha^{*}$ exclusively (ergo $Q_{2B}$ is the only
relevant $Q_{2}$ for $r\le1$).

\begin{table*}[t]
\begin{centering}
\fontsize{10}{12}\selectfont%
\begin{tabular}{cccr@{\extracolsep{0pt}.}lr@{\extracolsep{0pt}.}lcccc}
\hline 
$x^{q}$ & $r=\frac{q}{2}-1$ & PT Divergence & \multicolumn{2}{c}{$\alpha_{c}$} & \multicolumn{2}{c}{$\alpha^{*}$} & $1-Q^{*}$ & $-\log_{10}Q^{*}$ & $\alpha_{num}^{*}$ & $-\log_{10}Q_{num}$\tabularnewline
\hline 
\hline 
$x^{3}$ & $\frac{1}{2}$ & $n^{n/2}$ & 0&613 & 0&981 & $5.05\times10^{-1}$ & $3.05\times10^{-1}$ & 1.10 & $5.38\times10^{-1}$\tabularnewline
$x^{4}$ & 1 & $n^{n}$ & \quad 0&976 & \quad 1&317 & $4.32\times10^{-1}$ & $2.45\times10^{-1}$ & 1.33 & $2.83\times10^{-1}$\tabularnewline
$x^{6}$ & 2 & $n^{2n}$ & 1&91 & 2&08 & $1.54\times10^{-1}$ & $7.29\times10^{-2}$ & 1.80 & $9.4\times10^{-2}$\tabularnewline
$x^{8}$ & 3 & $n^{3n}$ & 2&34 & 2&38 & $5.63\times10^{-2}$ & $2.52\times10^{-2}$ & 2.25 & $2.6\times10^{-2}$\tabularnewline
$x^{10}$ & 4 & $n^{4n}$ & 2&750 & 2&761 & $1.52\times10^{-2}$ & $6.64\times10^{-3}$ & 2.70 & $7.1\times10^{-3}$\tabularnewline
$x^{12}$ & 5 & $n^{5n}$ & 3&1585 & 3&1605 & $3.05\times10^{-3}$ & $1.33\times10^{-3}$ & 3.20 & $1.5\times10^{-3}$\tabularnewline
$x^{2r+2}$ & $r\gg1$ & $n^{rn}$ & \multicolumn{2}{c}{$\frac{r+\ln r+1}{e}$} & \multicolumn{2}{c}{} & $\frac{1}{er}\left(\frac{e}{r}\right)^{r}$ &  &  & \tabularnewline
\hline 
\end{tabular}\normalsize
\par\end{centering}

\caption[Table \thetable]{\label{tab:critical_alphas}Convergence of the SCE for varying anharmonicity
$g|x|^{q}$. The columns are: power of perturbation $x^{q}$; $r=\frac{q}{2}-1$;
the large-order divergence of the summand in the standard PT expansion,
$a_{n}\propto\Gamma\left(\frac{qn}{2}\right)/n!\sim n^{rn}$; the
upper bound on the minimal value of $\alpha$ necessary for convergence,
$\alpha_{c}$, as predicted by our analytical error bound; the optimal
value for our bound, $\alpha^{*}$; the base of the uniform exponential
error at $\alpha^{*}$, $Q^{*}$ (i.e. $R\sim\left(Q^{*}\right)^{\N}$),
specified by its distance from unity; $\log_{10}Q^{*}$, showing the
minimal amount of significant digits obtained per increment of the
SCE order $\N$; $\alpha_{num}^{*}$, the numerically-obtained approximate
true value of the optimum; and $\log_{10}Q_{num}$, the value obtained
by numerical evaluation and least-squares fitting, for $\alpha_{num}^{*}$.
Note that as $r$ is increased, $\alpha^{*}$ approaches $\alpha_{c}$,
and thus the base $Q^{*}$ approaches unity. Since the analytical
bound is not very tight for $q\protect\neq4$, the true optimum value
$\alpha_{num}^{*}$ is actually smaller than $\alpha_{c}$. Lastly,
the final row shows the large $q\gg1$ leading order approximations
of $\alpha_{c}$ and $Q^{*}$.}
\end{table*}

In order to asses the applicability of SCE for anharmonicities with
$q\neq4$, we list the values of the critical $\alpha_{c}$ and the
optimal $\alpha^{*}$ for the first few integer perturbation powers
$q>2$ in \tabref{critical_alphas}. $\alpha_{c}$ is given by \eqref{critical_alpha_D2_large_r_Q2}
for $r>1$, or obtained numerically by equating $Q_{2B}$ in \eqref{Q2B_small_r}
to unity for $r<1$. $\alpha^{*}$ is found numerically by equating
the appropriate $Q_{2}$ to the corresponding $Q_{1}$, either that
obtained by setting $K/\N'\rightarrow\alpha$ in \eqref{Quotient_D1_general_q},
or $\frac{\alpha}{\alpha+1}$.

For larger orders of anharmonicity, the optimal value $\alpha^{*}$
tends towards $\alpha_{c}$. This is because $\alpha_{c}$ grows large,
restricting us to larger values of $\alpha$. However, at large $\alpha$
we find that $Q_{1}\rightarrow1^{-}$, so the conditions $Q_{2}=Q_{1}$
and $Q_{2}=1$ become virtually identical. We may use this intuition
to estimate the convergence rate for very high anharmonicities $r\gg1$:
By substituting $K/\N'\rightarrow\alpha$ and putting $v_{0}=\frac{1}{\alpha}-\delta v$
in \eqref{optimal_tangent}, we can find the post-leading correction
$\delta v$, 
\begin{gather*}
\alpha\delta v=\frac{r}{r^{2}+\alpha^{r}}\,.
\end{gather*}
Plugging this into \eqref{Quotient_D1_general_q},
\[
Q=\alpha v_{0}e^{\frac{r-1}{r}\left(1-\alpha v_{0}\right)}=\left(1-\alpha\delta v\right)e^{\frac{r-1}{r}\alpha\delta v}\approx e^{-\frac{\alpha\delta v}{r}}=e^{-\frac{1}{r^{2}+\alpha^{r}}}\,,
\]
and we now assume $\alpha=\alpha_{c}$, given by \eqref{critical_alpha_D2_large_r_Q2}.
This gives values in good agreement with the values in \tabref{critical_alphas},
especially for $r\ge4$. For the $q,r\gg1$ limit, we discard the
$r^{2}$ factor. Since $\alpha_{c}\sim r$, we require all corrections
up to $\mathcal{O}\left(1\right)$ to find the limit of $\alpha^{r}$.
The asymptotic form of \eqref{critical_alpha_D2_large_r_Q2} is 
\begin{equation}
\alpha_{c}=\left(r+\ln r+1\right)/e+\mathcal{O}\left(\left(\ln r\right)^{2}/r\right)\,.\label{eq:critical_alpha_scaling}
\end{equation}
This yields 
\begin{align*}
1-Q & \sim1-e^{\frac{1}{\alpha_{c}^{r}}}\sim\twocolbr\left[\frac{\left(r+\ln r+1\right)}{e}\right]^{-r}\approx\frac{1}{re}\left(\frac{r}{e}\right)^{-r}\,.
\end{align*}

While in principle the SCE is convergent for all $q$, \tabref{critical_alphas}
shows that the exponential convergence rate decreases upon increasing
$q$. However, it should be noted that these bounds are not particularly
tight as compared with the numerically fitted values (as demonstrated
in \secref{numerical_results_q_4} for $q=4$; additional numerical
results are listed in the table). Furthermore, they also do not reflect
the additional convergent factor of $\left(1-1/G\right)^{\N}$, giving
a stretched exponential behavior which can assist the convergence
rate for weak couplings $g\lesssim1$. \revised{In contrast, we note that for $q>4$, the standard PT series coefficients
grow more rapidly than factorially, and thus are not amenable to usual
Borel resummation.}

Lastly, we note that if $M\left(\N\right)\sim\N^{p}$ with $1<p<1+\frac{1}{r}$,
then $\alpha$ is increasing, and again the error is dominated by
domain $\mathcal{D}_{1}$. We may use our $r\gg1$ estimate for $Q_{1}$,
since its derivation only relied on $\alpha\gg1$, which will be satisfied
for sufficiently large $\N$ for $p>1$. We then find that 
\[
Q^{\N}\approx e^{-\N/\alpha^{r}}=e^{-\N\times\N^{-r\left(p-1\right)}}=e^{-\N^{1-r-rp}}\,,
\]
and convergence is attained if $1<p<1+\frac{1}{r}$. Conversely, the
$1-1/G$ factor scales as 
\[
\left(1-\frac{1}{G}\right)^{\N}\sim\left(1-\frac{1}{M^{1-\frac{2}{q}}}\right)^{\N}\sim e^{-\N\times M^{\frac{2}{q}-1}}\sim e^{-\N^{1-p\left(1-2/q\right)}}\,,
\]
which adds an additional $g$-dependent convergent factor for $1<p<\frac{1}{1-2/q}=1+\frac{1}{r}$,
that is, for the same range of $p$. This proves case 2a of Proposition
\ref{Proposition_Combined} for arbitrary $q$.

\section{\label{sec:SCE-of-airy-function-Ai}SCE in the Complex Plane: Oscillatory
Integrals and Stokes Phenomenon}

Following the success of the SCE in treating the anharmonic oscillator,
we wish to elucidate other properties of this technique. We would
like to explore how the SCE carries over to complex functions, and
in particular, oscillatory integrals. Using the results of the previous
section for $q=3$, we will treat the Airy function $\mathrm{Ai}\left(z\right)$,
which has many applications across the fields of optics, quantum mechanics,
fluid mechanics, and elasticity \cite{OlivierOaresAiryApplications2010}. 

Our SCE of $\mathrm{Ai}\left(z\right)$ will be performed around the
limit $z\rightarrow\infty$, much like its standard asymptotic expansion.
We will argue for several different behaviors of the SCE which will
depend on where $z$ lies in the complex plane with respect to the
Stokes lines \cite{Stokes1864} of the Airy function \cite{OlivierOaresAiryApplications2010},
defined below. This will be reflected in the behavior of the SCE parameter
$G$ and a factor $\left(1-\sqrt{z}/\sqrt{G}\right)^{\N}$, which
is analogous to the stretched exponent of the anharmonic oscillator.
We will see three regimes, depending of the argument of $z$: monotonic
and uniform exponential convergence, reduced-rate convergence, and
initial explosion before eventual convergence. 

A previous LDE treatment of a similar problem \cite{BlencoweJonesKorteAiryLDE1998},
originally in the context of non-Hermitian Hamiltonians, was restricted
to $z=0$, and thus did not observe this rich phenomenology. Furthermore,
here will show that the SCE criteria naturally gives rise to a framework
which explains these distinct behaviors. Qualitatively, the different
regimes in the different Stokes sectors will arise due to two different
causes: the first transition, to non-monotonic convergence, is a feature
of the solutions for $G$ as a function of $M$ and $z$, and the
SCE transitions it smoothly for all $\left|z\right|$. The third behavior
is born by a mutual incompatibility between two demands: the self-consistency
of an SCE moment $M$, and that at zeroth order $G$ should approach
$z$. This discrepancy is a generalization of the double-well behavior
of \secref{double_well} to the complex plane.

For fixed $\left|z\right|$, all three cases above eventually converge
exponentially for large enough $\N\sim\left|z\right|^{\frac{3}{2}}$.
In particular, the result of the SCE will vary smoothly across the
Stokes lines for any argument $z'$ within a disk of radius $\left|z\right|$
around the origin.

\subsection{The SCE of $\mathrm{Ai}\left(z\right)$}

The Airy function can be put into the integral representation \cite[Eq. (9.5.7)]{NIST:DLMF}
\begin{align}
\mathrm{Ai}\left(z\right) & =\frac{e^{-\frac{2}{3}z^{3/2}}}{\pi}\int_{0}^{\infty}e^{-z^{1/2}t^{2}}\cos\left(\frac{t^{3}}{3}\right)dt\twocolbr\sim\frac{e^{-\frac{2}{3}z^{3/2}}}{\pi z^{\frac{1}{4}}}\int_{-\infty e^{\frac{1}{4}\arg z}}^{\infty e^{\frac{1}{4}\arg z}}e^{-x^{2}+\frac{i}{3}z^{-\frac{3}{4}}x^{3}}dx\label{eq:Airy_representation_integral_cosine}
\end{align}
for $\left|\arg\,z\right|<\pi$. A change of integration variables
was used to illustrate the correspondence between this case and the
previous sections: separating the cosine into its two complex oscillating
components, $z^{-\frac{3}{4}}$ can take the role of the nonlinear
coupling coefficient $g$ of a cubic perturbation. Crucially, the
PT of $\mathrm{Ai\left(z\right)}$ corresponds to its asymptotic expansion
for large $z$. We purposefully do not employ the new variables in
the calculation below, for two reasons: First, it would change the
integration contour, depending on the phase of $z$. Second, this
transformation would produce a $z^{-\frac{1}{4}}$ Jacobian which
will diverges at the origin; this is to be avoided, as we would like
to demonstrate the SCE's success even for $z=0$, which corresponds
to $g=\infty$.

We would now like to introduce the self-consistent variable $G$,
so that we expand around the Gaussian weight $e^{-\sqrt{G}t^{2}}$.
One may be tempted immediately to take $\cos\left(\frac{t^{3}}{3}\right)=\frac{1}{2}\left(e^{\frac{it^{3}}{3}}+e^{-\frac{it^{3}}{3}}\right)$,
so that we now need to expand $\int e^{-\sqrt{G}t^{2}-\left(\sqrt{z}-\sqrt{G}\right)t^{2}\pm it^{3}}$
and then expand in powers $\sum_{n}\left(\left(\sqrt{z}-\sqrt{G}\right)t^{2}\pm\frac{it^{3}}{3}\right)^{n}$.
However, This will clearly lead all terms with an odd power of $t$
to vanish identically upon integration. In particular, the expansion
would not contain any first-order correction in the non-linearity
$t^{3}$. This would cause the first-order correction of any observable
to be proportional to $\left(\sqrt{z}-\sqrt{G}\right)$, and self-consistency
would only be achieved if $G=z$, thus defeating the purpose of SCE.
An alternative would be to demand self-consistency to second order;
however, this would lead to the loss of an attractive feature of a
first-order consistency scheme: the independence of internal summations
from the argument $z$. As we saw in the case of the anharmonic oscillator,
it was this independence that allowed us to demonstrate uniform and
exponential convergence. A lesson that is learned from this is that
one must take care not to introduce artificial symmetries into the
problem when applying the SCE: Originally, the relation between the
two components of the cosine is complex conjugation, but upon expansion
becomes that of parity.

Instead, we expand each component of the cosine separately: 
\begin{align}
\mathrm{Ai}\left(z\right) & =\frac{e^{-\frac{2}{3}z^{3/2}}z^{-\frac{1}{4}}}{2\pi}\left[\tilde{\mathrm{Ai}}_{+}\left(z\right)+\tilde{\mathrm{Ai}}_{-}\left(z\right)\right]\,,
\end{align}
with $\tilde{\mathrm{Ai}}_{\pm}\left(z\right)\equiv z^{\frac{1}{4}}\int_{0}^{\infty}e^{-z^{1/2}t^{2}\pm\frac{i}{3}t^{3}}dt$.
Our strategy now is to SCE-expand each reduced Airy function in a
separate expansion. Henceforth, it is important to take roots of $z$
and $G$ carefully, with respect to the principal branch of the root
function, defined such that $\arg\sqrt{z}=\frac{1}{2}\arg z$ for
$-\pi<\arg z\le\pi$, and with a branch cut along the negative real
axis. Letting $\Delta=\pm1$, we find \bw
\begin{align}
\tilde{\mathrm{Ai}}_{\Delta}^{\left(\N\right)}\left(z\right) & =z^{\frac{1}{4}}\sum_{n=0}^{\N}\frac{1}{n!}\sum_{l=0}^{n}\binom{n}{l}\int_{0}^{\infty}e^{-\sqrt{G_{\Delta}}t^{2}}\left(\left(\sqrt{G_{\Delta}}-\sqrt{z}\right)t^{2}\right)^{n-l}\left(\frac{i\Delta}{3}t^{3}\right)^{l}dt\nonumber \\
 & =z^{\frac{1}{4}}\sum_{n=0}^{\N}\frac{1}{n!}\sum_{l=0}^{n}\binom{n}{l}\left(\sqrt{G_{\Delta}}-\sqrt{z}\right)^{n-l}\left(\frac{i\Delta}{3}\right)^{l}\int_{0}^{\infty}e^{-\sqrt{G_{\Delta}}t^{2}}t^{2n+l}dt\,.
\end{align}
 \ew Let us assume that $\arg G_{\pm}\neq\pi$, so that this integral
converges. Defining $t'=G_{\Delta}^{\frac{1}{4}}t$, the integration
path is rotated in the complex plane to $\left[0,\infty e^{\frac{1}{4}\arg G_{\Delta}}\right)$.
The integrand $e^{-\left(t'\right)^{2}}$ is entire and decays to
zero at infinity for $\left|\arg t'\right|<\frac{\pi}{4}$, and since
we assumed $\left|\arg G_{\Delta}\right|<\pi$, we may deform the
integration path to again run over the positive real axis. This leads
to \bw 
\begin{align}
\tilde{\mathrm{Ai}}_{\Delta}^{\left(\N\right)}\left(z\right) & =\frac{1}{2}\left(\frac{z}{G_{\Delta}}\right)^{\frac{1}{4}}\sum_{n=0}^{\N}\frac{1}{n!}\left(1-\frac{\sqrt{z}}{\sqrt{G_{\Delta}}}\right)^{n}\sum_{l=0}^{n}\binom{n}{l}\left(-3i\Delta G_{\Delta}^{\frac{3}{4}}\left(1-\frac{\sqrt{z}}{\sqrt{G_{\Delta}}}\right)\right)^{-l}\Gamma\left(\frac{2n+l+1}{2}\right)\,.
\end{align}
 \ew

For the moments of the variable $t$, we find that the first order
integral is \bw 
\begin{align*}
\left\langle t^{2M}\mathrm{Ai}\left(z\right)\right\rangle ^{\left(1\right)} & =\frac{e^{-\frac{2}{3}z^{3/2}}z^{-\frac{1}{4}}}{2\pi}\sum_{\Delta=\pm}\frac{1}{2}\left(\frac{z}{G_{\Delta}}\right)^{\frac{1}{4}}G_{\Delta}^{\frac{M}{2}}\Gamma\left(\frac{1}{2}+M\right)\times\\
 & \left\{ 1+\left(1-\frac{\sqrt{z}}{\sqrt{G_{\Delta}}}\right)\left[\left(\frac{1}{2}+M\right)+\left(-3i\Delta G_{\Delta}^{\frac{3}{4}}\left(1-\frac{\sqrt{z}}{\sqrt{G_{\Delta}}}\right)\right)^{-1}\frac{\Gamma\left(2+M\right)}{\Gamma\left(\frac{1}{2}+M\right)}\right]\right\} \,.
\end{align*}
\ew It is clear that setting $G_{+}=G_{-}$ would cancel the entire
first order contribution as discussed above, i.e., that no first-order
consistency can be established. Conversely, finding the net correction
to the moment $\left\langle t^{2M}\right\rangle $ is laborious due
to the required summation over $\Delta$. A solution would be to regard
the two integrals $\tilde{\mathrm{Ai}}_{\pm}$ as independent, so
that each generates its own moments which must be conserved independently.
In other words, we treat the SCE for the Airy function as the sum
of two separate SCEs, whose combined numeric value gives $\mathrm{Ai}\left(z\right)$.
The condition for each $G_{\Delta}$ is now 
\begin{align}
0 & =M\left[1-\left(3i\Delta G_{\Delta}^{\frac{3}{4}}\left(1-\frac{\sqrt{z}}{\sqrt{G_{\Delta}}}\right)\right)^{-1}\frac{C_{3}\left(M\right)}{M}\right]\,,
\end{align}
with $C_{3}\left(M\right)\sim M^{\frac{3}{2}}$ defined as in \eqref{G_equation_general_q}
for $q=3$. With $y_{\Delta}=G_{\Delta}^{\frac{1}{4}}$ (principal
value implied), we then have
\begin{equation}
3i\Delta y_{\Delta}\left(y_{\Delta}^{2}-\sqrt{z}\right)-C_{3}\left(M\right)/M=0\,,\label{eq:Complex_Airy_G_equation}
\end{equation}
whose root of interest is the one which tends \emph{continuously}
$y\rightarrow z^{\frac{1}{4}}$ for large $\left|z\right|$. The immediate
consequence is that $3i\Delta G_{\Delta}^{\frac{3}{4}}\left(1-\frac{\sqrt{z}}{\sqrt{G_{\Delta}}}\right)=\frac{C_{3}\left(M\right)}{M}$
which is a constant (in $z$) once again. This finally yields \bw
\begin{align*}
\tilde{\mathrm{Ai}}_{\Delta}^{\left(\N\right)}\left(z\right) & =\frac{1}{2}\left(\frac{z}{G_{\Delta}}\right)^{\frac{1}{4}}\sum_{n=0}^{\N}\frac{1}{n!}\left(1-\frac{\sqrt{z}}{\sqrt{G_{\Delta}}}\right)^{n}\sum_{l=0}^{n}\binom{n}{l}\left(-\frac{C_{3}\left(M\right)}{M}\right)^{-l}\Gamma\left(\frac{2n+l+1}{2}\right)\,.
\end{align*}
 \ew We argue that the same moment $M$ should be fixed symmetrically
for $\Delta=1$ and $\Delta=-1$, so we have \bw
\begin{align}
\tilde{\mathrm{Ai}}^{\left(\N\right)}\left(z\right) & =z^{\frac{1}{4}}\sum_{n=0}^{\N}\frac{1}{3^{n}n!}\left[\sum_{l=0}^{n}\binom{n}{l}\left(-\frac{C_{3}\left(M\right)}{M}\right)^{n-l}\Gamma\left(\frac{2n+l+1}{2}\right)\right]\times\frac{i^{n}}{2}\left\{ \left(G_{+}\right)^{-\frac{3n+1}{4}}+\left(-1\right)^{n}\left(G_{-}\right)^{-\frac{3n+1}{4}}\right\} \,,\label{eq:SCE_Complex_Airy_With_G}
\end{align}
 \ew where we performed another substitution of \eqref{Complex_Airy_G_equation}
after isolating $1-\frac{\sqrt{z}}{\sqrt{G_{\Delta}}}$.

Furthermore, for real $z$ it is apparent from \eqref{Complex_Airy_G_equation}
that the equations for $y_{\pm}$ can be transformed from one to the
other by mapping $i\mapsto-i$. This implies that the roots $y_{\pm}$
satisfy the same relation, namely that $G_{-}^{\frac{1}{4}}=\left(G_{+}^{\frac{1}{4}}\right)^{*}$.
Thus, $\frac{i^{n}}{2}\left[\left(G_{+}\right)^{-\frac{3n+1}{4}}+\left(-1\right)^{n}\left(G_{-}\right)^{-\frac{3n+1}{4}}\right]$
gives (up to a sign) either the real or imaginary part of $G_{+}^{-\frac{3n+1}{4}}$,
according to the parity of $n$, so one can see that the above expansion
is manifestly real for real $z$, by construction. For general $z$,
one finds that $G_{-}^{\frac{1}{4}}\left(z\right)=\left(G_{+}^{\frac{1}{4}}\left(z^{*}\right)\right)^{*}$,
which quickly leads to the conclusion that the expansion satisfies
$\tilde{\mathrm{Ai}}^{\left(\N\right)}\left(z^{*}\right)=\left[\tilde{\mathrm{Ai}}^{\left(\N\right)}\left(z\right)\right]^{*}$
for all $\N$.

\subsection{Analytic Properties and Solutions for $G_{\pm}$ Across Stokes Lines}

Recall that each of the Airy SCEs $\tilde{\mathrm{Ai}}_{\pm}\left(z\right)$
is in essence a complex extension of the partition function of an
anharmonic oscillator with an $x^{3}$ perturbation, as analyzed in
\secref{General-q-convergence}. In particular, this implies that
the coefficients of the SCE expansion, when viewed as power series
in $\left(1-\sqrt{z}/\sqrt{G}\right)$, decay exponentially fast for
a proper choice of $M$. The $q=3$ entry in \tabref{critical_alphas}
implies that the Airy SCE converges for $M$ linear in $\N$ and $M/\N\ge0.613$,
and recovers at least $\approx0.305\N$ decimal places at the estimated
optimum $M/\N\approx0.981$. This bound is not particularly tight,
and in practice the SCE appears to converge for $\alpha$ as low as
$0.4$, and exhibits its optimum at roughly $M/\N\approx1.1$. 

This convergence holds so long as the exponential convergence of these
coefficients is not overwhelmed by any divergence due to the stretched
exponent $\left(1-\sqrt{z}/\sqrt{G}\right)^{\N}$. We must now explore
the behavior of this factor as a function of the order $\N$, and
more importantly, of the location of $z$ in the complex plane.

The only $z$ dependence in the Airy SCE enters through $G_{\pm}$
and consequently the factors $\left(1-\sqrt{z}/\sqrt{G_{\pm}}\right)^{n}$,
which are governed by the solutions of \eqref{Complex_Airy_G_equation}
as a function of $M$ and $z$. We will now see that these factors
posses three different behaviors, depending on where $z$ lies in
the complex plane, relative to the Stokes lines \cite{Stokes1864}
of the Airy function \cite{OlivierOaresAiryApplications2010}. These
lines represent the dynamics and role reversal of the saddles of the
integral representation (\ref{eq:Airy_representation_integral_cosine}):
The integrand has two saddles, located at $t=0$ and $t=-2i\sqrt{z}$,
producing the leading-order exponentials $e^{-\frac{2}{3}z^{\frac{3}{2}}}$
and $e^{+\frac{2}{3}z^{\frac{3}{2}}}$, respectively. For real $z$,
only the first saddle is in the $t$ integration path, which cannot
be deformed to pass through the second; thus, the second saddle makes
no contribution, and $\mathrm{Ai\left(z\right)}$ is exponentially
small. Increasing the phase of $z$ to $\frac{\pi}{3}$, now the integration
contour may pass through both saddles, which are both imaginary exponentials
with norm $1$. Increasing the phase further, the magnitude of the
second saddle diminishes while the first's is enlarged, reaching maximal
dominance at $\arg z=\frac{2\pi}{3}$. Lastly for negative $z$, both
saddles are again oscillatory. The situation is similar in the bottom
half of the plane. Thus, the lines of phase $\arg z=0,\pm\frac{\pi}{3},\pm\frac{2\pi}{3}$
and $\pi$, which are called Stokes lines\footnote{We will not draw the distinction between Stokes and anti-Stokes lines.}
define six different wedges in the complex plane with three distinct
behaviors of $\mathrm{Ai}\left(z\right)$ --- exponential decay, growth,
and oscillation. 

\begin{figure*}[tp]
\centering{}\subfloat{\includegraphics[width=\figurewidth]{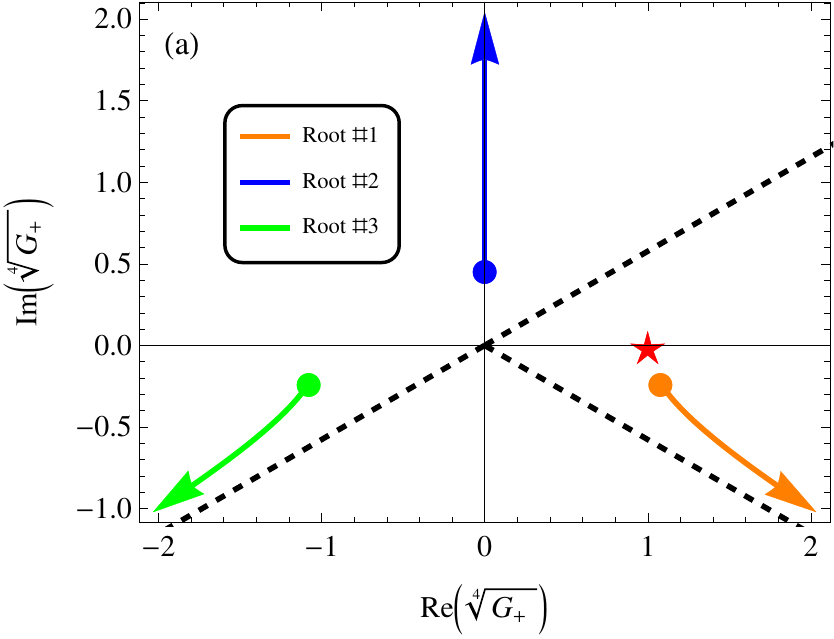}\label{fig:SCE_Airy_roots_of_G}}\hspace*{\fill}\subfloat{\includegraphics[width=\figurewidth]{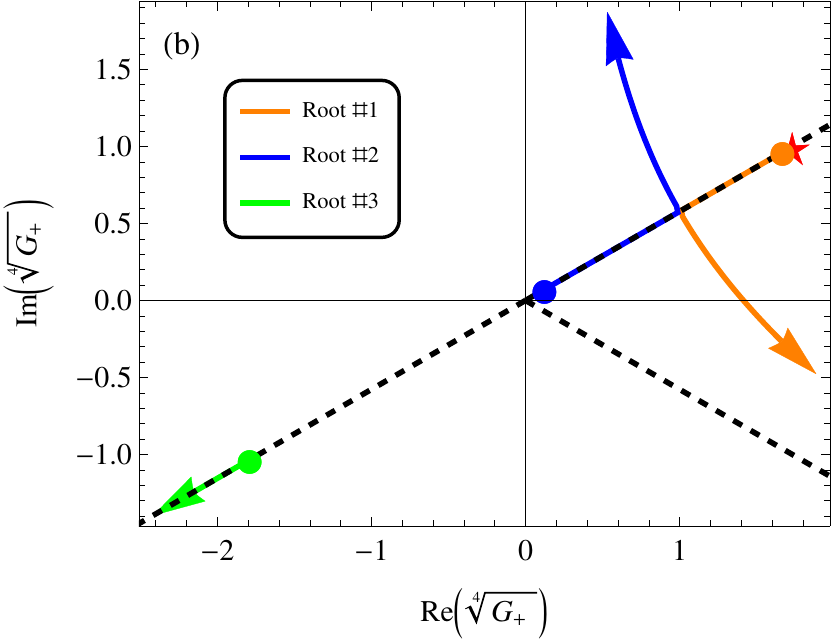}\label{fig:SCE_Airy_roots_of_G_arg_120}}\caption[Figure \thefigure]{Trajectories of the three solutions of \eqref{Complex_Airy_G_equation}
for $y_{+}=G_{+}^{\frac{1}{4}}$ and varying $M$. (a) The trajectories
for $z=1$. The arrowheads point in the direction the solution moves
as $M$ is increased, up to $1000$; the base of the arrow is the
value of each root at $M=1$. Note that only one particular root is
in the vicinity of $z^{\frac{1}{4}}$ (marked by a red star), tending
to it asymptotically if $\left|z\right|$ is increased with $M$ held
fixed. As $M$ increases the solutions tend towards the three cubic
roots of $C_{3}\left(M\right)/3iM$, which lie on the positive imaginary
axis and along the $\pm e^{\mp\frac{i\pi}{6}}$ directions, marked
by the dashed lines. Note that the solution of interest (orange) tends
to the principal branch of the cubic root, $\sqrt[3]{C_{3}\left(M\right)/3iM}$;
this behavior experiences a crossover when transitioning across the
Stokes line at $\arg z=\frac{2\pi}{3}$, after which the root which
stems from $z^{\frac{1}{4}}$ tends to the imaginary axis, cf. the
right panel. (b) The same trajectories for $z=16e^{\frac{2\pi i}{3}}$.
Note that initially all roots lie on the line parallel to $e^{\frac{i\pi}{6}}$.
The ``primary'' root coming from $z^{\frac{1}{4}}$ ``collides''
with the root coming from $0$, after which they become a complex-conjugate
pair of solutions for $r=\left(\frac{G}{z}\right)^{\frac{1}{4}}$,
and continue to tend to the same asymptotes as before. For $\arg z$
any larger, the asymptotes of these two roots will be swapped.}
\end{figure*}

Let us then explore the behavior of $1-\sqrt{z}/\sqrt{G}$ as a function
of $\arg z$. Beginning with real $z$, the resolved trajectories
of the three solutions for $G_{+}^{\frac{1}{4}}$ as $M$ is increased
are depicted in \figref{SCE_Airy_roots_of_G}, for $z=1$. Note the
structure of the solutions: one imaginary and a pair which is symmetric
about the imaginary axis. This is directly deduced by the fact that
for real $z$, substituting $y=ix$ into \eqref{Complex_Airy_G_equation},
the resulting equation for $x$ is a cubic equation with strictly
real coefficients, so is has one real root and a conjugate pair of
complex roots. After rotating back to $y$, we obtain the aforementioned
structure.

Next, we observe that indeed one particular root of $G^{\frac{1}{4}}$,
that colored orange in \figref{SCE_Airy_roots_of_G}, tends continuously
to $z^{\frac{1}{4}}$ for large $\left|z\right|$. This identifies
the root of interest among the three which should be inserted into
\eqref{SCE_Complex_Airy_With_G}. Note that as $M\rightarrow\infty$
this root tends to the principal root of $\sqrt[3]{C_{3}\left(M\right)/3iM}$.
However, this property is suddenly violated once we cross the Stokes
line at $\arg z=\frac{2\pi}{3}$. To see this, we substitute $y=rz^{\frac{1}{4}}$
into \eqref{Complex_Airy_G_equation} to obtain 
\[
3\Delta ir(r^{2}-1)z^{\frac{3}{4}}=C_{3}\left(M\right)/M\,.
\]
Examining $z=\left|z\right|e^{\frac{2\pi}{3}i+i\delta\varphi}$ which
is close to the Stokes line, we then have 
\begin{equation}
r(r^{2}-1)=-\frac{\Delta C_{3}\left(M\right)}{3\left|z^{\frac{3}{4}}\right|M}e^{-\frac{3}{4}i\delta\varphi}\,,\label{eq:Airy_equation_scatter_on_2pi3_line}
\end{equation}
 so on the line ($\delta\varphi=0$), the equation for $r$ is also
cubic with real coefficients. There are three roots which converge
to $r=0,-1$ and $+1$ for large $\left|z\right|$. For $\Delta=+1$,
as $M$ is increased, a negative real root $r<-1$ exists, while the
roots originating at $0,1$ must eventually reach argument $\pm\frac{\pi}{3}$,
as they are the two complex-conjugate cubic roots of $-\frac{C_{3}\left(M\right)}{M}\sim-\sqrt{M}$.
Clearly, there is a critical value $M\left(z\right)$ where these
two roots meet on the real line before jumping off it and becoming
a conjugate pair. Since the roots meet, it is ill-defined to ask to
which side of the line the root originally at $r=1$ scatters; the
collision mixes their identity. See \figref{SCE_Airy_roots_of_G_arg_120}
for details. The point at which this collision occurs can be found
by equating the discriminant of \eqref{Airy_equation_scatter_on_2pi3_line}
to zero with $\delta\varphi=0$, yielding 
\begin{equation}
\sqrt{\frac{4}{3}}\left|z\right|^{\frac{3}{4}}=\frac{C_{3}\left(M\right)}{M}\sim\sqrt{M}\,.\label{eq:Airy_M_Scatter_point}
\end{equation}

If $z^{\frac{1}{4}}$ is slightly below the line $e^{+\frac{i\pi}{6}}$(i.e.
$\delta\varphi=\epsilon<0$), then this symmetry is broken, and the
solution $r=1$ is ``scattered'' below $r=0$, tending to the line
$e^{-\frac{i\pi}{3}}$, so $G_{+}^{\frac{1}{4}}$ tends to $e^{-\frac{i\pi}{3}+\frac{1}{4}\frac{2\pi i}{3}}=e^{-\frac{i\pi}{6}}$.
This is the case which happens with real and positive $z$. If instead
$\delta\varphi>0$, then our privileged root scatters from $r=1$
up to $r\propto e^{+\frac{i\pi}{3}}$ and $G_{+}^{\frac{1}{4}}$ tends
to the line $e^{+\frac{i\pi}{3}+\frac{1}{4}\frac{2\pi i}{3}}=e^{+\frac{i\pi}{2}}$.
The same logic applies for the root $G_{-}$, with all the signs of
the arguments above negated. 

We now come to the following interpretation of the behavior of the
Airy SCE as the argument of $z$ is increased: Either one of the $z$-continuous
roots of $G_{+}^{\frac{1}{4}}$ or $G_{-}^{\frac{1}{4}}$ tends from
$z^{\frac{1}{4}}$ to the imaginary line as $\N$ (and consequently
$M$) is increased. Note that when deriving the SCE for $\tilde{\mathrm{Ai}}$,
we had assumed $\arg G\neq\pi$, so that $\mathrm{Re}\sqrt{G}>0$
and the integral $\int e^{-\sqrt{G}t^{2}}dt$ converged. This implied
that the arguments of the quartic roots $G^{\frac{1}{4}}$ must be
in the range $\left[-\frac{\pi}{4},\frac{\pi}{4}\right]$. However,
as the argument of one of the roots now tends from $\arg z$ to $\pm\frac{\pi}{2}$,
it at one point leaves this range and renders our assumption wrong,
and the Gaussian identities that we had used are invalid. Conceptually,
looking back to the original integral, now $\mathrm{Re}\sqrt{G}<0$
so one may argue that the Gaussian weight $e^{-\sqrt{G}t^{2}}$amplifies
the $t^{3}$ perturbations which is dominant far from from the origin,
contrapuntal to our perturbative approach. The only remedy to this
situation is to replace the offending root with another solution of
\eqref{Complex_Airy_G_equation} which is contained inside the $\pm\frac{\pi}{4}$
phase cone. For $G_{\pm}^{\frac{1}{4}}$, this is the root whose argument
approaches $\mp\frac{i\pi}{6}$. This means that the SCE cannot reconcile
two conflicting goals: self-consistently eliminating first order corrections,
and reproducing the original integral representation at the low-$\N$\textbackslash{}large-$z$
limit. This is similar to the incompatibility observed in the discussion
of the double-well case in \secref{double_well}.

\begin{figure*}[tp]
\centering{}\subfloat{\includegraphics[width=\figurewidth]{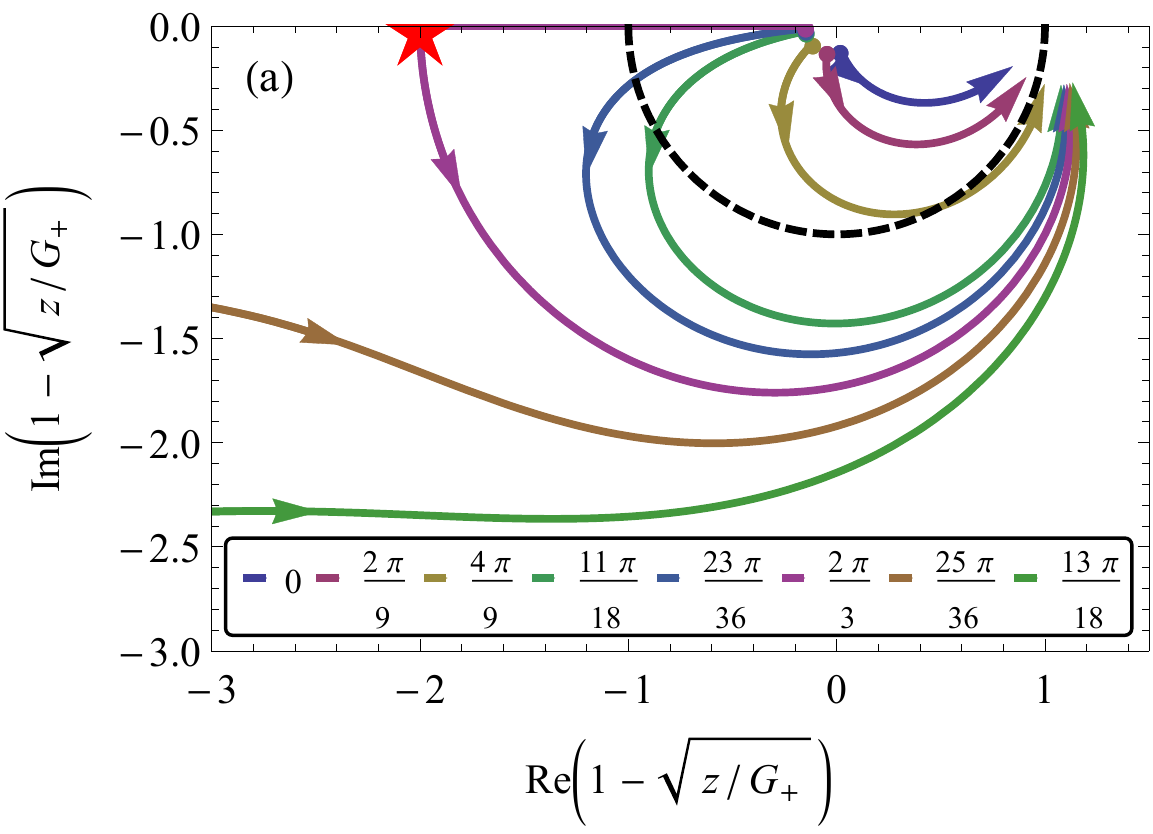}\label{fig:SCE_Airy_near_Stokes_Line_G_Trajectory}}\hspace*{\fill}\subfloat{\includegraphics[width=\figurewidth]{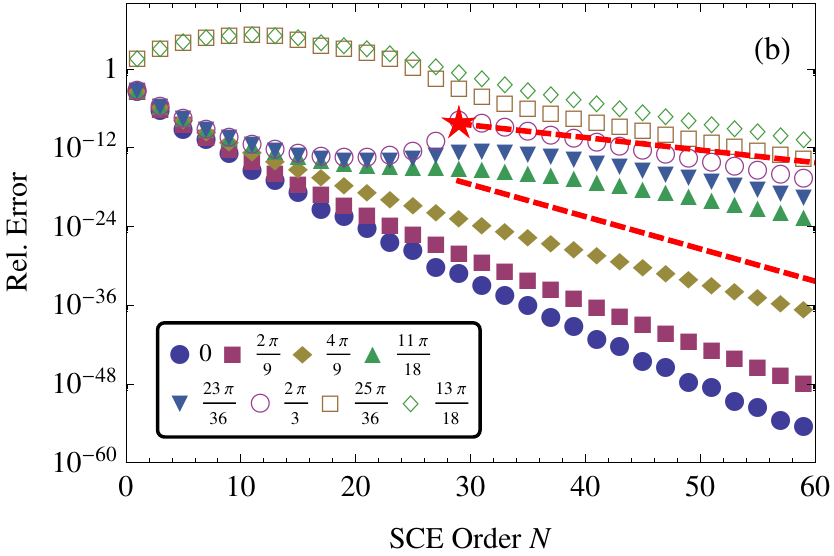}\label{fig:SCE_Airy_Eye}}\caption[Figure \thefigure]{Behavior of the SCE in the complex plane, for fixed $\left|z\right|$
and varying $\arg z$. (a) Trajectories of the factor $1-\sqrt{z}/\sqrt{G_{+}}$,
as $M$ is increased up to $10^{4}$, for the listed values of $\arg z$
and $\left|z\right|=8$. The unit circle is plotted for reference.
As the argument of $z$ is increased, the trajectories sweep outwards.
For $\left|\arg z\right|<\frac{\pi}{3}$, the trajectories are contained
inside the unit circle. For $\frac{\pi}{3}<\left|\arg z\right|<\frac{2\pi}{3}$,
they exit it and eventually converge to $+1$ from outside. However,
the closer $\arg z$ is to $\frac{2\pi}{3}$, the ``sharper'' the
angle of protrusion through the circle, and the further outside the
trajectory reaches before attaining its maximal magnitude. At $\arg z=\frac{2\pi}{3}$,
the trajectory heads radially outwards on the negative real axis before
turning sharply (red star) at $\left(-2\right)$, a value independent
of $\left|z\right|$ which corresponds to the scattering point visible
in \figref{SCE_Airy_roots_of_G_arg_120}. This sharp turn occurs at
$M\approx28.09$, as predicted by \eqref{Airy_M_Scatter_point} for
$\left|z\right|=8$. This point corresponds to the cusp in the right
panel (red star). Lastly, for $\left|\arg z\right|>\frac{2\pi}{3}$,
the origins of the trajectories jump abruptly to far away from the
unit circle. Another arrowhead was placed on the points corresponding
to $M=30$ for reference; for any larger $M$, the trajectories are
continuous as a function of $\arg z$. (b) The relative error of the
SCE versus the order $\N$ for the same values of $z$ as in the left
panel, with $M\left(\N\right)=\N$. For all arguments under $\frac{2\pi}{3}$,
the error is roughly identical until order $\sim10$, at which point
the convergence rate is slowed down with increasing phase. As the
argument is increased towards $\frac{2\pi}{3}$, a local bump in the
error forms. For $\mbox{\ensuremath{\arg}}z=\frac{2\pi}{3}$, a cusp
occurs at $\N=29$ (red star). This is the first integer value of
$\N$ for which $M$$\left(\N\right)$ is greater than the value at
which the collision in \figref{SCE_Airy_roots_of_G_arg_120} occurs,
corresponding to the red star in the left panel. After this bump,
the expansion returns to converge exponentially for all $\arg z$.
The errors for $\arg z>\frac{2\pi}{3}$ exhibit the diverging nature
of the Airy SCE in this Stokes wedge at low orders, but eventually
the SCE starts to converge for sufficiently large $\N$. Note the
eye formed for $\N<30$; for $\N>30$, the error transitions smoothly
across both Stokes lines, showing that the SCE smooths out the Stokes
lines at large order. Lastly, the lower and upper dashed lines represent
the uniform exponential convergence component in the $\left|\arg z\right|<\frac{\pi}{3}$
and $\frac{\pi}{3}<\left|\arg z\right|<\frac{2\pi}{3}$ Stokes sectors,
respectively. The lower line was obtained by fitting an exponential
profile to the SCE of $\mathrm{Ai}\left(0\right)$, which does not
exhibit a stretched-exponential component (cf. \figref{SCE_Airy_at_0_error}).
This line bounds the error for all $\left|\arg z\right|<\frac{\pi}{3}$,
while for $\frac{\pi}{3}<\left|\arg z\right|<\frac{2\pi}{3}$, the
error will eventually exceed it once the corresponding trajectory
leaves the unit circle; for example, the yellow diamonds with $\arg z=\frac{4\pi}{9}$
cross this line at order $\N\approx200$. The upper dashed line was
obtained by multiplying the lower line by $2^{\N}$, as a bound on
the factor $\left(1-\sqrt{z}/\sqrt{G_{+}}\right)^{\N}$ in the $\frac{\pi}{3}<\left|\arg z\right|<\frac{2\pi}{3}$
wedge. Note that to a good approximation, the cusp lies on this line
(cf. the left panel). }
\end{figure*}

Now that a single limit is chosen for $G_{\pm}$, we can analyze the
behavior of the factor of $1-\sqrt{z}/\sqrt{G}$, and its impact on
the SCE accuracy, when $M\rightarrow\infty$. These are demonstrated
in Figs. \ref{fig:SCE_Airy_near_Stokes_Line_G_Trajectory} and \ref{fig:SCE_Airy_Eye},
respectively. 

Recalling that $C_{3}\left(M\right)/3M\sim\sqrt{M}$, it asymptotically
behaves as 
\begin{equation}
\left(1-\frac{\sqrt{z}}{\sqrt{G_{\Delta}}}\right)^{\N}\sim\left(1-\frac{\sqrt{z}}{M^{\frac{1}{3}}3^{-\frac{2}{3}}e^{-\frac{\Delta}{3}\pi i}}\right)^{\N}\sim e^{-\left(\frac{9}{\alpha}\right)^{\frac{1}{3}}e^{\frac{\Delta}{3}\pi i}\sqrt{z}N^{\frac{2}{3}}}\,,\label{eq:Airy_stretched_exponent}
\end{equation}
giving us again a stretched exponent. This factor is convergent only
if $\mathrm{Re}\left\{ e^{\frac{\Delta}{3}\pi i}\sqrt{z}\right\} >0$,
or if $\left|\frac{\varphi}{2}+\Delta\frac{\pi}{3}\right|<\frac{\pi}{2}$,
with $\varphi=\arg z$. This would imply that the factors $1-\sqrt{z}/\sqrt{G_{\pm}}$,
which clearly approach $1$ as $M\rightarrow\infty$, do so from within
the unit circle in the complex plane. To satisfy this simultaneously
for both $\Delta=\pm1$, we require that $\left|\varphi\right|<\frac{\pi}{3}$,
revealing the other Stokes line of the Airy function. For larger arguments,
this factor approach $1$ from outside the circle, and competes against
the exponentially convergent series coefficients. However, we note
a difference between the two stokes wedges: 

For $\frac{\pi}{3}<\left|\varphi\right|<\frac{2\pi}{3}$ and large
$z$ (small $M$), $\sqrt{G}$ originate in the vicinity of $\sqrt{z}$,
and this $1-\sqrt{z}/\sqrt{G}$ starts close to zero. This implies
that it steadily grows in magnitude as $M$ increases, eventually
protruding out of the unit circle, after which it reaches some maximal
magnitude greater than unity, and then starts approaching $1$ from
outside. However, this maximal magnitude is bounded by its value for
$\varphi=\pm\frac{2\pi}{3}$. Substituting \eqref{Airy_M_Scatter_point}
back into \eqref{Airy_equation_scatter_on_2pi3_line}, we find that
at the collision $r=3^{-\frac{1}{2}}$, or $\sqrt{G_{+}}=\sqrt{z}/3$,
so that at this point $\left(1-\sqrt{z}/\sqrt{G_{+}}\right)^{\N}=\left(-2\right)^{\N}$
universally for all $\left|z\right|$. Consulting \tabref{critical_alphas},
note that the convergent coefficients are bounded by $10^{-0.305\N}\approx2.02^{-\N}$,
so the error of the expansion at the collision (the cusp visible in
\figref{SCE_Airy_Eye}) is bounded by $(2/2.02)^{\N}\approx10^{-0.004\N}$.
However, the actual exponential convergence rate is roughly $10^{-0.5\N}$,
so the scaling of the cusp is closer to $10^{-0.20\N}$, or $10^{-0.27\left|z\right|^{\frac{3}{2}}}$
if we substitute $\N=M/\alpha$, with $M$ given approximately by
\eqref{Airy_M_Scatter_point}, and $\alpha=\alpha^{*}\approx1$. This
implies that the exponential convergence is always dominant in this
sector, regardless of $\left|z\right|$. 

In stark contrast, for $\frac{2\pi}{3}<\left|\varphi\right|<\pi$,
$\sqrt{G}$ originates near zero and so the factor $1-\sqrt{z}/\sqrt{G}$
is very large for small $M$, before it tends to $1$ at large order.
The larger $\left|z\right|$, the more egregious this problem becomes.
This is essentially an extension to the complex plane of the situation
observed in \eqref{Double_Well_G}: Once again, the SCE cannot simultaneously
reconcile both the demand for self-consistency of $\left\langle t^{2M}\right\rangle $
and the limit of $G\rightarrow z$. Thus, we expect that for $\frac{2\pi}{3}<\left|\varphi\right|<\pi$,
the SCE starts by diverging for small $\N$. At larger orders, the
stretched-exponential behavior of \eqref{Airy_stretched_exponent}
is recovered, which competes against the convergent coefficients,
and the exponential convergence becomes dominant (i.e., the error
becomes monotonically decreasing thereafter) at order $\N\sim\left|z\right|^{\frac{3}{2}}$.

Lastly, since we picked a consistent large-$M$ limit for $G_{\pm}$,
we note this implies that at sufficiently large order, the SCE transitions
smoothly across the $\varphi=\pm\frac{2\pi}{3}$ Stokes line. This
is because at $\N\rightarrow\infty$, the factors $\left(1-\sqrt{z}/\sqrt{G}\right)^{\N}$
are continuous as a function of $z$, producing the stretched exponent
in \eqref{Airy_stretched_exponent} which is entire. It is only once
$\N$ is reduced, that the roots split discontinuously, with $\sqrt{G}\rightarrow\sqrt{z}$
or $\sqrt{G}\rightarrow0$, depending on whether $\arg z$ is larger
or smaller than $\frac{2\pi}{3}$. This occurs abruptly, at the value
of $\N$ at which the ``collision event'' depicted in \figref{SCE_Airy_roots_of_G_arg_120}
takes place, and predicted by \eqref{Airy_M_Scatter_point} to be
at $\N=\frac{1}{\alpha}\times\frac{4}{3}\left|z\right|^{\frac{3}{2}}$.
Of course, the Airy function is smooth across the lines, and it was
also shown \cite{Berry1989a,Berry1989b,olver1990stokes} that the
usual asymptotic expansion is smooth if sufficiently magnified. In
Ref. \cite{Berry1989a}, this smoothing is attained by truncating
the series at its least term, which occurs at $\N=\frac{4}{3}\left|z\right|^{\frac{3}{2}}$
(given by the large-$n$ summand in \eqref{Airy_tilde_standard_asymptotic_form},
or by the difference between the exponents in the two saddles mentioned
above, $\pm\frac{3}{2}z^{\frac{3}{2}}$, as prescribed by \cite{berry1991hyperasymptotics}).
This implies that by increasing the ratio $\alpha$, the SCE can provide
a smooth approximation faster than the standard asymptotic expansion. 

To summarize, we find the following behavior of the SCE, based on
the location of $z$ with respect to the Stokes lines of the Airy
function:

(i) For $\left|\arg z\right|<\frac{\pi}{3}$, the SCE correctly captures
the behavior of $\mathrm{Ai}\left(z\right)$ at any order if $M=\alpha\N$
is chosen appropriately. Convergence is exponential and uniform, with
a rate bounded by $10^{-0.305\N}$ {[}cf. \tabref{critical_alphas}{]},
and supplemented by a convergent $z$-dependent stretched-exponential
$\left|1-\sqrt{z}/\sqrt{G}\right|^{\N}\sim\exp\left\{ -\mathrm{Re}\left(3^{2/3}\alpha^{-1/3}e^{i\pi/3}z^{1/2}\N^{2/3}\right)\right\} $
{[}cf. \eqref{Airy_stretched_exponent}{]}. 

(ii) For $\frac{\pi}{3}\le\left|\arg z\right|\le\frac{2\pi}{3}$,
SCE provides better approximations with increasing $\N$ from as soon
as $\N=0$, but the rate of convergence is diminished once $1-\sqrt{z}/\sqrt{G}$
exits the unit circle. If this protrusion is at a sharp angle, then
the error experiences a bump. However for any $z$, it may be bounded
by $\left|1-\sqrt{z}/\sqrt{G}\right|^{\N}<2^{\N}\approx10^{0.30\N}$
{[}see \figref{SCE_Airy_near_Stokes_Line_G_Trajectory}{]}, which
is weaker than the exponentially convergent component (both its bound
and its rate in practice), and convergence is formally still uniform. 

(iii) For $\frac{2\pi}{3}<\left|\arg z\right|\le\pi$, SCE produces
an initially increasing error at leading orders due to a very large
$\left|1-\sqrt{z}/\sqrt{G}\right|^{\N}$, but once a critical order
$\N_{c}\left(z\right)$ is surpassed, after which the exponentially-convergent
coefficients become dominant, the SCE is again convergent. $\N_{c}$
is monotonic in $\left|\arg z\right|$ and scales as $\sim\left|z\right|^{\frac{3}{2}}$,
as implied by \eqref{Airy_stretched_exponent}.

(iv) At large enough order, the SCE is smoothed, and varies continuously
between the three regimes above. This occurs once $M\left(\N\right)=\frac{4}{3}\left|z\right|^{\frac{3}{2}},$
analogous to the least-term Stokes smoothing at $\N=\frac{4}{3}\left|z\right|^{\frac{3}{2}}$
of traditional asymptotics.

All four regimes are depicted in \figref{SCE_Airy_Eye}. The smoothing
and subsequent exponential convergence at sufficiently large order
imply that the SCE converges uniformly on any disk of finite radius
in the complex plane, i.e. for $\left|z\right|\le R<\infty$. Due
to the duality $g\Leftrightarrow z^{-\frac{3}{4}}$, this corresponds
the to the uniform convergence of the double-well SCE for any finite
$g\ge\epsilon>0$ in \secref{double_well}.

\subsection{Further Numerical Results}

\begin{figure}[tp]
\begin{centering}
\includegraphics[width=\figurewidth]{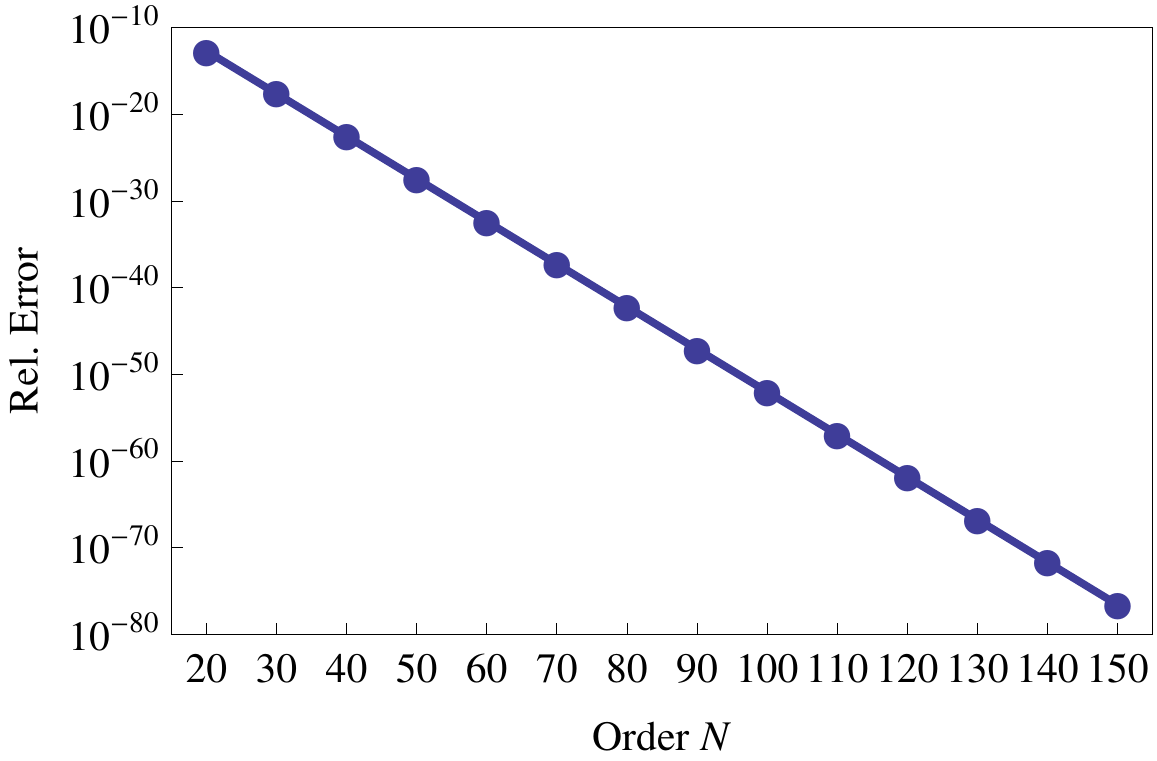}
\par\end{centering}

\caption[Figure \thefigure]{\label{fig:SCE_Airy_at_0_error}Convergence of the SCE to $\mathrm{Ai}\left(0\right)$.
Shown is the relative error of the expansion (dots) versus the order
$\N$, along with an exponential fit, $R^{\left(\N\right)}=10^{B-A\N}$
(solid line). We evaluate the series with $M=\N$. The fitted value
for $A$ is $0.492$, showing that the SCE recovers another significant
digit of $\mathrm{Ai}\left(0\right)$ every other order, greatly exceeding
the convergence rate estimated in \tabref{critical_alphas} for $q=3$. }
\end{figure}

Recalling that $z=0$ corresponds to an infinite anharmonicity {[}cf.
the rhs of \eqref{Airy_representation_integral_cosine}{]}, and so
represents an extreme test case, we first examine the convergence
of the SCE to $\mathrm{Ai}\left(0\right)$. We take $z=0$ as having
phase zero, that is, lying on the positive real line, and so it is
contained in the first Stokes wedge. In fact, evaluating the expansion
at $z=0$ allows us to isolate the uniform exponential convergence
rate without the additional stretched-exponential component, since
$\left(1-\sqrt{z}/\sqrt{G_{\pm}}\right)^{\N}=1$ identically. This
is shown in \figref{SCE_Airy_at_0_error}. Remarkably, while the SCE
is formulated around large $\left|z\right|$, it produces a convergent
sum even at $z=0$, while the comparable asymptotic expansion \eqref{Airy_tilde_standard_asymptotic_form}
has a convergence radius of zero. The relative error of the expansion
is fitted with an exponential trend $\log_{10}R^{\left(\N\right)}=B-A\N$,
reaching a minimal $\chi^{2}=0.03$ for the values $A=0.492$ and
$B=-2.85$. This implies that SCE reproduces a significant digit of
$\mathrm{Ai}\left(0\right)$ every two orders. Furthermore, this rate
is then assisted by a convergent stretched-exponent in the first Stokes
wedge, $\left|\arg z\right|<\frac{\pi}{3}$. 

\begin{figure}[tp]
\begin{centering}
\includegraphics[width=\figurewidth]{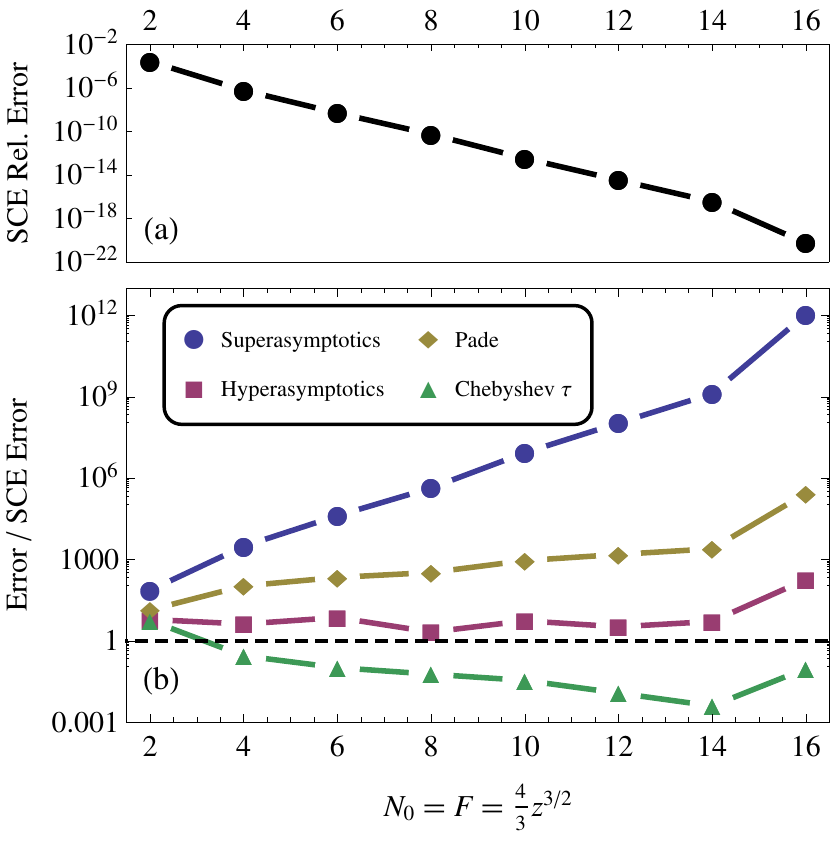}
\par\end{centering}

\caption[Figure \thefigure]{\label{fig:Airy_SCE_vs_Hyperasymptotics}Comparison of the Airy SCE
with other asymptotic methods. (a) The relative error of the Airy
SCE. The horizontal axis is the singulant $F=\frac{4}{3}z^{\frac{3}{2}}$,
and at each point the SCE is evaluated to order $\N_{0}=F$, which
is the same truncation as the superasymptotic scheme. The solid lines
are a guide to the eye. (b) The errors produced by superasymptotics,
hyperasymptotics, Padé approximants and the $\tau$ method, relative
to that produced by the SCE (i.e., the data was normalized by that
shown in the top panel). The solid lines are a guide to the eye. Evaluated
at the same order, the SCE consistently outperforms superasymptotics
and the Padé approximants, and again produces an error very similar
to hyperasymptotics (cf. \figref{SCE_vs_hyperasymptotics} and subsequent
discussion), although the hyperasymptotic series is formally of order
$2\N_{0}$. The Chebyshev $\tau$ method appears to give a few additional
significant digits, but the SCE again converges faster in the large-$\N$
limit --- this is demonstrated in \figref{SCE_Airy_vs_Lanczos_vs_N}.}
\end{figure}

Let us again compare the SCE against the methods of \appref{Competing_Methods}.
The numerical evaluation of the SCE versus super and hyperasymptotics,
as well as the Padé and $\tau$ approximations, is depicted in \figref{Airy_SCE_vs_Hyperasymptotics}.
All methods are evaluated at the same order as the superasymptotic
least-term truncation $\N_{0}=\frac{4}{3}z^{\frac{3}{2}}$, and the
hyperasymptotic trans-series is evaluated up to level three, or less
if it terminates earlier. Similarly to \secref{numerical_results_q_4},
the SCE exhibits performance much better then superasymptotics, and
comparable with hyperasymptotics, though in principle the hyperasymptotic
expansion is of order $2\N_{0}$. 

\begin{figure*}[tp]
\begin{centering}
\subfloat{\includegraphics[width=\figurewidth]{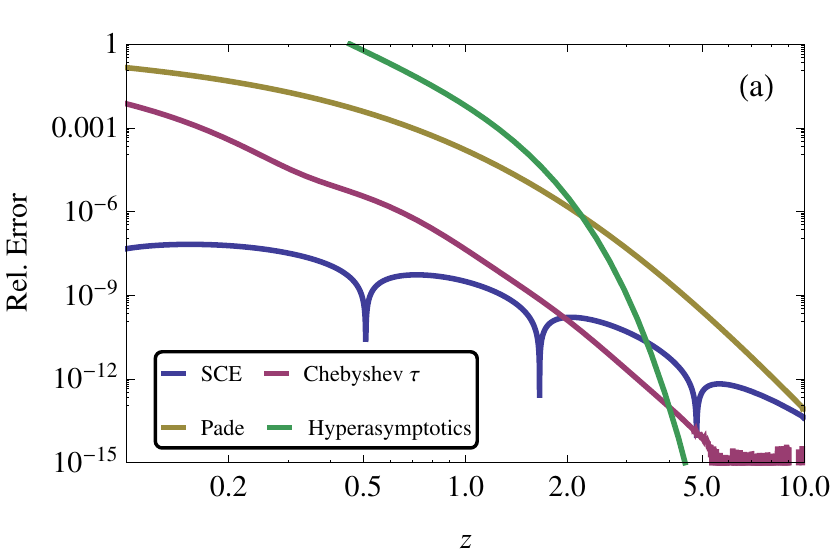}\label{fig:SCE_Airy_vs_Pade_Lanczos_vs_g}}\hspace*{\fill}\subfloat{\includegraphics[width=\figurewidth]{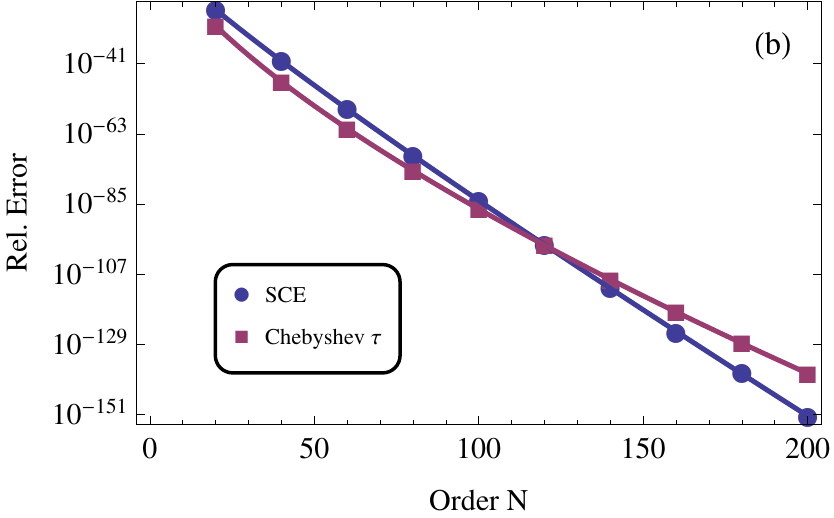}\label{fig:SCE_Airy_vs_Lanczos_vs_N}}
\par\end{centering}

\caption[Figure \thefigure]{Comparison of the Airy SCE versus the Padé and Chebyshev $\tau$
approximations (a) The relative accuracy of the three methods as a
function of $z$. All are evaluated at fixed order $\N=7$. Also plotted
is the estimated best achievable hyperasymptotic error, as predicted
by \eqref{Airy_Hyperasymptotic_Error_Bound}. By construction, the
hyperasymptotic approximation order is $2\N_{0}=2\times\frac{4}{3}z^{\frac{3}{2}}$,
which gives $2\N_{0}=7$ at $z\approx1.9$. Much like the anharmonic
case, SCE dominates the other methods for ``stronger'' couplings
--- this time, at smaller $z$. The noise seen near $z=10$ in the
$\tau$ method is numerical noise due to evaluation at machine precision.
The sharp dips in the SCE error are points where the error switches
signs, i.e. points where $\mathrm{Ai}^{\left(\N\right)}\left(z\right)$
and $\mathrm{Ai}\left(z\right)$ intersect. (b) The relative error
of the SCE and the Lanczos Chebyshev $\tau$ method versus $\N$,
for $z=8$. As before (cf. \figref{SCE_vs_Pade_Lanczos_vs_N}), the
SCE surpasses the $\tau$ method for sufficiently large $\N$. Both
methods were fitted with trend lines of the form $10^{C-A\N-B\N^{\frac{2}{3}}}$.
The SCE yields the values $A=0.492$ and $B=1.446$, greatly exceeding
our minimal bounds of uniform $A=0.305$ and \textbf{$z$}-dependent
\textbf{$B=1.28$}. The $\tau$ method yields a negligible $A=0.007$,
suggesting that it only contains the $10^{-B\N^{\frac{2}{3}}}$ component,
with $B=4.00$.}
\end{figure*}

The SCE is compared more directly with the Padé and $\tau$ schemes
in \figref{SCE_Airy_vs_Pade_Lanczos_vs_g}, where they are all evaluated
at fixed order $\N=7$. The SCE is consistently more accurate than
the Padé approximants, and overtakes the Chebyshev $\tau$ method
for stronger non-linearities (smaller $z$). We again find that even
for larger $z$ the SCE is more rapidly convergent than the Chebyshev
$\tau$ method, as shown in \figref{SCE_Airy_vs_Lanczos_vs_N}.

\section{\label{sec:Multiple-DOF}Extension to Multiple Degrees of Freedom}

In the case of an oscillator in more than one spatial dimension, or
alternatively that of several coupled oscillators, we may examine
a potential of the generic form
\[
V\left(\left\{ x_{i}\right\} \right)=\frac{1}{2}\sum\gamma_{ij}x_{i}x_{j}+\sum g_{ijkl}x_{i}x_{j}x_{k}x_{l}\,,
\]
for which we assume that $\gamma_{ij}$ is a positive definite quadratic
form and $g_{ijkl}$ is a non-negative form, and each coordinate $\left\{ x_{i}\right\} _{i=1}^{\NN}$
takes values on the entire real line\footnote{If $\gamma_{ij}$ is not positive definite, then $g_{ijkl}$ must
never vanish, and the accuracy of the expansion would be minimal at
the ray along which $\gamma\left(\Omega\right)$ is negative and $\frac{g\left(\Omega\right)}{\gamma^{2}\left(\Omega\right)}$
is minimal, which is difficult to estimate in the general case. Uniform
convergence (over the space of possible $g_{ijkl}$) would be lost,
but for a given mass tensor $\gamma_{ij}$, exponential convergence
is attained for any (strictly positive) $g_{ijkl}$ at sufficiently
large $\N$. }. The total partition function for the system would be
\[
\Xi=\int_{-\infty}^{\infty}\prod_{i=1}^{\NN}dx_{i}e^{-\beta\left(\frac{1}{2}\sum\gamma_{ij}x_{i}x_{j}+\sum g_{ijkl}x_{i}x_{j}x_{k}x_{l}\right)}\,.
\]
However, we may transform this integral to radial and angular coordinates
in the $\NN$-dimensional $x$-space, to get 
\begin{align}
\Xi & =\int d^{\NN-1}\Omega\int_{0}^{\infty}dr\,r^{\NN-1}e^{-\beta\left(\frac{1}{2}\gamma\left(\Omega\right)r^{2}+g\left(\Omega\right)r^{4}\right)}\nonumber \\
 & =\int d^{\NN-1}\Omega\left(\beta\gamma\left(\Omega\right)\right)^{-\frac{\NN}{2}}\Z_{\NN-1}\left(\frac{g\left(\Omega\right)}{\beta\gamma^{2}\left(\Omega\right)}\right)\,,\label{eq:many_body_partition_function}
\end{align}
with $\Omega$ representing all the angular dependencies, and $\Z_{\NN-1}$
defined similarly to $\Z$ in \eqref{Z_definition}, apart from the
change of integration limit and measure by $\int_{-\infty}^{\infty}dx\rightarrow\int_{0}^{\infty}x^{\NN-1}dx$.
$\gamma\left(\Omega\right)$ and $g\left(\Omega\right)$ are the quadratic
and quartic coefficients of the potential along the ray defined by
$\Omega$, which by assumption are positive and non-negative, respectively.
By satisfying first-order self-consistency for $\left\langle r^{2M}\right\rangle $,
we find that 
\[
\Z_{\NN}^{\left(\N\right)}\left(g\right)=\frac{1}{2}\left(\frac{2}{G}\right)^{\frac{\NN+1}{2}}\sum_{n=0}^{\N}\left[1-\frac{1}{G}\right]^{n}\sum_{l=0}^{n}\binom{n}{l}\left(-1\right)^{l}\frac{\Gamma\left(n+l+\frac{\NN+1}{2}\right)}{n!\left(M+\NN+2\right)^{l}}\,,
\]
and the SCE of the partition function is 
\begin{equation}
\Xi^{\left(\N\right)}=\int d^{\NN-1}\Omega\left(\beta\gamma\left(\Omega\right)\right)^{-\frac{\NN}{2}}\Z_{\NN-1}^{\left(\N\right)}\left(\frac{g\left(\Omega\right)}{\beta\gamma^{2}\left(\Omega\right)}\right)\,.\label{eq:many_body_SCE}
\end{equation}

For finite $\NN$ (i.e., $\NN$ which does not scale with $\N$),
the convergence of the expansion will not be adversely impacted by
the additional $\NN$-dependent factors. In particular, if the SCE
parameter $M$ is chosen such that $\Z_{\NN}^{\left(\N\right)}$ is
exponentially convergent (e.g., $M+\NN=\alpha\N$), then the angular
integration in \eqref{many_body_SCE} adds a numeric factor\footnote{Some $\left(\text{const.}\right)^{\NN}$ corresponding to the area
of the $\NN$-sphere; note that the area of a unit $\NN$-sphere scales
as $\sim\frac{\pi^{\NN/2}}{\Gamma\left(\NN/2\right)}\sim\left(\frac{2\pi e}{\NN}\right)^{\NN/2}$
which is actually decreasing with $\NN$.} which depends on $\NN$, but the error would still be exponentially
decaying in $\N$ for any strength of the anharmonicity. Indeed, the
numerically efficacy of a similar scheme was recently demonstrated
without proof for the $O\left(n\right)$ model in zero dimensions
\cite{RosaFariasRamosOnZeroDim}.

The case of many-body problems (i.e., $\NN\gg1$) requires greater
care, as we expect that the rate of convergence of the expansion with
$\N$ would depend on $\NN$. We leave this subject for future work.

\section{\label{sec:Conclusions-and-Outlook}Conclusions and Outlook}

In this paper we have investigated the analytical properties of the
SCE by applying it to the toy model of the classical anharmonic oscillator
in thermal equilibrium. We utilized the benefit of an explicit closed-form
expansion to show that for this model the SCE is exponentially and
uniformly convergent for any positive power $q$ and strength $g$
of the anharmonicity $gx^{q}$, compared with standard perturbation
theory \revised{which is rapidly divergent and must rely on resummation}.
We put analytical bounds on the remainder of the expansion at any
given order, allowing us to identify the space of expansion parameters
that guarantees convergence. Remarkably, we argued that the expansion
remains (non-uniformly) convergent for double-well (negative quadratic
coupling) potentials, \revised{even if it is not performed about the double-well minima, but rather
around a harmonic reference}. We also provide an estimate for the optimal choice of a self-consistent
quadratic coupling, and the minimal rate of convergence obtained for
this value. Lastly, we briefly explored the complex-plane behavior
of the SCE by applying it to the Airy function $\mathrm{Ai}\left(z\right)$,
where we have seen the effect of the Airy function's Stokes lines,
and their expected smoothing at large order. In both cases, the SCE
compares favorably against other numerical and asymptotic methods,
most strikingly in the strong-coupling regime. We have also shown
that convergence carries over to any arbitrary finite number of coupled
oscillators.

Due to its convergence for any $q$, we expect that the SCE should
still be convergent for any analytic perturbing potential $gV\left(x\right)$
which scales polynomially at infinity. It can then be shown that an
$x^{2M}$-consistent SCE is only an implicit function of $g$, depending
explicitly only on $G\left(M,g\right)$. The large-order convergence
rate would then be determined by the asymptotic scaling of $V\left(x\right)$
for $x\rightarrow\infty$. We concede that for all but the simplest
systems, obtaining corrections beyond the first few leading terms
is impractical; despite this fact, we believe that we have demonstrated
the appeal of SCE even at low orders, with its superb numerical accuracy
at small $g$ and improved behavior at large $g$.

We have also elucidated the relationship between the SCE and related
variational schemes, such as ODM, OPT, and LDE. In particular, we
have observed the equivalence between the scaling of $G\left(\N\right)$
in the SCE with that given by the PMS and FAC conditions. However,
the SCE offers a few advantages: (i) The PMS or FAC criteria might
not be solvable exactly, most notably at high order; furthermore,
it is non-trivial that either of them has a unique solution for $G\left(\N\right)$,
or any at all (for example, in Ref. \cite{BuckleyDuncanJonesZeroDimension}
the PMS condition has no solution for $\N$ even, despite the fact
that the approximation is optimal for even $\N$, as shown in \figref{Error_vs_alpha}).
The SCE condition, which is always first order, is easily implemented
and solved. (ii) The SCE condition is physically motivated, allows
desired physical features to be built-in into the approximation, and
permits flexibility in the formulation of the expansion, such as in
the choice of conserved observables. Its strength is exemplified by
the remarkable result that in the SCE, optimal convergence is achieved
repeatedly in the linear scaling $M\left(\N\right)\sim\N$, for any
possible anharmonicity.

These results provide fertile grounds for additional inquiry: We have
seen that for a given system, the SCE is not unique --- neither the
exact scheme of the expansion nor the selection of the self-consistent
parameters are such. Is there any physical significance to preserving
the spatial moments $x^{2M}$? If so, then one would require an \emph{a
priori }method of estimating a suitable choice for $M$ (or perhaps
even $G\left(M\right)$ directly) when the self-consistency criteria
cannot be evaluated explicitly. If not, are there other observables
that would yield better results? A deeper question is whether the
values of $\alpha$ found above are universal, depending only on the
scaling of $V\left(x\right)$ for large $x$; for example, repeating
the calculations for a $d$-dimensional isotropic oscillator with
$r^{q}$ anharmonicities would produce the same values. \iftrue

The next logical extensions to the method are apparent: One is an
expansion in many mutually-coupled degrees of freedom, up to the thermodynamic
limit, in which an expansion of intensive quantities, instead of the
macroscopic partition function, may be preferable. Another is investigating
the SCE of the quantum anharmonic oscillator, for which expanding
the partition function is easiest in the imaginary-time path integral
formulation, making it equivalent to a 1D classical string. This immediately
raises additional ambiguities in the scheme: \else Furthermore, extending
the method to the quantum-mechanical case mentioned above by an imaginary
path integral, several additional ambiguities immediately arise: \fi
The choice of basis for the path-integral (i.e., spatial vs. coherent
representation); the addition of a modified mass as another self-consistent
parameter; and an extended selection of observables to optimize, such
as $\hat{x}^{2M}$, $\hat{p}^{2M}$, and $\hat{\mathcal{H}}_{0,1}^{2M}$,
most of which were equivalent in the classical case, up to a redefinition
of $M$. \iftrue However, in the high-temperature limit, all these
schemes should reduce to the same classical results presented here.
In particular, since the SCE could be formulated for any temperature,
it might bridge the regimes of zero and finite $T$, whose connection
was not apparent in previous treatments of the problem \cite{BenderDuncanJonesLogZ1994,ArvantisJonesParkerLogZ1995}.
Ideally, a combination of both extensions will allow the application
of the SCE to various field-theoretical models. \else Subsequently,
the SCE might be extended to various field-theoretical models. \fi
\revised{Given the success of low-order SCE or the related methods (ODM, OPT,
and LDE) for such systems \cite{schwartz2002stretched,Edwards2002Lagrangian,Li1996vortex,mattinglyStevenson1994,stevenson2013optimization,abdallaSUSYLDE2009,kneur2007emergence,kneur2010NambuJonaLasinio,BraatenFadescuOn2002,YukalovMelting1985,Hirofumi2007LDEIsingNLSM,kneur2003OPTBoseEinsteinBEC,deSouzeCruz2001TransitionTemperatureBoseGas,RosaFariasRamosOnZeroDim},
the prospects for this seem promising.}

Or course, the notion of modifying the studied physical system to
mitigate divergences is nothing new. Examples are the masking of a
bare mass by renormalization \cite{PeskinSchroeder1995}, or the shift
of the oscillation frequency in secular perturbation theory \cite{strogatz}.
While conceptually distinct, in the sense that it removes a series
divergence instead of reining in isolated terms, we would like to
see if the SCE, when applied in these contexts, might shed new light
on these methods. 

\revised{Lastly, similar approaches of expansion about a shifted zeroth-order
system were used for related purposes, such as ensuring the convergence
of the resulting PT \cite{Turbiner1984,Turbiner2005}. Most recently,
Serone, Spada, and Villadoro introduced Exact Perturbation Theory
(EPT) \cite{serone2017power,SeroneSpadaVilladoro2017} an approach
in which the problem is framed as a particular realization in the
parameter space of a more general model with a Borel-resummable PT,
and the coupling-dependent interpolation within this space provides
the non-perturbative contributions. This is comparable to the SCE
parameter $G$, whose $g$ dependence encodes into the expansion information
which is non-perturbative in $g$. However, as this technique relies
on the underlying resummability of the interpolated model, it requires
knowledge of the large-order terms of its PT, whereas the SCE provides
a convergent sequence of approximations. Furthermore, the formulation
of such an EPT is limited by the availability of a resummable analogue,
which is much more restrictive than the flexibility afforded by the
SCE. Nevertheless, these approaches might have a deeper connection
with the SCE. Indeed, the asymptotic nature of PT is well understood,
providing physical insights such as tunneling and instantons \cite{ZinnJustinQFT2002},
resurgent trans-series \cite{Cherman2015}, and so forth. It yet remains
to be seen how these reflect in the analytical structure of the SCE,
which we hope will become clearer in future work.}

We thank Michael V. Berry, Leonid I. Glazman, Eytan Katzav, Michael
Kroyter, and Moshe Schwartz for very helpful discussions. Support
from the Israeli Science Foundation (grant number 227/15), the Israel
Ministry of Science and Technology (under contract number 3-12419),
the German-Israeli Science Foundation (grant I-1259-303.10/2014),
and the US-Israel Binational Science Foundation (grants 2014262 and
2016224), is gratefully acknowledged.

\appendix

\section{\label{app:Direct_1F1_estimation}Direct Estimation of the Quartic
SCE Coefficients}

The specific case of the quartic anharmonicity admits a direct error
estimation by inspection of the SCE series coefficients. Given by
the sum over $l$ in \eqref{SCE_partition}, these coefficients may
be expressed as 
\begin{align*}
S_{n,K} & =\sum_{l=0}^{n}\binom{n}{l}\frac{\Gamma\left(n+l+\frac{1}{2}\right)}{n!\left(-K\right)^{l}}\twocolbr=\frac{\left(-1\right)^{n}\pi}{K^{n}n!\Gamma\left(\frac{1}{2}-2n\right)}{}_{1}F_{1}\left(-n,\frac{1}{2}-2n;-K\right)\,,
\end{align*}
where $_{1}F_{1}\left(a,b;z\right)$ is Kummer's confluent hypergeometric
function \cite[Chap. 13]{NIST:DLMF}. Note that when the parameter
$a$ is a negative integer, this is a finite-order polynomial in $z$.
As this is an alternating polynomial in the variable $K$, it is difficult
to estimate directly; delicate cancellations occur between terms of
similar magnitude and opposite sign. However, we may remedy this situation
by utilizing Kummer's transformation rule \cite[Eq. (13.2.39)]{NIST:DLMF},
\[
_{1}F_{1}\left(a,b;z\right)=e^{z}{}_{1}F_{1}\left(b-a,b;-z\right)\,,
\]
so we find 
\[
S_{n,K}=\frac{\left(-1\right)^{n}\pi e^{-K}}{K^{n}n!\Gamma\left(\frac{1}{2}-2n\right)}{}_{1}F_{1}\left(\frac{1}{2}-n,\frac{1}{2}-2n;+K\right)\,.
\]
This, however, comes at a cost: since now the parameter $a$ is not
a negative integer, this hypergeometric function constitutes an infinite
series. Despite this, it is useful to note that for $a$ that is not
a negative integer, we have an asymptotic expansion for large arguments
\cite[Eq. (13.2.23)]{NIST:DLMF},
\[
_{1}F_{1}\left(a,b;z\rightarrow\infty\right)\sim\frac{\Gamma(b)}{\Gamma\left(a\right)}e^{z}z^{a-b}\,,
\]
 so for very large $K\gg n$, we get the leading order behavior
\[
S_{n,K}\sim\frac{\left(-1\right)^{n}\pi}{\Gamma\left(\frac{1}{2}-n\right)n!}=\frac{\Gamma\left(n+\frac{1}{2}\right)}{n!},
\]
where we have used the reflection property of the gamma function.
This reproduces the bounding form we used to prove case 3 of Proposition~\ref{Proposition_Combined}
at the end of \secref{convergence_q_4}. 

Writing down the sum represented by $_{1}F_{1}$ explicitly, we have
\begin{align*}
S_{n,K} & =\frac{\left(-1\right)^{n}\pi e^{-K}}{K^{n}n!\Gamma\left(\frac{1}{2}-n\right)}\sum_{s=0}^{\infty}\frac{K^{s}\Gamma\left(-n+s+\frac{1}{2}\right)}{s!\Gamma\left(-2n+s+\frac{1}{2}\right)}\twocolbr=\frac{\left(-1\right)^{n}\Gamma\left(n+\frac{1}{2}\right)}{K^{n}n!e^{K}}\sum_{s=0}^{\infty}\frac{K^{s}\Gamma\left(2n-s+\frac{1}{2}\right)}{s!\Gamma\left(n-s+\frac{1}{2}\right)},
\end{align*}
again by the reflection property. Note the sign of the terms entering
the sum as $s$ is increased: Starting from $s=0$, we start with
an initial sign of $\left(-1\right)^{n}$ which holds steady, as all
other factors are positive. Beginning at $s=n+1$, the gamma function
in the denominator starts alternating signs at every increment of
$s$. Lastly, after $s=2n+1$, the gamma function in the numerator
also starts alternating signs, thus canceling the sign oscillations
in the denominator. In total, exactly $n$ sign flips occur --- meaning
that for all $s>2n+1$, the terms are strictly positive. 

Also note that the magnitude of the ratio of gamma functions attains
a minimum at $s\sim\frac{3n}{2}$, since by reflection this ratio
can be expressed as $\sim\Gamma\left(\frac{1}{2}+\frac{n}{2}-\left(s-\frac{3n}{2}\right)\right)\times\Gamma\left(\frac{1}{2}+\frac{n}{2}+\left(s-\frac{3n}{2}\right)\right)$.
This ratio diverges symmetrically about this point. The factor $\frac{K^{s}}{s!}$,
however, is peaked at some finite $s$ and attenuates the gamma ratio
for $s\rightarrow0,\infty$. Thus, we claim that the summand above
has two peaks. They occur for integer values of $s$ which are closest
to the two solutions of
\[
\frac{K\left(n-s-\frac{1}{2}\right)}{\left(s+1\right)\left(2n-s-\frac{1}{2}\right)}=1\,,
\]
 which are given by \bw 
\begin{align}
s_{max}^{\pm} & =\frac{1}{4}\left[\left(4n+1\right)+\left(2K-1\right)-3\pm\sqrt{\left(4n+1\right)^{2}+\left(2K-1\right)^{2}-1}\right]\approx n+\frac{K}{2}\pm\sqrt{n^{2}+\left(\frac{K}{2}\right)^{2}}\,.\label{eq:1F1_smax_approx}
\end{align}
 \ew 

Note that $s_{max}^{-}<n$ and $s_{max}^{_{+}}>2n$, so they are both
outside the interval of alternating signs. Since the peaks are very
narrow (as the summand depends on $s$ exponentially, due to all the
gamma functions), we may approximate the total summation by the contribution
of the summand only at the two peak indices, giving us
\[
S_{n,K}\approx P_{+}+\left(-1\right)^{n}P_{-}\,,
\]
with 
\[
P_{\pm}=\frac{\Gamma\left(n+\frac{1}{2}\right)}{K^{n}n!e^{K}}\frac{K^{s_{max}^{\pm}}\Gamma\left(+\frac{n}{2}\mp\left(s_{max}^{\pm}-\frac{3n}{2}\right)\right)}{s_{max}^{\pm}!\Gamma\left(-\frac{n}{2}\mp\left(s_{max}^{\pm}-\frac{3n}{2}\right)\right)}\,,
\]
and where we have used the reflection property of the gamma function
so that its arguments are positive for both sign cases. 

Since the SCE in \eqref{SCE_partition} is a power series in $\left(1-1/G\right)$,
we argue that its remainder should be on the order of the last term
in the sum. Let us then specifically examine the last coefficient,
$S_{\N,K}$. Recalling that $K=M+2$ and substituting $M=\alpha\N$
into the approximate roots in \eqref{1F1_smax_approx}, we find the
following limits:
\begin{align*}
\ln Q_{-} & \equiv\lim_{\N\rightarrow\infty}\frac{1}{\N}\ln P_{+}\twocolbr=-\frac{1}{2}\left[\alpha+\sqrt{\alpha^{2}+4}+4\mathrm{tanh}^{-1}\left(\frac{\alpha}{2}-\frac{1}{2}\sqrt{4+\alpha^{2}}\right)\right]\,,
\end{align*}
\begin{align*}
\ln Q_{+} & \equiv\lim_{\N\rightarrow\infty}\frac{1}{\N}\ln P_{-}\twocolbr=-\frac{1}{2}\left[2\ln\left(\frac{\alpha}{\sqrt{\alpha^{2}+4}-2}\right)+\alpha-\sqrt{\alpha^{2}+4}\right]\,.
\end{align*}
This implies that the scaling of $S_{\N,K}$ is exponential in $\N$.
To have $S_{\N,K(\N)}\rightarrow0$ for $\N\rightarrow\infty$, we
require that both the above logarithms be negative, i.e. that $Q_{\pm}<1$.
$Q_{+}<1$ for any $\alpha>0$, while $Q_{-}<1$ only for $\alpha$
sufficiently large. This is satisfied for $\alpha>\alpha_{c}$ with
the threshold value 
\[
\alpha_{c}\approx0.895\,,
\]
found numerically. Similarly to \secref{convergence_q_4}, one of
the peaks alternates in sign while the other does not. However, their
cancellation now represents the FAC condition, and not the PMS. Solving
the condition $Q_{-}=Q_{+}$ for $\alpha$ numerically yields 
\[
\alpha^{*}\approx1.325,\quad Q_{\pm}^{\N}=10^{-0.288\N}\,.
\]

The critical values we have obtained are much closer to the actual
values fitted numerically, $\alpha_{c}\approx0.8$, $\alpha^{*}\approx1.33$
and $\log_{10}Q^{*}=-0.283$ (cf. \tabref{critical_alphas}), than
those obtained from our analytical bound, and are essentially the
same values obtained in Ref. \cite{BuckleyDuncanJonesZeroDimension}
for the OPT/LDE, showing that the SCE matches these methods for the
optimal choice of $M(\N)$. Indeed, the last term in the summation
proves to be a much tighter error estimator than the analytical bounds
we have deduced in \secref{convergence_q_4}; this is demonstrated
in \figref{Error_vs_alpha}.

In the case that $M\left(\N\right)\sim\N^{p}$ with $0\le p<1$, we
note that that $\alpha=M/\N$ is monotonically decreasing. For some
$\N$ sufficiently large, $\alpha<\alpha_{c}$ which implies that
$Q_{-}$ becomes larger then unity, and $\left|\left(-1\right)^{n}P_{-}\right|\approx Q_{-}^{\N}\gg1$.
However, since $Q_{+}<1$ for any $\alpha$, then $P_{+}\approx Q_{+}^{\N}\ll1$,
and it cannot cancel out the other peak, regardless of the parity
of $\N$. Thus, the expansion must diverge for this scaling of $M$
with $\N$.

\section{\label{app:Competing_Methods}Summary of Competing Asymptotic and
Numerical Methods}

In \secref{numerical_results_q_4} we compared the SCE with other
asymptotic methods for the case of $g\ll1$, while in the strongly
coupled regime $g\gg1$ we compared it against numerical approximation
schemes. In this Appendix, we will briefly describe each.

\subsection{\label{sub:Superasymptotics}Superasymptotics}

The superasymptotic expansion of $\Z$ is defined by terminating its
usual asymptotic series at its least term \cite{boyd1999devil}. This
truncation usually depends on the value of $g$. We can find the general
superasymptotic form of $\Z$ by standard PT,
\begin{align}
\Z_{SA}\left(g\right) & =\sum_{n=0}^{\N_{0}}\int_{-\infty}^{\infty}e^{-\frac{1}{2}x^{2}}\frac{\left(-gx^{4}\right)^{n}}{n!}dx\nonumber \\
 & =\sqrt{2}\sum_{n=0}^{\N_{0}}\frac{\left(-4g\right)^{n}}{n!}\Gamma\left(2n+\frac{1}{2}\right)\,,\label{eq:Z4_Superasymptotic_Series}
\end{align}
where the cutoff $\N_{0}$ is the index of the least term, which is
the integer nearest to
\[
\frac{\sqrt{64g^{2}+32g+1}-16g+1}{32g}\sim\frac{1}{16g}\,.
\]
This truncation rule will rise naturally in the next scheme, hyperasymptotics. 

A slight improvement can be made to this approximation by considering
the discarded sum beyond $\N_{0}$. Using the large-$\N$ asymptotic
form of the summand, Borel resummation \cite{costin2008asymptotics}
of the remainder may be performed, as worked out in Ref. \cite{Negele1998}. 

For the Airy function, one can use the standard asymptotic expansion
of $\mathrm{Ai}\left(z\right)$ (\cite[Eq. (10.4.59)]{abramowitz1964handbook},
after some manipulation of the Gamma function and removal of the prefactors
between $\mathrm{Ai}$ and $\tilde{\mathrm{Ai}}$),
\begin{equation}
\tilde{\mathrm{Ai}}\left(z\right)\sim\sum_{n=0}^{\infty}\frac{\left(-1\right)^{n}}{9^{n}\left(2n\right)!}\Gamma\left(3n+\frac{1}{2}\right)z^{-\frac{3n}{2}}\,,\label{eq:Airy_tilde_standard_asymptotic_form}
\end{equation}
and again truncate the series at its least term.

\subsection{\label{sub:Hyperasymptotics}Hyperasymptotics of $\Z\left(g\right)$}

Past the optimal truncation of superasymptotics, one may find an asymptotic
expansion for the remainder. Truncation of this series at its least
term will yield a new, smaller remainder. This process can be iterated
systematically to improve upon superasymptotics. This technique it
called hyperasymptotics, and was developed by Berry and Howls \cite{berry1990hyperasymptotics,berry1991hyperasymptotics}.
For a more recent review of resurgent transseries, see Ref. \cite{aniceto2018primer}.

We shall now apply hyperasymptotics to the quartic anharmonic oscillator.
Following \cite{berry1991hyperasymptotics} and using the same notation,
let us compute the integral 
\[
\Z\left(g\right)=\int_{-\infty}^{\infty}e^{-\left(\frac{1}{2}z^{2}+gz^{4}\right)}dz=\int g\left(z\right)e^{-kf\left(z\right)}dz\,,
\]
with $g\left(z\right)=1$, $k=1$, and $f\left(z\right)=\frac{1}{2}z^{2}+gz^{4}$.
The saddle points $z_{n}$ satisfy 
\[
4gz_{n}^{3}+z_{n}=0\,,
\]
and are 
\begin{gather*}
z_{1}=-\frac{i}{2\sqrt{g}},\qquad z_{2}=0,\qquad z_{3}=\frac{i}{2\sqrt{g}}\,,
\end{gather*}
where we have retained a numbering of the saddles similar to \cite[ Sec. 5]{berry1991hyperasymptotics}.
The function $f$ attains the values
\begin{gather*}
f_{1}=f_{3}=-\frac{1}{16g},\qquad f_{2}=0\,.
\end{gather*}
The singulants, defined as $F_{nm}=f_{m}-f_{n}$, are then
\begin{gather*}
F_{21}=F_{23}=-\frac{1}{16g},\qquad F_{13}=0\,.
\end{gather*}

\begin{figure*}[tp]
\begin{centering}
\subfloat{\includegraphics[width=\figurewidth]{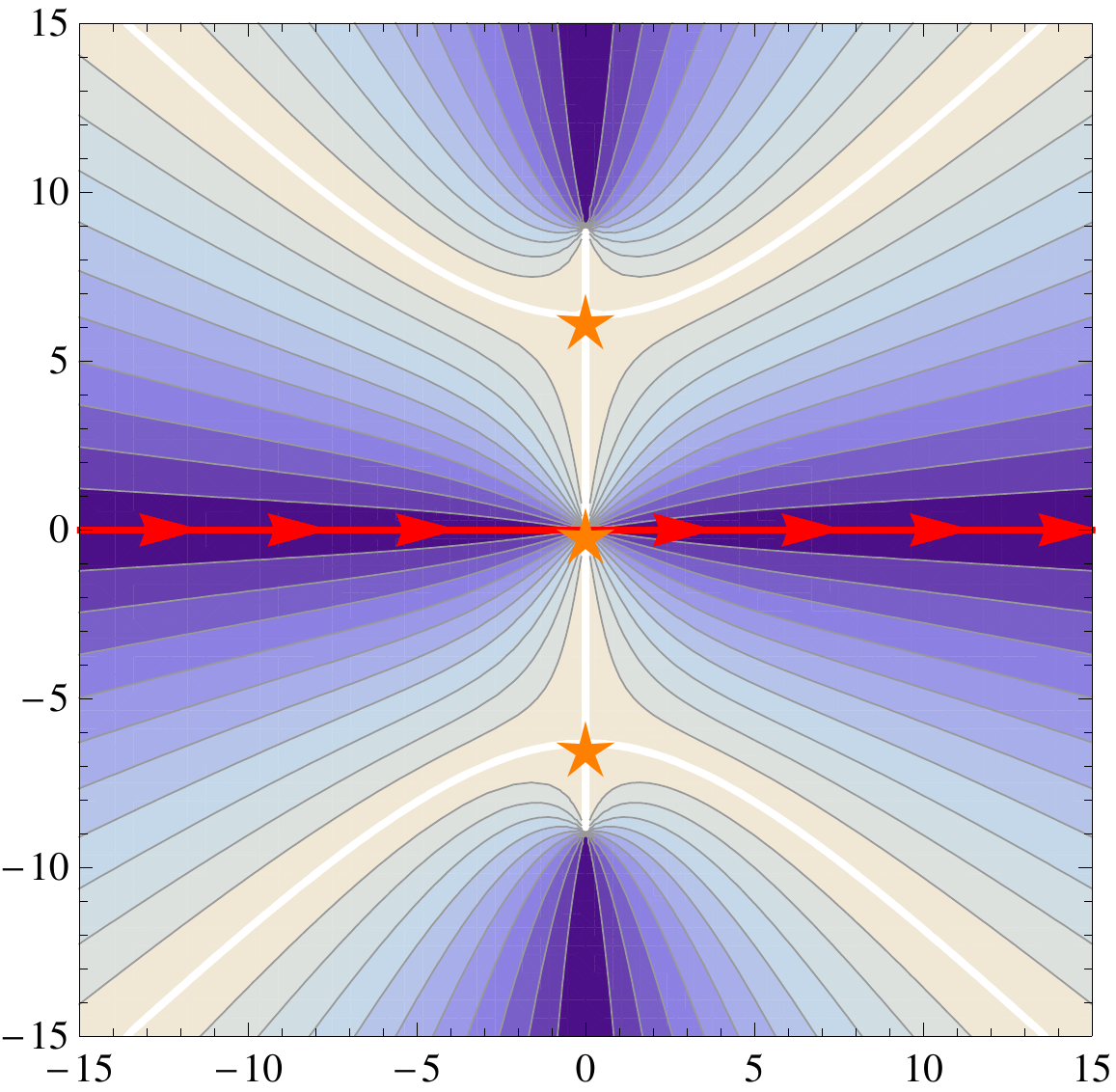}}\hspace*{\fill}\subfloat{\includegraphics[width=\figurewidth]{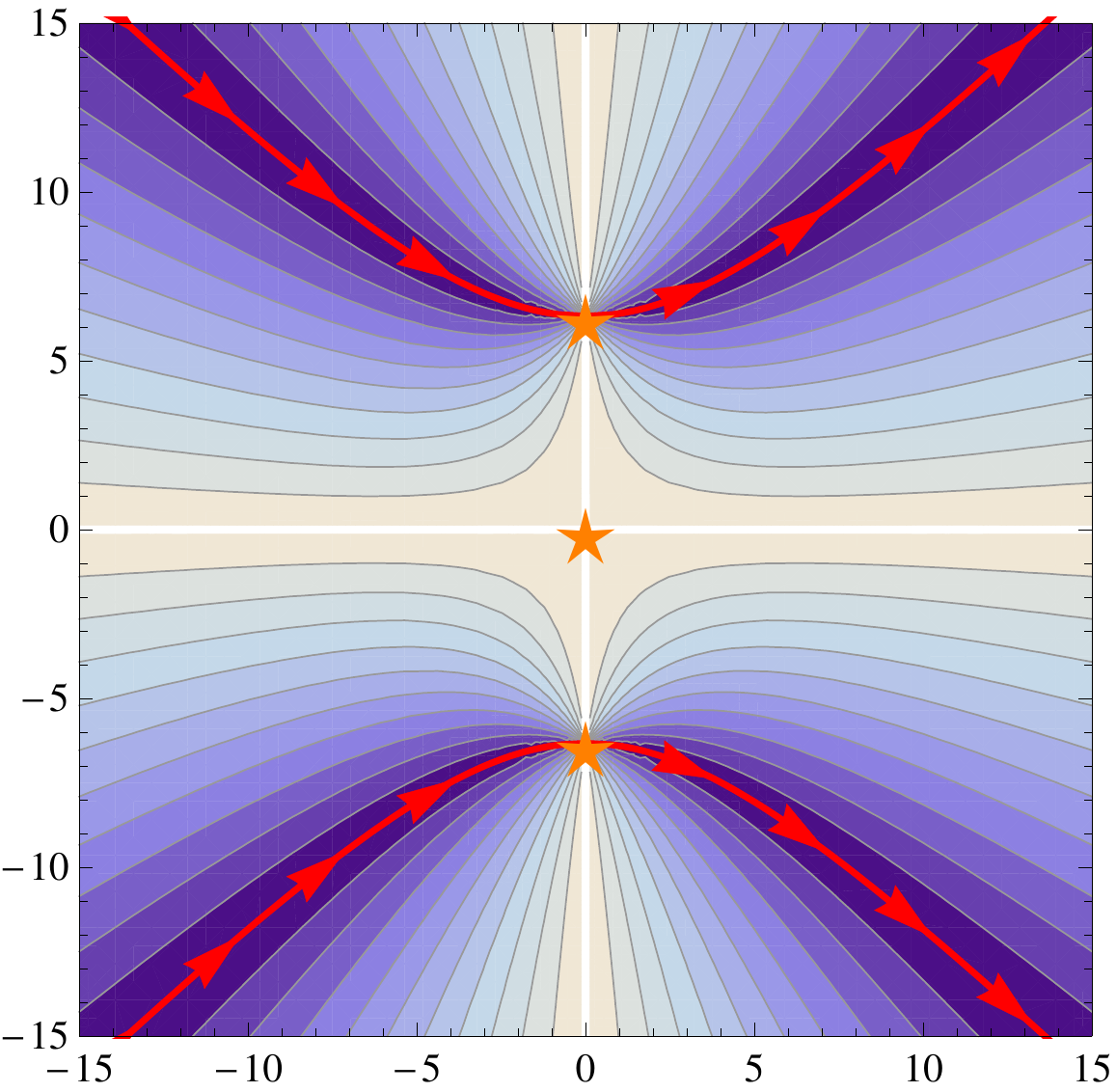}}
\par\end{centering}

\caption[Figure \thefigure]{\label{fig:Hyperasymptotics_Steepest_Paths}Phase contours of $k\left(f\left(z_{n}\right)-f_{n}\right)$.
The orange stars are the saddles $z_{n}$, numbered from bottom to
top. The red lines depict the steepest descent paths through $z_{n}$,
where $k\left(f\left(z_{n}\right)-f_{n}\right)$ is real and positive.
The colored contours map the absolute value of the argument of this
quantity. Left panel: The steepest descent path through $z_{2}$,
with $\theta_{k}=0$. Right panel: The paths through $z_{1,3}$ with
$\theta_{k}=\pm\pi.$ All contours point from left to right. Saddle
$2$ is adjacent to both $1$ and $3$, which are not mutually adjacent. }
\end{figure*}

Note that saddle $z_{2}$ produces the largest ``action'' $f_{2}$.
Inspecting the steepest-descent path for varying $k$ (and assuming
real and positive $g$) going through each saddle, defined by $k\left(f\left(z_{n}\right)-f_{n}\right)=u\in\mathbb{R}^{+}$
(and obtained by solving the ODE $\frac{dz}{du}f\left(z\left(u\right)\right)=1$
with $z\left(0\right)=z_{n}$ as a boundary condition), we find \bw
\begin{gather}
z_{1}\left(u\right)=-\frac{1}{2}\sqrt{\frac{\pm4\sqrt{gu/k}-1}{g}},\qquad z_{2}\left(u\right)=\pm\frac{1}{2}\sqrt{\frac{\sqrt{16gu/k+1}-1}{g}},\qquad z_{3}\left(u\right)=+\frac{1}{2}\sqrt{\frac{\pm4\sqrt{gu/k}-1}{g}}\,,\label{eq:hyperasymptotic_contours}
\end{gather}
 \ew where the top and bottom signs correspond to the two halves
of the steepest path, though the directions of the contours are yet
undetermined. Taking the limit of $u\rightarrow+\infty$ with $k=1$,
we see that all paths end at complex infinity,
\begin{gather*}
z_{1}\left(u\right)\rightarrow\left\{ -\infty,-i\infty\right\} ,\ z_{2}\left(u\right)\rightarrow\left\{ -\infty,\infty\right\} ,\ z_{3}\left(u\right)\rightarrow\left\{ \infty,i\infty\right\} \,.
\end{gather*}
The original integration contour matches the steepest-descent path
of saddle $2$. This means that we will approximate $\Z\left(g\right)$
by the hyperasymptotic expansion about this point (i.e., all ``multiple
scattering paths'', as described in \cite{berry1991hyperasymptotics},
begin at saddle $2$). The steepest-descent paths are illustrated
in \figref{Hyperasymptotics_Steepest_Paths}. 

We shall now follow the formalism of \cite{berry1991hyperasymptotics}
to compute the hyperasymptotic expansion. With $\theta_{k}\equiv\arg k$,
we define our desired contour $C_{2}\left(\theta_{k}=0\right)$ as
running on the real axis from left to right. In order to calculate
the asymptotic series $T_{r}^{\left(2\right)}$ about $z_{2}$ (i.e.,
\cite[Eq. (8)]{berry1991hyperasymptotics}), we must define the root
$\left[k\left(f\left(z\right)-f_{2}\right)\right]^{\frac{1}{2}}=\left[kz^{2}\left(\frac{1}{2}+gz^{2}\right)\right]^{\frac{1}{2}}$
so that it has phase $0$ on the outgoing edge and $\pi$ on the incoming
edge of $C_{2}\left(\theta_{k}\right)$. We thus define
\begin{equation}
\forall z\in C_{2}:\qquad\left[k\left(f\left(z\right)-f_{2}\right)\right]^{\frac{1}{2}}\equiv k^{\frac{1}{2}}z\sqrt{\frac{1}{2}+gz^{2}}\,,\label{eq:sqrt(f(z)-f2)_on_C2}
\end{equation}
where $k^{\frac{1}{2}}$ is taken on the smooth manifold of the square
root (i.e., it is continuous in $k$ and experiences no branch cuts,
and thus is multi-valued), while $\sqrt{\frac{1}{2}+gz^{2}}$ is taken
with respect to the principal branch of the square root, with a cut
along the negative real axis. As such, as a function of $z$, the
expression above is analytic on $C_{2}$, and specifically, only has
branch cuts stemming from $\sqrt{2}z_{1}$ and $\sqrt{2}z_{3}$ along
the imaginary line, both pointing away from $z_{2}$. 

The asymptotic expansion around $z_{2}$ is then 
\begin{align}
T_{r}^{\left(2\right)} & =\frac{\left(r-\frac{1}{2}\right)!}{2\pi i}\oint_{z_{2}}\frac{\left[\left(f\left(z\right)-f_{2}\right)\right]^{\frac{1}{2}}dz}{\left[f\left(z\right)-f_{2}\right]^{r+1}}\nonumber \\
 & =\frac{\left(r-\frac{1}{2}\right)!}{2\pi i}\oint_{0}\frac{z\sqrt{\frac{1}{2}+gz^{2}}dz}{\left[z^{2}\left(\frac{1}{2}+gz^{2}\right)\right]^{r+1}}\nonumber \\
 & =\frac{\left(r-\frac{1}{2}\right)!}{2\pi i}2^{\frac{1}{2}+r}\oint_{0}\sum_{t=0}^{\infty}\binom{-\frac{1}{2}-r}{t}\left(2g\right)^{t}z^{-1+2t-2r}dz\nonumber \\
 & =\sqrt{2}\left(-4g\right)^{r}\frac{\Gamma\left(2r+\frac{1}{2}\right)}{r!}\,.
\end{align}
As expected, $T_{r}^{\left(2\right)}$ reproduces the asymptotic expansion
obtained by a direct calculation in Subsection \ref{sub:Superasymptotics}.

The first hyperseries iteration looks at the remainder of the sum
$T^{\left(2\right)}\left(k\right)=\sum_{r}k^{-r}T_{r}^{\left(2\right)}$.
This requires the adjacency topology of the saddles, that is, the
saddle parings $n,\,m$ for which exists some $k$ that rotates $n$'s
steepest-descent path so it reaches $m$. Since $2$'s steepest-descent
path splits the complex plane into its top and bottom halves, no such
path can exist between saddles $1$ and $3$, so they are not adjacent.
However, $2$ is adjacent to both. It is easily verified that for
$k=-1$ (or in general, $k$ with argument $\pm\pi$), the path $z_{2}\left(u\right)$
in \eqref{hyperasymptotic_contours} coincides partially with $z_{1,3}\left(u'\right)$
for some real $u'<u$ (i.e. $z_{2}\left(u\right)$ ``arrives late''
at the contours $C_{1,3}\left(\theta_{k}=\pm\pi\right)$). For these
arguments of $k$, the path $C_{1}$ has endpoints at $\left(\infty e^{-\frac{i\pi}{4}},\infty e^{-\frac{3i\pi}{4}}\right)$
and $C_{3}$ has endpoints $\left(\infty e^{+\frac{i\pi}{4}},\infty e^{+\frac{3i\pi}{4}}\right)$.
Care must be taken when defining the direction of the contours: $C_{2}$
coincides with $C_{1}$ if we rotate it clockwise, so $k$ has to
rotate in the other direction, meaning we encounter $C_{1}\left(\theta_{k}=+\pi\right)$,
which points from left to right. Due to a similar argument, we encounter
$C_{3}\left(\theta_{k}=-\pi\right)$, also pointing from left to right,
which is obtained by rotating $C_{2}$ counterclockwise. In order
to be consistent with the definition of $\arg k=-\arg F_{nm}$, we
identify 
\[
\arg F_{21}=-\pi,\qquad\arg F_{23}=+\pi
\]
despite the fact that they are numerically equal. The next step, performed
in \cite[Eq. (15)]{berry1991hyperasymptotics}, sees us deforming
the integration contour $\Gamma_{2}\left(\theta_{k}=0\right)$, a
narrow strip that encircles $C_{2}\left(\theta_{k}=0\right)$ counterclockwise,
into the sum of the contours $C_{1,3}$. The orientation anomalies
$\gamma_{nm}$ are defined as $1$ if the deformation of $\Gamma_{n}$
requires reversing the direction of $C_{m}$, and $0$ otherwise.
Both $C_{1,3}$ run from left to right, implying that $C_{1}$ encircles
$C_{2}\left(0\right)=\mathbb{R}$ counterclockwise but $C_{3}$ does
not, so $\gamma_{21}=0$ and $\gamma_{23}=1$. 

The variable transformation \cite[Eq. (16)]{berry1991hyperasymptotics}
$u=v\left(f\left(z\right)-f_{2}\right)/F_{2m}$ does not yet uniquely
define the inverse transformation of $\left[f\left(z\right)-f_{2}\right]^{\frac{1}{2}}$.
Using our definition (\ref{eq:sqrt(f(z)-f2)_on_C2}), one observes
that the following holds:
\begin{align*}
\forall z\in C_{1}\left(+\pi\right):\qquad\arg\left\{ z\sqrt{\frac{1}{2}+gz^{2}}\right\} =-\frac{\pi}{2}\ ,\\
\forall z\in C_{3}\left(-\pi\right):\qquad\arg\left\{ z\sqrt{\frac{1}{2}+gz^{2}}\right\} =+\frac{\pi}{2}\ .
\end{align*}
Thus, since $F_{23}$ and $F_{21}$ have opposite arguments, one has
on $C_{3}\left(-\pi\right)$ that $\left[f\left(z\right)-f_{2}\right]^{\frac{1}{2}}=i\sqrt{\frac{u}{v}\left|F_{23}\right|}=+\sqrt{\frac{u}{v}}\times\sqrt{F_{23}}$,
while on $C_{1}\left(\pi\right)$ one obtains $\left[f\left(z\right)-f_{2}\right]^{\frac{1}{2}}=-i\sqrt{\frac{u}{v}\left|F_{21}\right|}=+\sqrt{\frac{u}{v}}\times\sqrt{F_{21}}$.
Indeed, no additional sign adjustments are necessary. 

Next, we require the coefficients $T_{r}^{\left(1,3\right)}$. For
this we now define
\[
\forall z\in C_{1}\left(\pi\right)\cup C_{3}\left(-\pi\right):\,\left[f\left(z\right)-f_{1,3}\right]^{\frac{1}{2}}\equiv\sqrt{g}\left(z-z_{1}\right)\left(z-z_{3}\right)\ .
\]
This can be verified to give the correct phases: On the outgoing edge
of $C_{1}$, $\arg\,z\sim-\frac{\pi}{4}$, so $k^{\frac{1}{2}}z^{2}$,
with $\arg\,k=\pi$ has phase 0. On the outgoing edge of $C_{3}$,
$\arg\,z\sim\frac{\pi}{4}$ and $\arg k=-\pi$, and again the required
condition holds. The coefficients at $z_{1}$ are then
\begin{align}
T_{r}^{\left(1\right)} & =\frac{\left(r-\frac{1}{2}\right)!}{2\pi i}\oint_{z_{1}}\frac{\left[\left(f\left(z\right)-f_{1}\right)\right]^{\frac{1}{2}}dz}{\left[f\left(z\right)-f\left(z_{1}\right)\right]^{r+1}}\nonumber \\
 & =\frac{\left(r-\frac{1}{2}\right)!}{2\pi i}\oint_{z_{1}}\frac{\sqrt{g}\left(z-z_{1}\right)\left(z-z_{3}\right)dz}{\left[g\left(z-z_{1}\right)^{2}\left(z-z_{3}\right)^{2}\right]^{r+1}}\nonumber \\
 & =\frac{\left(r-\frac{1}{2}\right)!}{2\pi i}g^{-r-\frac{1}{2}}\oint_{0}z{}^{-2r-1}\left(z+z_{1}-z_{3}\right)^{-2r-1}dz\nonumber \\
 & =\frac{\left(r-\frac{1}{2}\right)!}{2\pi i}g^{-r-\frac{1}{2}}\oint_{0}\sum_{t=0}^{\infty}\binom{-2r-1}{t}z{}^{t-2r-1}\left(z_{1}-z_{3}\right)^{-t-2r-1}dz\nonumber \\
 & =g^{-r-\frac{1}{2}}\left(z_{1}-z_{3}\right)^{-4r-1}\frac{\Gamma\left(2r+\frac{1}{2}\right)4^{r}}{r!}\,.
\end{align}
Clearly, for $T^{\left(3\right)}$, the roles of $z_{1}$ and $z_{3}$
are reversed, so we gain a $\left(-1\right)^{4r+1}=-1$ relative sign.
$z_{1}-z_{3}=-\frac{i}{\sqrt{g}}$, so we finally obtain 
\[
T_{r}^{\left(1\right)}=-T_{r}^{\left(3\right)}=i\frac{\left(4g\right)^{r}\Gamma\left(2r+\frac{1}{2}\right)}{r!}\,.
\]
Resurgence is now readily apparent: Up to prefactors, $T_{r}^{\left(1,3\right)}$
are the alternating version of $T_{r}^{\left(2\right)}$. 

Lastly, for the post-leading hyper-iteration, we need to examine the
modified integrals $\int_{C_{1,3}\left(\pm\pi\right)}e^{-f\left(z\right)}dz$.
The saddles $z_{1}$ and $z_{3}$ are not adjacent to each other:
The contour $C_{3}\left(\theta_{k}\right)$ never leaves the top half
of the complex plane, except for $\arg\,k$ which is a multiple of
$2\pi$, when it coincides with half of the real line. The same applies
to $C_{1}$ in the lower half. Thus, they could only meet on the real
line, but this just means that both of them coincide with $C_{2}\left(0\right)$.
This implies that we would deform $\Gamma_{1,3}$ in both cases to
the real line. With both cases $C_{2}$ is obtained by ``rotating''
$C_{1,3}$ back, so both encounter $C_{2}\left(\theta_{k}=0\right)$.
Thus, now the two singulants have the same phase, $\arg F_{12}=\arg F_{32}=0$.
We observe that
\[
\forall z\in C_{2}\left(0\right):\qquad\arg\left\{ \sqrt{g}\left(z-z_{1}\right)\left(z-z_{3}\right)\right\} =0\,,
\]
as the expression in the parenthesis is $z^{2}+\frac{1}{4g}$, which
is real and positive, and no manual sign corrections are necessary,
as $\left[f\left(z\right)-f_{m}\right]^{\frac{1}{2}}=+\sqrt{\frac{u}{v}}\times\sqrt{F_{m2}}$
for $m=1$ and $3$. However, during the deformation of $\Gamma_{1}$
into $C_{2}\left(0\right)$, we had to reverse the direction of $\Gamma_{1}$
($C_{2}\left(0\right)$ encircles $C_{1}\left(\pi\right)$ clockwise).
Thus, we again have an orientation anomaly of $\gamma_{12}=1$. To
summarize,
\begin{gather*}
\gamma_{23}=\gamma_{12}=1,\qquad\gamma_{21}=\gamma_{32}=0\,.
\end{gather*}

The hyperasymptotic trans-series is composed of several hyperseries,
or iterations. Each iteration is the asymptotic expansion of the previous
iteration's remainder, performed around the saddles adjacent to the
saddle about which the previous hyperseries was expanded. This successive
expansion around adjacent saddles is a called a ``scattering path,''
and the trans-series is the summation over all such paths. 

Each hyperseries is terminated earlier than the preceding one, so
the scheme naturally halts after a finite number of terms. We choose
to demonstrate the expansion for $g=\frac{1}{160}$. \cite[Eq. (30)]{berry1991hyperasymptotics}
allows us to determine the truncation of each hyperseries: Saddles
$1$ and $3$ are not adjacent to each other, all the scattering paths
consist of scattering back and forth from saddle $2$. Since the singulants
$\left|F_{12}\right|=\left|F_{32}\right|=\frac{1}{16g}$ are of equal
magnitude, all paths with the same number of scatterings will terminate
at the same order, and at every level the order is halved. We have
\begin{gather}
N\left(2\right)=\frac{1}{16g}=10,\quad N\left(21\right)=N\left(23\right)=5,\quad N\left(212\right)=N\left(232\right)=2,\nonumber \\
N\left(2121\right)=N\left(2123\right)=N\left(2321\right)=N\left(2323\right)=1\,.
\end{gather}
where $N\left(nm\ldots\right)$ is the truncation of the scattering
path that originates at $z_{n}$, scatters to $z_{m}$, and so on. 

Additionally, we note an emerging pattern: (i) For an odd number of
scatterings from saddle $z_{2}$, the scattering chain ends at either
$z_{1}$ or $z_{3}$. At $z_{3}$ we incur a minus sign due to traversing
$\gamma_{23}$, but as $T_{r}^{\left(1\right)}=-T_{r}^{\left(3\right)}$,
these two paths add constructively. (ii) For an even number of scatterings,
the chain ends at $z_{2}$. No matter the path, each edge has been
traversed the same number of times back and forth, so all paths have
gained the same amount of minus signs. Ergo, paths related by exchanges
$1\leftrightarrow3$ again add constructively.

This suggests that one can only sum over the length of scattering
paths, assuming only saddle $z_{3}$ or $z_{1}$ exists, and multiply
by the symmetry factor $2^{\text{number of scatterings to \ensuremath{z_{3}}}}$.

\begin{figure}[tp]
\begin{centering}
\includegraphics[width=\figurewidth]{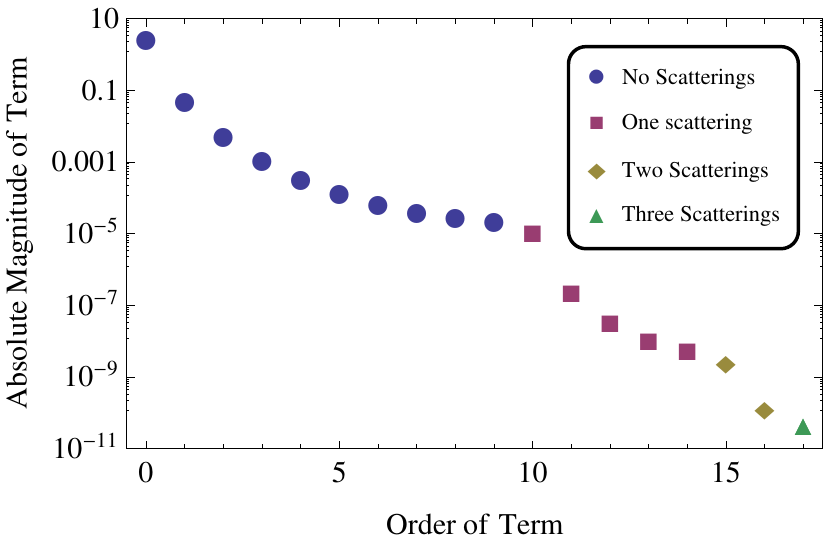}
\par\end{centering}

\caption[Figure \thefigure]{\label{fig:Magnitude_of_terms_in_hyperseries}Plotted are the absolute
values of the successive terms in the hyperasymptotic approximation
of $\Z$, after multiplication by their symmetry factors as in (\ref{eq:Z4_Hyperasymptotics_Hyperseries}),
for $g=\frac{1}{160}$. For this value, the hyperseries naturally
halts after three scatterings. The zeroth and first levels are calculated
exactly, the second level by numerical integration, and the last stage
is approximated. The final relative error of the trans-series after
it is summed is $5.7\times10^{-12}$.}
\end{figure}

The last ingredients we require are the terminants $K_{r}$. The first-level
terminants can be expressed as \cite[Appendix B]{berry1990hyperasymptotics}
\begin{align}
K_{r}^{\left(01\right)} & =\frac{\left(-1\right)^{\gamma_{01}}}{2\pi iF_{01}^{N_{0}-r}}\int_{0}^{\infty}dv_{0}\frac{v_{0}^{N_{0}-r-1}e^{-v_{0}}}{1-v_{0}/F_{01}}\nonumber \\
 & =\frac{\left(-1\right)^{\gamma_{01}+r+N_{0}}}{2\pi i}e^{-F_{01}}\Gamma\left(N_{0}-r\right)\Gamma\left(1+r-N_{0},-F_{01}\right)\nonumber \\
 & =\frac{\left(-1\right)^{\gamma_{01}}}{2\pi i}\left(-e^{-F_{01}}\text{E}_{1}(-F_{01})-\sum_{m=0}^{N_{0}-r-2}\frac{m!}{F_{01}^{m+1}}\right)\,,
\end{align}
with the indices $01$ representing the starting saddle of the path
and the first scattering, so $0\,=\text{saddle 2}$ and $1\,=\text{either \ensuremath{1} or 3}$.
The next terminant $K^{\left(012\right)}$ is found by numerical integration,
while the last required terminant, $K^{\left(0123\right)}$ is approximated
by the same method as in Ref. \cite{berry1990hyperasymptotics}. We
thus obtain
\begin{align}
\Z_{HA} & =\sum_{r=0}^{N\left(2\right)}T_{r}^{\left(2\right)}+2\sum_{r=0}^{N\left(23\right)}K^{\left(23\right)}T_{r}^{\left(3\right)}\twocolbr+2\sum_{r=0}^{N\left(232\right)}K^{\left(232\right)}T_{r}^{\left(2\right)}+4\sum_{r=0}^{N\left(2323\right)}K^{\left(2323\right)}T_{r}^{\left(3\right)}\,.\label{eq:Z4_Hyperasymptotics_Hyperseries}
\end{align}
The relative size of each term in this expansion is depicted in \figref{Magnitude_of_terms_in_hyperseries}. 

For $g=\frac{1}{160}$, the ultimate relative error achieved by this
approximation is about $5.7\times10^{-12}$. This may be compared
with the error predicted by \cite[Eqs. (32) and (34)]{berry1991hyperasymptotics}
\footnote{\cite[Eq. (34)]{berry1991hyperasymptotics} is given for the case
of only two saddles. However, this is obtained by simplification of
a general relation given in terms of the closest pair of adjacent
saddles at each level. Since in our case saddle $z_{2}$ is ``equidistant''
from $z_{1}$ and $z_{3}$ (i.e., $F_{21}=F_{23}$), which are not
mutually adjacent, then this relation holds.}
\begin{equation}
R_{HA}\left(S\right)=\frac{e^{-F}}{\sqrt{2\pi F}}\times\prod_{s=1}^{S}\frac{2^{\frac{s}{2}}}{\sqrt{\pi F}}e^{-\frac{\ln2}{2^{s-2}}F}\times2^{\left\lfloor \frac{S+1}{2}\right\rfloor }\,,\label{eq:Anharmonic_Z4_Hypersymptotic_Error}
\end{equation}
where $F=\left|F_{21}\right|=\frac{1}{16g}$, $S$ the number of scattering
stages, and $k=1$ is assumed. The three factors correspond to the
remainder after the zeroth stage, the successive improvements of each
hyperseries level, and the number of (numerically identical) paths
at the last stage, whose remainders we expect will add constructively.
$S$ of course is determined by $g$ (roughly $\log_{2}\frac{1}{16g}$,
as we saw above). Setting $g=\frac{1}{160}$ and $S=3$ (and dividing
by the exact $\Z\left(g\right)$), we obtain an expected relative
error of $4.0\cdot10^{-12}$, in close agreement with our numerical
result.

It is worth noting that for $g$ much smaller than $\frac{1}{160}$,
evaluation of the complete hyperseries becomes impractical: The approximation
used to determine $K^{\left(0123\right)}$ (and all subsequent terminants)
only gives about 3 correct significant digits of the terminant; thus,
only a couple of additional decimal places of the final result may
be extracted from the third and subsequent stages before this approximation
``washes out'' the true value of $\Z$. 

The hyperasymptotic treatment of the Airy function is simpler since
it only has two saddles. The exponential function over which we integrate
is $f\left(t\right)=\sqrt{z}t^{2}-\frac{i}{3}t^{3}$ {[}per the integral
representation (\ref{eq:Airy_representation_integral_cosine}){]},
which has two saddles, at $t_{1}=0$ and $t_{2}=-2i\sqrt{z}$, and
the singulants are $F_{21}=-F_{12}=\frac{4}{3}z^{\frac{3}{2}}$. The
first saddle is the starting point for the hyperasymptotic scattering
paths, and the expansion coefficients $T_{r}^{(1)}$ are those given
in \eqref{Airy_tilde_standard_asymptotic_form}. It can be shown that
at the other saddle, $T_{r}^{(2)}=i(-1)^{r}T_{r}^{(1)}$ and that
$\gamma_{12}=0$ and $\gamma_{21}=1$. The terminants are calculated
in the same manner as before.

Furthermore, in their introductory paper to hyperasymptotics \cite{berry1990hyperasymptotics},
Berry and Howls specifically analyze the Airy function and give an
error estimate \cite[Eq. (35)]{berry1990hyperasymptotics} ,
\begin{equation}
R_{Hyp}^{\left(\N\right)}=\sqrt{\frac{2}{\pi}}\left|F\right|^{\frac{1}{4}\log_{2}\left|F\right|+\frac{3}{4}-\log_{2}\left(3\sqrt{2\pi}\right)}e^{-\left(1+2\ln2\right)\left|F\right|}\,,\label{eq:Airy_Hyperasymptotic_Error_Bound}
\end{equation}
with $F=\frac{4}{3}z^{\frac{3}{2}}$ the singulant.

\subsection{\label{sub:Lanczos_Tau}Chebyshev Polynomial Approximation by Lanczos's
$\tau$ Method}

With the definition \eqref{Z_definition}, we have 
\[
\frac{d\Z}{dg}=-\int_{-\infty}^{\infty}x^{4}e^{-\left[\frac{1}{2}x^{2}+gx^{4}\right]}dx\,.
\]
 However, performing the rescaling $y=g^{\frac{1}{4}}x$, we have
\[
\Z\left(g\right)g^{\frac{1}{4}}=\int_{-\infty}^{\infty}e^{-\left[\frac{1}{2}g^{-\frac{1}{2}}y^{2}+y^{4}\right]}dy\,,
\]
so we can differentiate w.r.t $w=\sqrt{g}$, obtaining 
\begin{gather}
w^{\frac{5}{2}}\frac{d\Z}{dw}+\frac{1}{2}w^{\frac{3}{2}}\Z=\int_{-\infty}^{\infty}\frac{y^{2}}{2}e^{-\left[\frac{1}{2}w^{-1}y^{2}+y^{4}\right]}dy\,,\nonumber \\
w^{\frac{5}{2}}\frac{d^{2}\Z}{dw^{2}}+3w^{\frac{3}{2}}\frac{d\Z}{dw}+\frac{3}{4}w^{\frac{1}{2}}\Z=\int_{-\infty}^{\infty}\frac{y^{4}}{4w^{2}}e^{-\left[\frac{1}{2}w^{-1}y^{2}+y^{4}\right]}dy\,,\nonumber \\
w^{2}\frac{d^{2}\Z}{dw^{2}}+3w\frac{d\Z}{dw}+\frac{3}{4}\Z=-\frac{1}{4}\frac{d\Z}{dg}=-\frac{1}{4}\frac{d\Z}{dw}\frac{1}{2w}\,\,.
\end{gather}
 One may repose this equation in terms of $\Z\left(g\right)$, which
yields 
\begin{equation}
16g^{2}\Z''\left(g\right)+\left(1+32g\right)\Z'\left(g\right)+3\Z\left(g\right)=0\,.\label{eq:Anharmonic_Z4_ODE}
\end{equation}

With an ODE for $\Z\left(g\right)$ in hand, we would like to integrate
this equation from $g=0$, subject to the boundary condition $\Z\left(0\right)=\sqrt{2\pi}$.
Note that $g=0$ is a singular point of \eqref{Anharmonic_Z4_ODE},
so we do not necessarily require a second boundary condition. 

Next, we outline the procedure of the $\tau$ method, due to Lanczos
\cite{boyd1999devil,boyd2001chebyshev}: To obtain the $\N$-th order
approximation of $\Z$, instead of solving \eqref{Anharmonic_Z4_ODE}
approximately, we will obtain an exact solution to an approximate
equation,
\begin{equation}
16g^{2}\Z''\left(g\right)+\left(1+32g\right)\Z'\left(g\right)+3\Z\left(g\right)=\tau T_{\N}^{*}\left(\frac{g}{s}\right)\,,\label{eq:Anharmonic_Z4_Lanczos_Equation}
\end{equation}
with $T_{\N}^{*}\left(\frac{g}{s}\right)=T_{\N}\left(2\frac{g}{s}-1\right)$
the shifted Chebyshev polynomials of order $\N$; $s\ge g$ a parameter
that stretches the polynomials, which are normally defined on the
interval $\left[0,1\right]$; and $\tau$ a new variable which we
need to solve for. 

This equation may be solved by a power series in $g$. Crucially,
introducing $\tau$ permits a solution which terminates after a finite
power $g^{\N}$. Denoting the coefficients of the power series by
$\left\{ a_{n}\right\} _{n=0}^{\N}$, we have $\N+2$ unknowns (including
$\tau$), and $\N+1$ equations obtained by equating the coefficients
of $\left\{ g^{n}\right\} _{n=0}^{\N}$ on both sides of \eqref{Anharmonic_Z4_Lanczos_Equation}.
Together with the boundary condition $a_{0}=\Z\left(0\right)=\sqrt{2\pi}$,
we have a system of $\N+2$ linear equations. The resulting coefficients
$\left\{ a_{n}\right\} $ are $s$-dependent, and thus we finish the
procedure by choosing $s=g$ so that the entire domain of $T_{\N}^{*}$
is utilized without any ``waste.'' The resulting solution for $\Z\left(g\right)$
is now a rational function of $g$. This $\tau$ approximation can
be computed systematically with ease by any computer algebra software.

In the case of the Airy function, we note that $\mathrm{Ai}\left(z\right)$
satisfies the second-order differential equation \cite[Eq. (10.4.1)]{abramowitz1964handbook}
\[
\frac{d^{2}\mathrm{Ai}\left(z\right)}{dz^{2}}=z\cdot\mathrm{Ai}\left(z\right)\,.
\]
Defining $\tilde{\mathrm{Ai}}\left(z\right)=2\pi z^{\frac{1}{4}}e^{\frac{2}{3}z^{\frac{3}{2}}}\mathrm{Ai}\left(z\right)$
as before, and changing variables by defining $f\left(x\right)=\tilde{\mathrm{Ai}}\left(x^{-\frac{2}{3}}\right)$
(so again $x=z^{-\frac{3}{2}}$), one obtains a new differential equation,
\begin{equation}
5f\left(x\right)+24\left(2+3x\right)f'\left(x\right)+36x^{2}f''\left(x\right)=0\,.\label{eq:Airy_tilde_diff_equation}
\end{equation}

With the boundary condition $f\left(0\right)=\tilde{\mathrm{Ai}}\left(z\rightarrow\infty\right)=\sqrt{\pi}$,
we may apply the $\tau$ method again, and $\mathrm{Ai}\left(z\right)$
is then recovered by the appropriate substitutions.

\subsection{\label{sub:Pad=0000E9-Approximants}Padé Approximants}

With the usual perturbative expansion of $\Z$ in Subsection \ref{sub:Superasymptotics},
we define the Padé approximation of $\Z$ of order $\N$ to be \cite{BenderAndOrszag}
\[
\Z_{\text{Padé}}^{\text{\ensuremath{\left(\N\right)}}}\left(g\right)=\frac{P\left(g\right)}{Q\left(g\right)}\,,
\]
with $P\left(x\right)$ and $Q\left(x\right)$ polynomials of order
$\frac{\N}{2}$ in $g$, such that the expansion of $P/Q$ in powers
of $g$ reproduces the first $\N+1$ terms of \eqref{Z4_Superasymptotic_Series}
(neglecting the truncation at $\N_{0})$. Note that his method can
only be evaluated at even orders: Each polynomial has $\frac{\N}{2}+1$
degrees of freedom, while a total rescaling of both by the same constant
does not change the approximation, so we lose a single degree of freedom
and are left with $\N+1$ degrees, corresponding to the $\N+1$ terms
of the $\N^{\text{th}}$-order asymptotic series. We obtain this approximation
by using \emph{Mathematica}'s \verb|PadeApproximant| routine \cite{Mathematica}. 

This method can be applied just as easily to the perturbative power
expansion of $\mathrm{Ai}\left(z\right)$ in \eqref{Airy_tilde_standard_asymptotic_form}.

\bibliographystyle{apsrev4-1}
\bibliography{sce_pre}
\end{document}